\documentclass{aa}  

\usepackage{graphicx}
\usepackage{txfonts}
\usepackage{lipsum}
\usepackage{subcaption}         % necessary for continued figures, example in section 3
\usepackage{lscape}             % to rotate a single page table, example in appendix.
\usepackage{placeins}           % useful with \FloatBarrier, to keep 
\usepackage{tikz}                               

\usepackage{xspace}
\usetikzlibrary{positioning,arrows.meta} 

\usepackage{multicol, multirow}

\usepackage{mathtools}

\usepackage{courier}
\usepackage{booktabs}
\usepackage{threeparttable}
\usepackage{hyperref}
\newcommand{\PF}{\hyperlink{cite.pezzulli_accretion_2016}{PF16}\xspace}

\newcommand {\Msun}{\,{\mathrm{M}}_\odot}
\newcommand {\kms}{\,{\mathrm{km}}\,{\mathrm s}^{-1}}

\newcommand {\yr}{\,{\mathrm{yr}}}
\newcommand {\Gyr}{\,{\mathrm{Gyr}}}
\newcommand {\pc}{\,{\mathrm{pc}}}
\newcommand {\kpc}{\,{\mathrm{kpc}}}

\newcommand {\dex}{\,{\mathrm{dex}}}

\let\linenumbers\nolinenumbers

\begin{document}

%%%%%%%%%%%%%%%%%%%%%%%%%%%%%%%%%%%%%%%%
% if you use custom commands in your title,
% ensure to check your title when submitting!
%%%%%%%%%%%%%%%%%%%%%%%%%%%%%%%%%%%%%%%%
   \title{The impact of hot mode accretion and angular momentum conservation on metallicity gradients of galactic discs}

\titlerunning{Hot mode accretion and metallicity gradients}

%%%%%%%%%%%%%%%%%%%%%%%%%%%%%%%%%%%%%%%%
% Please separate each author with the \and command
%
% Use the \corrauth to provide the corresponding
% author address. It will be automatically inserted as 
% footnote in the PDF output.
%
% Please DO NOT include ORCIDs next to author names.
% Instead, please provide an active address for each coauthor:
% it will be automatically extracted by EDPS editorial system, 
% and co-authors will be be able to authenticate their ORCID.
%
% Only authenticated ORCIDs will be taken into account.
% ORCIDs included here will be removed.
%%%%%%%%%%%%%%%%%%%%%%%%%%%%%%%%%%%%%%%%

   \author{Jennifer K. S. Friske\inst{1}\corrauth{friske@astro.rug.nl},       % use \corrauth for the corresponding author
         Filippo Fraternali\inst{1} \and
         Gabriele Pezzulli\inst{1}}
        
   \institute{Kapteyn Astronomical Institute, University of Groningen, Landleven 12, 9747 AD Groningen, NL}

   \date{Received, 20XX}

% \abstract{}{}{}{}{}
% 5 {} token are mandatory
 
  \abstract{
 The hot circumgalactic medium (CGM) that surrounds a galactic disc inevitably rotates more slowly than the cold gas in the disc, due to a substantially higher pressure support against gravity. If such a medium accretes vertically onto the disc, it does so with a local deficit of specific angular momentum, thus angular momentum conservation leads to radial flows, advection of metals inwards and the creation or steepening of a metallicity gradient. 
The observed gradient, hence, carries information on the kinematics of the accreting gas.
  Previous models built on this premise parametrised the angular momentum mismatch as a simple function of radius, constant in time, rather than derived from the properties of the CGM itself.
  Here, we present a semi-analytic chemical evolution model of a disc galaxy with a cosmologically motivated evolution, a self-consistent composite, 3D and time-evolving potential and an isothermal rotating CGM. The model includes inside-out formation, the radial flows induced by the deepening of the potential well (acting on gas and stars alike) and those caused by accretion, following their combined effect on the gas-phase metallicity gradient over cosmic time.
 We use a Bayesian framework to fit simultaneously the structural properties and metallicity gradient of a galaxy, inferring both the required inside-out growth and the CGM temperature needed to drive the required radial flows and abundance gradient.
  Fitting our model to the Milky Way, we find a mild preference for a subvirial corona ($T_\mathrm{CGM}/T_{200} = 0.82^{+0.44}_{-0.38}$), in mild tension with observations, and we discuss possible resolutions.
  The predicted CGM rotation velocity varies with radius between $\approx 100-160 \kms$ at the present time and is substantially lower at earlier epochs.
   We further recover an evolution of the gradient over the past $\approx 8 \Gyr$ consistent with observations. Our approach is not restricted to the Milky Way and can be directly applied to external galaxies, offering a new handle to constrain the properties of hot mode accretion onto galactic discs. }


   \maketitle

%%%%%%%%%%%%%%%%%%%%%%%%%%%%%%%%%%%%%%%%%%%%%%%%%%%%%%%%%%%%%%

\section{Introduction}
Nearly all star-forming galaxies form stars at rates that would deplete their gas reservoir in far shorter times than their age, both in the nearby Universe \citep{roberts_content_1963, kennicutt_rate_1983, leroy_star_2008, saintonge_cold_2022} and at high-redshift \citep[e.g.][]{tacconi_phibss_2013, saintonge_validation_2013}. Additionally, analytic galactic chemical evolution (GCE) models have shown that a closed-box model (i.e., without any inflow of gas) leads to a stellar metallicity distribution incompatible with observations \citep{pagel_metal_1975, tinsley_stellar_1979}. Hence, both of these findings imply that galaxies need to accrete gas in order to keep up their star formation.

The source of gas accretion onto galaxies is ultimately the intergalactic medium (IGM), from which gas falls into dark matter halos as they grow and their potential wells deepen \citep[e.g.][]{white_Core_1978,fall_formation_1980}. 
However, the path from there to the central galaxy is still open to debate. 
Relatively small halos with shallow potential wells, typically with virial masses $M_{\rm vir}<10^{12} \,M_{\odot}$, are thought to experience cold-mode accretion, where the filamentary structure of the IGM stretches all the way to the centre of the potential well and feeds the galaxy. 
Conversely, the infalling gas in larger halos gets heated by virial shocks and has to cool down again before it can be accreted by the disc \citep[hot mode accretion; ][]{birnboim_virial_2003, keres_how_2005}.

These virial shocks cause the infalling gas to settle into a hot, ionised corona: the hot circumgalactic medium (CGM). 
The existence of such a medium around the Milky Way (MW) was proposed by \citet{spitzer_possible_1956} and has since been confirmed by X-ray emission and absorption observations, which indicate a temperature of order $2 \times 10^6\,$K \citep{gupta_huge_2012, henley_xmmnewton_2013, miller_constraining_2015} and an extent beyond $100\,$kpc, with some evidence for additional gas phases \citep{zheng_circumgalactic_2015, das_hot_2021}. 
Similar coronae have been detected around external galaxies of comparable or larger masses \citep{zhang_hot_2024}. 
Some of the properties of the hot CGM are still uncertain, in particular its density and metallicity, which are degenerate with one another.
\citet{miller_constraining_2015} found that $Z_\mathrm{CGM} \geq 0.3\,Z_\odot$ is required for the MW's CGM to be compatible with pulsar dispersion measures towards the Large Magellanic Cloud, while, recently, \citet{ponti_abundance_2023} found significantly lower values of $0.05 - 0.07\,Z_\odot$, inferred from the soft X-ray background. 
Despite these uncertainties, it is now established that MW-type galaxies are surrounded by hot media extending to their virial radii \citep{zhang_hot_2024} and containing a large fraction of the baryonic matter associated with their dark matter halos \citep{bregman_extended_2018, faerman_massive_2020, nicastro_xray_2023}.
In the MW, there is also evidence that the hot CGM was in place already at $z\approx 1$--$2$ \citep{gronnow_density_2024}.

Putting together the above pieces of evidence, a natural question regarding gas accretion onto MW-type galaxies is whether it originates from cold clouds that penetrate the hot CGM or from the cooling of the latter.
Cool ($T\sim 10^4\,{\rm K}$) gas clouds are ubiquitously observed in the CGM of nearby galaxies spanning a range of masses and types, through absorption-line observations towards background quasars \citep[e.g.][]{tumlinson_circumgalactic_2017,chen_cosmic_2020}.
While in low-mass (dwarf) galaxies such clouds could be evidence that gas accretion as (parts of) cosmological filaments might survive the passage through the halo (``cold in hot'' mode accretion), this seems improbable for more massive disc galaxies, where infalling clouds are, instead, likely destroyed and incorporated into the hot CGM \citep[e.g.][]{heitsch_fate_2009, afruni_clouds_2023}. 
Several works, including \citet{sancisi_cold_2008}, \citet{diteodoro_Gas_2014}, and more recently \citet{marasco_hi_2025} indeed show that directly observed gas accretion from infalling cold clouds is not enough to sustain the observed star formation. 

In this work, we focus on the remaining possibility: that gas accretion onto disc galaxies takes place from the cooling of the hot CGM. 
Traditionally, it was assumed that the CGM cools spontaneously, forming small clouds via thermal instability \citep[e.g.][]{kaufmann_cooling_2006}.  
The ability of a MW-like halo to host these thermal instabilities has been questioned \citep{binney_highvelocity_2009, hobbs_thermal_2013}, but its overall viability is still a matter of debate \citep{voit_global_2017, sormani_effect_2019}.
More likely, cooling takes place closer to the galactic disc. 
One possible mechanism is the so-called cooling flow, in which the outer parts of the corona lose entropy through radiation, and hence pressure support, causing the gas to flow inwards while approximately conserving its angular momentum \citep[e.g.][]{stern_cooling_2019, hafen_hotmode_2022}.
Alternatively, accretion could be driven by the galactic fountain. Here, cold clouds ejected by supernova feedback mix with the hot gas in the CGM, creating a layer where cooling becomes efficient, so that the newly cooled gas falls back onto the disc together with the fountain material \citep[e.g.][]{fraternali_accretion_2008, marinacci_mode_2010, armillotta_efficiency_2016}.

The above scenarios differ both in the angular momentum that the cooled gas carries when it reaches the disc and in the radial distribution of the accretion, but neither is directly accessible to observations.
They do, however, leave an imprint on the disc that assembles from the accreted gas \citep[e.g.][]{pezzulli_accretion_2016}.
We therefore adopt an indirect approach to better understand the impact of accretion: we build a disc evolution model with varying properties for the accreting CGM and use it to predict easily observable disc properties, such as metallicity gradients.
Fitting these predictions to observations then allows us to infer the properties of the accreting gas.

With the aim of building such a model, we start by listing a number of general expectations: 
\begin{enumerate}
    \item[i)] Being the CGM mostly hot, up to supervirial temperatures of $\approx 2 \times 10^6\,\mathrm{K}$ for the MW \citep{henley_xmmnewton_2013}, it is at least partly pressure supported and hence must be rotating slower than the galactic disc.
    \item[ii)] If the accreting gas that reaches the galactic disc has a lower angular momentum than the disc gas already present at the same radius, angular momentum conservation implies that as accretion mixes with the pre-existing gas, the specific angular momentum at that radius is reduced and this drives radial inflows \citep[already shown e.g. by][]{mayor_effect_1981}.
    \item[iii)] As shown by several GCE models, radial flows have a strong effect on the chemical composition of a galaxy, as the flows advect produced metals with them towards the galactic centre and drive, in conjunction with inside-out growth, the build-up of a metallicity gradient \citep[see e.g.][]{tinsley_evolution_1980, lacey_chemical_1985, goetz_abundance_1992, portinari_radial_2000,  schonrich_chemical_2009, wang_gasphase_2021}.
\end{enumerate}

While these expectations are in general not disputed, to our knowledge, no model so far self-consistently implemented the full chain from i) to iii). Early models that studied accretion with angular momentum conservation focused on assuming a profile of mismatch between the rotation of the disc and the CGM, without taking into account what CGM would produce these. 
\citet{mayor_effect_1981} only considered accretion of non-rotating gas. \cite{lacey_chemical_1985} generalised their equations but did not solve them to derive self-consistent radial flows.
\citet{pitts_chemical_1989}  significantly improved on this and also included in their derivation the option of a rotation curve neither constant in time nor in radius, but in their final calculation only considered models with a spatially linearly increasing CGM rotation curve and constant in time.
Additionally, they noted that radial flows are induced by a deepening gravitational well. As the potential grows, the circular velocity at a given radius increases, and stars and gas conserving their angular momentum have to move inwards to remain on circular orbits. However, they did not include this process in their final calculations.
\citet{bilitewski_radial_2012} revisited these classical works with a numerical chemical evolution model, but parametrised the flows in a linear way as well.  Finally, \citet{pezzulli_accretion_2016}, hereafter \PF, developed a semi-analytic method, extending the analytical approach and integrating the resulting equations along characteristic lines. They did however also focus on an angular momentum mismatch independent of time and radius. 
More recently, \citet{johnson_constraints_2025} studied radial flows induced by angular momentum mismatch (in addition to other sources of radial flows), but again used a constant mismatch in time and radius. They also included potential well deepening as a source of radial flows, but with a simple parametric prescription and did not combine the different mechanisms. 
\citet{vincenzo_chemical_2026} used the method of characteristics in MW chemical evolution models (called there the "Lagrangian approach"), but prescribed a fixed radial velocity instead of an angular momentum conserving approach. 

Most importantly for this paper, none of the above works infer back on the intrinsic properties of the accreting gas itself.
In this work, we take a first step towards this, putting together the analytic derivations of \PF with a cosmologically motivated CGM, as derived in \citet{pezzulli_angular_2017}, in order to constrain the angular momentum of the accreting gas in terms of physical parameters of the CGM. For this first paper, we assume for simplicity that the hot CGM is isothermal and has an exponential angular momentum distribution (AMD).
The latter is a simplified form of the AMDs found using cosmological hydrodynamical simulations by \citet{sharma_angular_2005}.
In addition, we follow the evolution of the full gravitational potential our galaxy through cosmic time. This is, to our knowledge, the first time this mechanisms has been self-consistently included in a chemical evolution model and treated together with other mechanisms driving radial inflows.

Including these sources of radial flows and applying our machinery to the MW data then allows us to infer the angular momentum properties of the CGM as a function of radius and time that best reproduce the properties of MW's disc, including the observed metallicity gradient.

This paper is organised as follows.
In \autoref{sec:model}, we are introducing the semi-analytic model we built and describe the main components as well as the observables it produces. Then we show how this model can be used to fit several parameters guiding the disc evolution in \autoref{sec:fitting}. We show the results of this fit in \autoref{sec:results} and compare it to observations. Finally, we discuss the effect different assumptions of our models and potential extensions in \autoref{sec:discussion} before concluding in \autoref{sec:conclusions}.

\section{Model}
\label{sec:model}
\begin{figure}
    \centering
    \includegraphics[width=1\linewidth]{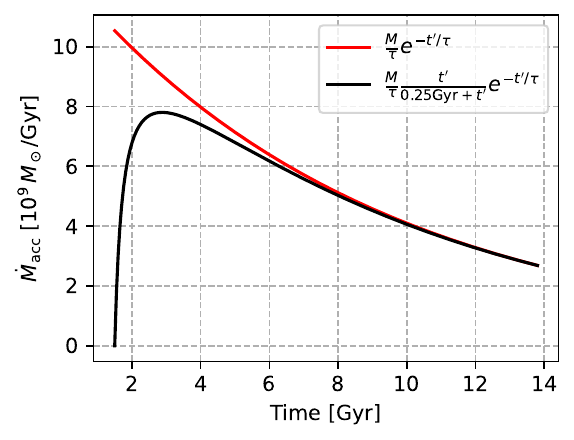}
    \caption{Comparison between the shape of the gas accretion history we chose (black curve) and a classic exponential accretion law (red). Here, $M=6\times10^{10}  M_{\odot}$ and $\tau=11 \Gyr$ For simplicity we set here $t' = (t-t_{0})$, $t_0=1.5\,{\rm Gyr}$.}
    \label{fig:M_acc}
\end{figure}

In this work, we are building a semi-analytic model of a disc galaxy, following the distribution of its stellar and gas components and its interstellar medium (ISM) metallicity over cosmic time.
For this, we start from an analytic description of the basic components of a galaxy (shape and growth of the gas and stellar disc) and make some assumptions about dark matter and baryonic halos in which it is embedded. Specifically, we choose an NFW halo \citep[][]{navarro_structure_1996}, with cosmologically motivated evolution and an isothermal CGM with an exponential AMD.   
A key component of this model is that we impose angular momentum conservation of the accreting gas.  Radial flows then appear naturally if the specific angular momentum of the CGM gas, as it joins the disc, is lower than the specific angular momentum of the disc gas at that radius. As the two gases mix, they flow inward to conserve angular momentum. 
Taking the star-formation history and radial flows together, we can then predict the metals produced at each timestep and the redistribution of these metals through the radial flows, leading to a metallicity gradient. 

The model is built on many of the same assumptions and derivations as \PF and we refer the reader there and to \citet{pezzulli_angular_2016} for more context on some of the underlying derivations.

\subsection{Disc morphology}
\label{sec:discMorph}

    Several theoretical arguments and observations indicate that galaxy discs tend to grow from the inside out \citep[e.g.][]{pezzulli_instantaneous_2015, garcia-benito_spatially_2017, frankel_insideout_2019}.
    Therefore, we must allow our evolving disc to form inside out
    if the data prefer it, while leaving room for the disc to stay constant in size or shrink.
    There are a number of ways to achieve this.
    
    A classic approach is to use accretion timescales that vary with radius \citep[e.g.][also used in \PF]{chiappini_abundance_2001}. This approach naturally produces a dip in the central gas surface density of the galaxy, such as that observed in the Milky Way, especially in the atomic gas phase \citep[e.g.][]{binney_galactic_1998, kalberla_global_2008, marasco_distribution_2017}. 
    Another option is to require the stellar disc to be exponential at all times and to vary the stellar scalelength. This enforces the observed shape of stellar discs, which are, in the majority of cases, close to exponential, and yields realistic star formation rate density profiles \citep[e.g.][]{pezzulli_instantaneous_2015}. Such models, however, admit only a limited parameter space: for sufficiently large radial growth rates, the star formation rate surface density becomes negative at small radii (see equation (2) in \citealt{pezzulli_instantaneous_2015}). 

    In this paper, we make a slightly different choice and require that the shape of the gas disc remains exponential at all times \citep[as proposed by][]{schonrich_chemical_2009}, i.e.
    \begin{equation}
        \label{eq:sigma_gas}
        \Sigma_\mathrm{g}(t,R)=\frac{M_g(t)}{2\pi R_g(t)^2}\,e^{-R/R_g(t)},
    \end{equation}
    where $R_\mathrm{g}(t)$ is the scalelength of the gas disc and $M_\mathrm{g}(t)$ the gas mass. 
    By construction, this always produces a gas distribution that peaks in the centre. 
    Observations of external galaxies tend to support this: while the \ion{H}{i} disc often shows a central dip, this is regularly filled in by $\mathrm{H_2}$, bringing the combined surface density profile close to an exponential \citep[e.g.][]{wong_relationship_2002, bigiel_universal_2012, yim_star_2016}.
    The Milky Way is somewhat of an exception, but it is possible that the inner gas dip there is caused by the Galactic bar rather than by a change in the accretion law. 
    In general, we caution that in galaxies with bar dominated centres, our approach may not fully represent the underlying surface density and should only be fitted to the data outside the bar region (cf. \autoref{sec:fitting}).

    Inside-out formation is then included in our model by varying the gas scalelength $R_g(t)$ over time.
    While a variety of scalelength evolutions could, in principle, be employed, in this work we focus on the simple case of a linearly increasing scalelength:
    \begin{equation}
    \label{eq:R_gas}
      R_g(t)=R_{g,0}+\frac{R_{g,f}-R_{g,0}}{t_f-t_0}\,(t-t_0),
    \end{equation}
    with two free parameters, $R_{g,0}$ and $R_{g,f}$, the scalelengths at the beginning of the evolution, $t=t_0$, and at the end, $t=t_f$, respectively.
    
    We note that times in our model are always in terms of cosmic time, i.e. $t=0$ denotes the Big Bang. We make this choice because we later introduce a dark matter halo with a cosmologically motivated evolution; hence, we need an unambiguous redshift–time correspondence. The starting time for the formation of the galaxies in our model is set to $t_0 = 1.5\,\Gyr$ and we end at $t_f = 13.8\Gyr$.

\subsection{Star formation law}
We are using a classical Kennicutt-Schmidt law \citep[][]{kennicuttjr._global_1998}
    \begin{equation}
        \label{eq:KSlaw}
        \dot{\Sigma}_\mathrm{SF} = A_\mathrm{KS}(1-\mathcal{R})\left(\frac{\Sigma_\mathrm{g}}{1 \Msun \pc ^{-2}}\right)^n \; \Msun \pc ^{-2} \Gyr^{-1},
    \end{equation}
    with a Schmidt-factor $n = 1.4$ and a prefactor $A_\mathrm{KS} = 0.1625$ after the helium correction, meaning that $\Sigma_\mathrm{g}$ in our model signifies the total gas mass. $\mathcal{R} = 0.3$ is the instantaneous return fraction. We are assuming instantaneous recycling in this work, i.e., splitting stars into immortal (70\%) and short-lived stars (30\%) and returning the gas and the produced yields directly to the ISM. These values depend on the choice of the initial mass function (IMF), where 0.3 is consistent with a Salpeter IMF \citep[][]{salpeter_luminosity_1955}. 
    Requiring instantaneous recycling also means that $\dot{\Sigma}_\mathrm{SF}$ is the reduced star formation rate and ${\Sigma}_\mathrm{SF}$ the mass of the gas locked up in stars. 
    
\subsection{Gas accretion}
\label{sec:mass_accretion}
    A key ingredient of our model is gas accretion from the CGM. We chose to parametrise the mass growth of the galaxy, i.e., the total \textit{amount} of gas the galaxy accretes at each timestep, $\dot M_{\mathrm{acc}}(t)$, while leaving the \textit{shape} of the accretion, $\dot \Sigma_{\mathrm{acc}}(t)$, free and determined by imposing angular momentum conservation of the gas. 
    
    In general, the gas mass of the galaxy is determined by the interplay between gas accretion and star formation, and is given by
    \begin{equation}
    \label{eq:Mgas}
    M_g(t)=M_\mathrm{g,0}+M_{\mathrm{acc}}(t)
    -\left(1+\frac{\eta}{1-\mathcal{R}}\right)\int_{t_0}^t\dot M_\star(t')\,\mathrm{d}t',
    \end{equation}
    where $M_\mathrm{g,0}$ is the initial gas mass, while $M_{\mathrm{acc}}(t)$ and $M_\star(t)$ are the total accreted gas mass and stellar mass produced between $t$ and $t+{\rm d}t$, respectively.
    $\eta$ is the mass-loading factor and parametrises a simple stellar feedback, proportional to star formation. Note that we need to include a factor $1/(1-\mathcal{R})$  in eq.~\eqref{eq:Mgas}, as $\eta$ is defined proportional to the total star formation rate, not the reduced one. For the fiducial model, we are only considering $\eta = 0$, i.e. no permanent loss of gas from the galactic disc. We discuss including a non-zero $\eta$ in \autoref{sec:mass_loading}. In principle, it is likely that $\eta$ changes both with time and with radius, but we leave this exploration for future work.

    Our mass accretion history follows a classical exponential law but with a linear increase at early times:
    \begin{equation}
    \label{eq:Macc}
    \dot M_{\mathrm{acc}}(t)=
    \frac{M_\mathrm{1}}{\tau_\mathrm{acc}}\frac{t - t_{0}}{0.25 \Gyr+(t- t_{0})}\,e^{-\frac{t- t_{0}}{\tau_\mathrm{acc}}}.
    \end{equation}
    The value of $0.25\Gyr$ was chosen to prevent a huge spike in gas accretion in the first few timesteps while being small enough that $\dot M_{\mathrm{acc}}$ approaches the exponential accretion history after $\approx {\rm a~few} \Gyr$ (see \autoref{fig:M_acc}).

    In the following, we aim to use observational data to constrain the accretion history of our model galaxy, which would mean fitting the two free parameters in eq.\eqref{eq:Macc}. However, fitting $M_1$ and $\tau_\mathrm{acc}$ directly is impractical, as both affect the total accreted mass at $t_f$ and will hence be partially degenerate. Instead, for each model, we solve for the value of $M_1$ that reproduces a given total baryonic mass of the disc at the end of the simulation, $M_\mathrm{bar} = M_{g}(t_f) + M_\star(t_f)$, and fit $M_\mathrm{bar}$ instead. This leaves $\tau_\mathrm{acc}$ free to solely describe the shape of the accretion. 
    Throughout this paper, 
    we set an initial gas mass $M_\mathrm{g,0}= 10^6 M_\odot$; this choice has only a negligible impact on our calculations.
    
    The above accreted mass is then distributed across the disc such that at any timestep, $\Sigma_\mathrm{g}(t)$ has the shape defined by eqs.\eqref{eq:sigma_gas} and \eqref{eq:R_gas}.
    To achieve this, it is useful to introduce the concept of effective accretion of the gas, as proposed by \citet{pitts_chemical_1989}:
    \begin{equation}
        \label{eq:sigma_eff}
        \dot{\Sigma}_{\mathrm{eff}} \coloneqq  \left(1+\frac{\eta}{1-\mathcal{R}}\right) \dot{\Sigma}_\mathrm{SF} + \dot{\Sigma}_\mathrm{g} 
    \end{equation}
    This quantity is the rate at which gas must be supplied to a given radius to fulfil a certain structural evolution of the disc, irrespective of where that gas comes from. In the absence of radial flows, $\dot{\Sigma}_{\mathrm{eff}} = \dot{\Sigma}_{\mathrm{acc}}$, and accretion alone supplies the surface density required at each radius. We discuss how radial flows change this in \autoref{sec:radial_flows}.

\subsection{Galactic potential and rotation curve}
    \label{sec:potential}
    Most GCE models, including \PF, make the simplifying assumption of a flat and non-evolving circular velocity of the galactic disc. Here, we improve upon these previous works by accounting for both the time and radial dependence of the rotation curve, derived self-consistently in the evolving gravitational potentials of dark matter, stars, and ISM.
    
    For the dark matter component, we assume that it follows an NFW profile \citep[][]{navarro_structure_1996} at each timestep, with mass growth following \cite{mcbride_mass_2009}:
    \begin{equation}
    \label{eq:McBrideMass}
        M_{\mathrm{200}}(z)=M_{200,\mathrm{z=0}}\,(1+z)^{\beta}\,e^{-\gamma z},        
    \end{equation}
    where $M_{200,\textrm{z=0}}$ is the present-day halo mass, $\beta=0.1$, and $\gamma =0.7$.
    
    The concentration parameter follows the parametrisation of \cite{dutton_cold_2014}:
    \begin{equation}
        \log_{10} c_{\mathrm{200}} = a(z) + b(z)\,\log_{10}\,\left(\frac{M_{\mathrm{200}}}{10^{12}\,h^{-1}\,M_\odot}\right),
    \end{equation}
    with $a(z)  = 0.520 + (0.905 - 0.520) \exp(-0.617 z^{1.21})$ and $b(z) = -0.101 + 0.026z$.
    The dark matter contribution to the rotation curve, assuming a spherical halo, is then calculated as 
    \begin{equation}
        v_{\mathrm{DM}}(t,R)=\sqrt{\frac{G\,M_{\mathrm{halo}}(<R)}{R}},
    \end{equation}
    where $M_{\mathrm{halo}}(<R)$ is the dark matter mass contained within a spherical radius equal to $R$, whose shape is given by the NFW profile \citep{navarro_structure_1996}.

    For ease of use, we calculate the gas and stellar contributions to the disc using the python package \texttt{vcdisk}\footnote{\url{https://vcdisk.readthedocs.io/}}, which solves Poisson’s equation for a given baryonic surface density and returns the circular velocity contribution, using the method from \cite{casertano_rotation_1983}.    
   The total circular velocity in the midplane is then given by $v_D^2(t,R) = v_\mathrm{DM}^2(t,R) + v_\mathrm{gas}^2(t,R) + v_\star^2(t,R)$,where $v_\mathrm{gas}^2(t,R)$ and $v_\star^2(t,R)$ denote the contributions from the gas and stellar components, respectively.
   
   Besides the circular speed, in order to set up a self-consistent CGM (section \ref{sec:CGMkin}), we also need to determine the full galactic potential close to and away from the disc. For this, we make direct use of the method laid out in \citet{casertano_rotation_1983} and calculate the full potential distribution of the baryonic disc.
   First, we assume a vertical structure of the gas and stellar disc. Following \citet{vanderkruit_Surface_1981}, we chose 
    \begin{equation}
    \label{eq:vertical_density}
    \rho(R,z) = \Sigma_\mathrm{bar}(R) \, \mathrm{sech}^2\left(-\frac{z}{h_z}\right),
    \end{equation}
    where we set the same radially constant vertical scaleheight for stars and gas: $h_\mathrm{z,gas} = h_\mathrm{z,\star} = 0.3 \mathrm{kpc}$.
    Then, the Poisson equation can be solved for the potential using a Hankel transformation of the density \citep[see the appendix of][]{casertano_rotation_1983}. \footnote{In order to keep the model runtime low enough to keep it computationally feasible to use it in an MCMC, we have to make our numerical grids coarser than desirable for such a transformation. This induces small numerical ringing in the discretised Hankel transform, most noticeably at large radii. We suppress these artefacts with an exponential high-wavenumber filter, ($D(k)=\exp[-(k/k_{\rm cut})^4]$), where ($k_{\rm cut}=0.29\,k_{\rm Nyq}$) and ($k_{\rm Nyq}=\pi/\Delta R$) is the radial Nyquist wavenumber. This is purely a numerical regularisation; we verified that reasonable variations of the filter have a negligible effect on the resulting CGM rotation curves.}

    \subsection{CGM kinematics}
    \label{sec:CGMkin}
    
   \begin{figure}
        \centering
        \includegraphics[width=1\linewidth]{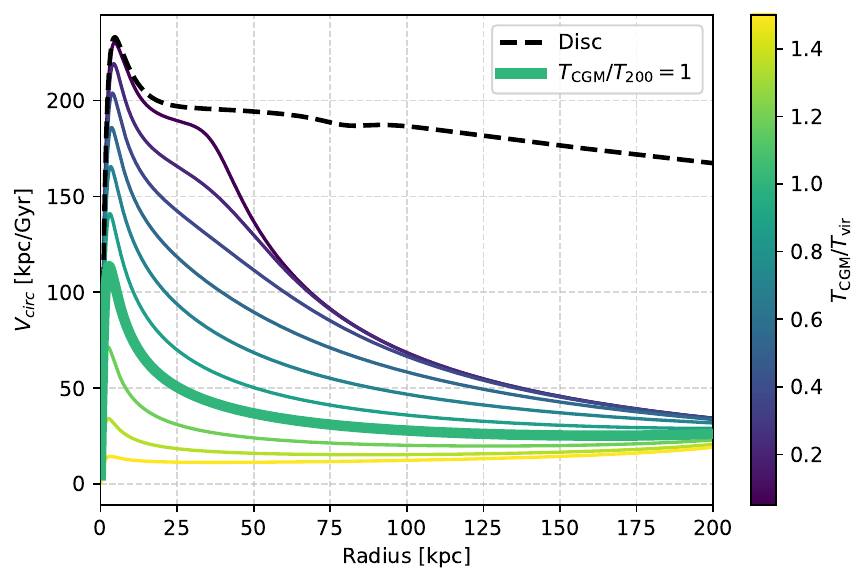}
        
        \caption{The in-plane rotation curve of the CGM at the present time obtained for a galaxy model resembling the MW and different CGM temperatures. The black dashed line shows the disc rotation curve for comparison and the thick green curve the values for $T_\mathrm{CGM} = T_\mathrm{200}$. The parameter values for the underlying model can be found in \autoref{tab:model_values}.} 
        \label{fig:cgm_rotation}
    \end{figure}

    The CGM kinematics are largely unconstrained by observations. 
    \citet{pezzulli_angular_2017} connected the thermal structure of the CGM to the ability of a galaxy disc to continue to grow inside-out. Their idea was that the angular momentum of the gas accreting from the CGM has to be larger than the average angular momentum of the disc for the disc to follow the observed inside-out growth. 
    They concluded that a simple, isothermal model is generally able to sustain such an evolution. We hence set up our CGM in the same way in this work, as an isothermal halo with an exponential AMD. We note that both of these assumptions are simplifications of the general problem, and we return to them in \autoref{sec:varying_the_cgm}.
    
    In the following, we recall and adapt the formulae derived in \citet{pezzulli_angular_2017}, in order to provide a full view of how their model can be adapted to a cosmologically motivated galactic model. 
   The first ingredient is the angular momentum of the CGM. We set its mean specific angular momentum, $\bar{j}_\mathrm{CGM}(t)$, proportional to that of the dark matter halo $\bar{j}_\mathrm{DM}(t)$, for which we adopt the standard relation \citep[e.g.][]{bullock_universal_2001}
      \begin{equation}
      \label{eq:j_CGM}
          \bar{j}_\mathrm{CGM}(t) = f_\mathrm{j}\, \bar{j}_\mathrm{DM}(t)
          = f_\mathrm{j}\, \lambda_\mathrm{spin} \sqrt{2\,G\,M_\mathrm{200}(t)\,R_\mathrm{200}(t)},
      \end{equation}
      where $\lambda_\mathrm{spin}$ is the halo spin parameter, for which we adopt a constant $\lambda_\mathrm{spin} = 0.035$, as found in cosmological simulations \citep[e.g.][]{bullock_universal_2001}. Since $M_\mathrm{200}$ and $R_\mathrm{200}$ grow with time, $\bar{j}_\mathrm{CGM}$ is likewise time-dependent. The proportionality constant $f_\mathrm{j}$ between the halo and the CGM is, in principle, not known. We set $f_\mathrm{j} = 1$ for our models \citep[see e.g.][showing that the angular momentum fraction of the baryons in the disc is clustered around  $f_\mathrm{j, disc} = 1$, with a slight dependence on halo gas fraction]{romeo_specific_2023}, but checked that varying it by a factor of order two does not change our results. 
    
    For the AMD of our CGM, we use a simple exponential prescription with a cut-off:
    \begin{equation}
        \label{eq:psi_of_j}
        \psi (j) = 
        \begin{cases}
        \psi_0  \exp\left({-\omega \frac{j}{\bar{j}_\mathrm{CGM}(t)}} \right) &j\leq j_\mathrm{max}\\
        0 &j>j_\mathrm{max}
        \end{cases}.
    \end{equation}
    \citet{pezzulli_angular_2017} showed that the three normalisation constants $\psi_0$, $j_\mathrm{max}$ and $\omega$ can be expressed in terms of a single parameter $\xi$, see their eq.\ (56), using the boundary condition that the CGM reaches $j_{\max}$ exactly at $R_{200}$. Here, we set $j_{\max} = 3.9 \bar{j}_\mathrm{DM}$, which is the average value found by \citet{bullock_universal_2001}.

    We note that the rotation curve of the CGM does not depend on its total mass: the normalisation $\psi_0$ cancels out in the solution, which is set entirely by the gravitational potential, the temperature of the corona, and the shape of the AMD. A CGM mass would be needed if we wanted to investigate the density profile of the CGM as well. However, as this is not relevant to this work, we choose not to make more assumptions than necessary.

    Finally, we set the temperature of the CGM, which we take to be isothermal at each timestep and equal to a constant fraction of $T_{200}$:
    \begin{equation}
    \label{eq:f_temp}
          f_\mathrm{temp} = \frac{T_\mathrm{CGM}(t)}{T_\mathrm{200}(t)}, \quad \textrm{where} \quad
          T_\mathrm{200}(t) = \frac{\mu_\mathrm{mol} m_\mathrm{p}}{2 k_\mathrm{B}}
                           \frac{G M_\mathrm{200}(t)}{R_\mathrm{200}(t)},
      \end{equation}
    with the Boltzmann constant $k_\mathrm{B}$, the proton mass $m_\mathrm{p}$ and the mean molecular weight of the gas $\mu_\mathrm{mol}= 0.59$.
    $f_\mathrm{temp}$ is the last free parameter of our model, and arguably the most important one, as it directly affects the kinematics of the CGM and hence of the accreting gas.
    Note that \citet{pezzulli_angular_2017} normalised $T_\mathrm{CGM}$ to the halo virial temperature instead. We use $T_\mathrm{200}$ so that all halo quantities are defined with respect to the same overdensity definition.

    Having specified the potential, the angular momentum distribution, and the temperature of our hot CGM, we can determine its rotation. Following \citet{pezzulli_angular_2017}, we assume it to be in hydrostatic equilibrium. Then, we can solve for the midplane density $\rho_0(R)$ and the specific angular momentum (their eq. (59)):
        \begin{align}
              \frac{\mathrm{d}j_\mathrm{CGM}}{\mathrm{d}R}(R) &= 4\pi \chi(R)\, R^2\,
                  \frac{\rho_0(R)}{\psi\big(j_\mathrm{CGM}(R)\big)}, \\
              \frac{\mathrm{d}\rho_0}{\mathrm{d}R}(R) &= \frac{1}{c_\mathrm{s}^2}\,
                  \frac{\rho_0(R)}{R}\left[\left(\frac{j_\mathrm{CGM}(R)}{R}\right)^2
                  - v_D^2(R)\right],
      \end{align}
      where the first equation follows from mass conservation, and the second encodes hydrostatic equilibrium as a balance between the internal pressure of the gas, rotation, and gravity. Here, $v_D$ is the circular speed in the plane as defined above, \mbox{$c_\mathrm{s} = \sqrt{k_\mathrm{B}T_\mathrm{CGM}/(\mu_\mathrm{mol} m_\mathrm{p})}$} is the isothermal sound speed and $\chi(R)$ contains the vertical structure of the corona:
      \begin{equation}
      \label{eq:chi_def}
          \chi(R) = \frac{1}{R} \int_0^{+\infty}
          \exp\left(-\frac{\Phi(R,z) - \Phi(R,0)}{c_\mathrm{s}^2}\right)\mathrm{d}z,
      \end{equation}
    where $\Phi(R,z)$ is the full three-dimensional potential derived in \autoref{sec:potential}, including the contribution of the disc. 
    We solve this system as a two-point boundary value problem, between a small inner radius $R_\mathrm{c}$ and $R_\mathrm{200}$, requiring that $j_\mathrm{CGM}(R_{200}) = j_\mathrm{max}$ and that the small amount of coronal mass enclosed within $R_\mathrm{c}$ is accounted for at the inner boundary \citep[see footnote 5 in][]{pezzulli_angular_2017}.\footnote{Numerically, we do not integrate for $j_\mathrm{CGM}$ directly, but for the fraction of the coronal mass with specific angular momentum below $j$, $C(j) = M_\mathrm{CGM}^{-1}\int_0^{j}\psi(j')\,\mathrm{d}j'$. By construction, $C$ then runs from $0$ to $1$, so the outer boundary condition becomes simply $C=1$, and $\psi$ cancels from the first equation. The two formulations are exactly equivalent and related by the inversion $j_\mathrm{CGM}(C) = -(\bar{j}_\mathrm{CGM}/\omega)\ln\left[1-C\left(1-\mathrm{e}^{-\xi}\right)\right]$.}

    The in-plane CGM circular velocity is then given as
    \begin{equation}
        \label{eq:vCGM}
        v_\mathrm{CGM}(R) = \frac{j_\mathrm{CGM}(R)}{R}. 
    \end{equation}
    We show this rotation curve for different temperatures for a MW-like halo at $z=0$ in \autoref{fig:cgm_rotation}. As expected, the corona rotates more slowly the hotter it is, as it becomes increasingly pressure supported, and its rotation curve flattens out once the corona is supervirial at $f_\mathrm{temp} > 1$.

    \subsection{Self-consistent radial flows}
    \label{sec:radial_flows}
    Angular momentum conservation leads to two distinct sources of radial flows within a galactic disc that evolves according to our assumptions, both of which are self-consistently accounted for in our model. 

    First, as the gravitational potential of the galaxy evolves with the dark matter halo and the disc grows, the circular speed at a certain radius increases. This means that a star or a gas cloud on an orbit with a certain specific angular momentum ($j=R v_\mathrm{rot}$) needs to move further towards the centre, until its rotational velocity $v_\mathrm{rot}$ matches the local circular speed again. This mechanism 
    %affects both stars and gas. It 
    was pointed out by \citet{pitts_chemical_1989}, but most works since then assumed a static potential and hence did not include this source of radial flows.
    Second, gas accreting from the CGM reaches the disc with a specific angular momentum $j_\mathrm{acc}(R)$ that generally is lower than that of the disc material at the radius where it lands. Mixing with the accreted gas therefore changes the specific angular momentum of the gas in the annulus, such that it no longer matches that of a circular orbit there and has to move inwards.
    We note that these two sources are independent: the first mechanism drives radial flows even in a galaxy with corotating disc and CGM, while the second one also works in a static potential.

    We note that in this work, we treat accretion directly from the CGM and thus, the properties of the accreting gas are essentially equal to the properties of the CGM as it joins the disc. 
    Therefore, we set $j_\mathrm{acc}(R,t) \coloneqq j_\mathrm{CGM}(R,t) = R v_\mathrm{CGM}(R,t)$ as defined in eq.~\eqref{eq:vCGM}. In a more general case, one would also have to decide the vertical height in the CGM at which accretion occurs, as the angular momentum distribution and rotation curve can vary with height above the disc. In our case, using an isothermal CGM, this is not the case: according to the Poincaré–Wavre theorem \citet{lebovitz_Rotating_1967, tassoul_Theory_1980}, an isothermal and hence barotropic fluid has to rotate cylindrically. 

    \subsubsection{Stellar redistribution}
    \label{sec:stellar_redistribution}
    Of the two mechanisms mentioned above, the evolving potential also affects the stellar component of the disc, which therefore gains a radial velocity $u_{R\star}$. The continuity equation for the stellar surface density, ${\Sigma}_\star$, then has two contributions: a star formation term, $\dot{\Sigma}_\mathrm{SF}$, and a radial flux term:
      \begin{equation}
          \dot{\Sigma}_\star = \dot{\Sigma}_\mathrm{SF} - \frac{1}{2 \pi R} \partial_R \mu_\star.
      \end{equation}
      Here, $\mu_\star = 2\pi R \Sigma_\star u_{R\star}$ is the radial mass flux of the stars and
        \begin{equation}
        \label{eq:ur_star}
        u_{R\star} = - \frac{\partial_t j_\mathrm{disc}(R,t)}{\partial_R j_\mathrm{disc}(R,t)}
    \end{equation}
    is their radial velocity \citep[see][]{pitts_chemical_1989}, with $j_\mathrm{disc}(R,t) = R \, v_D(R,t)$ . We note that the flux only redistributes stars but does not eliminate them, so the total stellar mass is still given by $M_\star(t) = \int_{t_0}^{t}\mathrm{d}t' \int_0^\infty 2\pi R\, \dot{\Sigma}_\mathrm{SF}(R,t')\,\mathrm{d}R$.
    Integrating over time then gives the stellar surface density,
    \begin{equation}
        \Sigma_\star(R,t) = \int_{t_0}^t \mathrm{d}t' \left[\dot{\Sigma}_\mathrm{SF} - \frac{1}{2 \pi R} \partial_R \mu_\star\right].
    \end{equation}

    Since $\Sigma_\star$ itself contributes to the potential, including this redistribution makes $\Sigma_\star(R,t)$ and $v_D(R,t)$ mutually dependent, and we iterate the model until the relative change in $v_D$ between iterations is less than $10^{-3}$. While this increases the cost of a model evaluation by a factor of $\sim 2$, it ensures that the disc and the potential are self-consistent. Overall, the resulting effect on the stars is small; the radial velocity of the stars peaks early on at about $\approx 1 \kms$ and stays below $0.5 \kms$ for most of the model evaluation (see also \autoref{fig:stellar_redistribution} where we show this effect for an example model).

\subsubsection{Radial flows of the gas}
    In \autoref{sec:mass_accretion}, we defined the effective accretion $\dot{\Sigma}_\mathrm{eff}(R,t)$ in terms of the structural evolution of the disc.\footnote{Note that, unlike in \PF, our definition of $\dot{\Sigma}_\mathrm{eff}$ is in terms of $\dot{\Sigma}_\mathrm{SF}$ and not $\dot{\Sigma}_\star$, as the two are no longer identical once the stars are redistributed.} It determines how much gas each annulus needs, but not where this gas comes from.
    In this section, we can now decompose it into actual accretion from the CGM and gas supplied through radial flows driven by the two mechanisms mentioned above: an evolving potential and the kinematics of the accreting gas.

    We start by expressing $\dot{\Sigma}_\mathrm{eff}$ in terms of the two sources that contribute to it, an accretion term and radial flows:
    \begin{equation}
          \label{eq:sig_eff_vs_sig_acc}
          \dot{\Sigma}_\mathrm{eff} = \dot{\Sigma}_\mathrm{acc} - \frac{1}{2\pi R} \partial_R \mu_\mathrm{g}.
      \end{equation}
    Here, equivalently to the stars above, we define the radial mass flux of the gas as
    \begin{equation}
      \label{eq:mu}
        \mu_\mathrm{g}(R,t) = 2 \pi R \Sigma_\mathrm{g} u_{R,\mathrm{g}},
    \end{equation}
    where $u_{R,\mathrm{g}}$ is the radial velocity of the gas.

    \citet{pitts_chemical_1989} showed that angular momentum conservation allows us to express the radial velocity as (their eq.~6, rewritten in our notation):
    \begin{equation}
        \label{eq:ur}
        u_{R,\mathrm{g}} = - \frac{\dot{\Sigma}_\mathrm{acc} (j_\mathrm{disc} - j_\mathrm{acc}) + \Sigma_\mathrm{g} \partial_t  j_\mathrm{disc}}{\Sigma_\mathrm{g} \partial_R  j_\mathrm{disc}}.
    \end{equation}
    Here, the first term carries the angular momentum mismatch between the disc, $j_\mathrm{disc}$, and the accreting CGM, $j_\mathrm{acc}$, and the second encodes the change of the potential, analogous to eq.~\eqref{eq:ur_star} for the stars. For a constant $v_D(t,R)$, we recover eq.~(20) in \PF.

    Inserting eqs.~\eqref{eq:ur} and \eqref{eq:sig_eff_vs_sig_acc} into eq.~\eqref{eq:mu} gives us a full PDE of the radial gas mass flux:

    \begin{equation}
        \label{eq:muPDE}
        \mu_\mathrm{g} = - 2 \pi R^2 \left[\dot{\Sigma}_\mathrm{eff} + \frac{1}{2\pi R} \partial_R \mu_\mathrm{g} \right]\left(\frac{j_\mathrm{disc}- j_\mathrm{acc}}{R\;\partial_R j_\mathrm{disc}} \right) - 2\pi R \frac{\Sigma_\mathrm{g} \partial_t j_\mathrm{disc}}{\partial_R j_\mathrm{disc}}.
    \end{equation}
    Here, we recognise in the first term the dimensionless factor introduced by \PF:
    \begin{equation}
        \label{eq:alpha}
        \alpha \coloneqq \frac{j_\mathrm{disc}- j_\mathrm{acc}}{R\;\partial_R j_\mathrm{disc}}, 
    \end{equation}
    which is a measure of the mismatch between the specific angular momentum of the disc and that of the accreting gas. In the case of a flat rotation curve (as in the \PF models), this reduces to $\alpha = 1-\frac{v_\mathrm{acc}}{v_\mathrm{disc}}$.
    
    Eq.\eqref{eq:muPDE} at fixed $t$, is a linear ODE in radius of the form
    \begin{equation*}
        \frac{\partial}{\partial R}\mu_\mathrm{g}(R,t) =  A(R,t)\,\mu_\mathrm{g}(R,t) + B(R,t),
    \end{equation*}
    which has the explicit solution
    \begin{equation*}
        \mu_\mathrm{g}(R,t) =  e^{\int_0^R A(R',t)\,\mathrm{d}R'} \left[ \int_0^R e^{-\int_0^{R'} A(R'',t)\,\mathrm{d}R''} B(R',t)\,\mathrm{d}R' + C(t) \right].
    \end{equation*}
    In this case
    \begin{align*}
        A(R,t) &=  - \frac{1}{R\; \alpha(R,t)} = -\frac{\partial_R j_\mathrm{disc}}{\left(j_\mathrm{disc} - j_{\mathrm{acc}}\right)}\qquad \mathrm{and}\\
        B(R,t) &= -2\pi\,R\, \left[   \dot{\Sigma}_{\mathrm{eff}}+  \frac{\Sigma_\mathrm{g}\partial_{t}j_\mathrm{disc}}{j_\mathrm{disc} - j_\mathrm{acc}}\right].
    \end{align*}
    The solution is thereby fully determined once the integration constant is fixed, which follows from requiring vanishing flows in the very centre, i.e. $\mu_\mathrm{g}(0,t) = 0$. The radial velocity of the gas and the accretion profile then follow from eqs.~\eqref{eq:mu} and \eqref{eq:sig_eff_vs_sig_acc} as
      \begin{equation}
        \label{eq:ur_and_sigma_acc}
          u_{R,\mathrm{g}} = \frac{\mu_\mathrm{g}}{2 \pi R \Sigma_\mathrm{g}} \qquad \mathrm{and} \qquad \dot{\Sigma}_\mathrm{acc} = \dot{\Sigma}_\mathrm{eff} + \frac{1}{2 \pi R} \partial_R \mu_\mathrm{g}.
      \end{equation}
      
    Finally, we note that towards the outer edge of the radial grid we integrate on, where $\Sigma_\mathrm{g}$ becomes very small, eq.~\eqref{eq:ur_and_sigma_acc} turns even a small mass flux into an arbitrarily large radial velocity. The mass involved is negligible, but such velocities make the calculation numerically unstable. We therefore suppress $\mu_\mathrm{g}$ smoothly once the gas surface density drops below $4\times 10^{4}\,\Msun\kpc^{-2}$, corresponding to an \ion{H}{I} column density of about $0.4\times 10^{19}\,\mathrm{cm}^{-2}$. This threshold is only reached well outside the star-forming disc ($R>5\,R_\mathrm{g}$ at all times), so the flows in the region we are interested in are unaffected. We describe this suppression and show that it does not affect our results in \autoref{sec:mu_suppression}.

\subsection{Metallicity gradients}
\label{sec:metallicity}
    Finally, we can write down the evolution of the metallicity of the gas, expressed as the abundance of an element $i$ normalised by its yield, $\tilde X_i = X_i / y_i$. Here $y_i$ is the mass of element $i$ returned to the ISM per unit of the mass locked up in long-lived stars. In these units, the evolution is the same for every element, so our treatment remains general rather than element-specific; the abundance of a given element follows from multiplying by its own yield.
    The evolution of $\tilde X_i$ is then given by
    \begin{equation}
    \label{eq:ODEmetallicity}
        \frac{\partial \tilde X_i}{\partial t}
        =
        \frac{\dot\Sigma_\mathrm{SF}}{\Sigma_\mathrm{g}}
        -(\tilde X_i - \tilde{X}_{i,\mathrm{CGM}})\frac{\dot\Sigma_{\mathrm{acc}}}{\Sigma_\mathrm{g}}
        -u_{R,g}\,\frac{\partial \tilde X_i}{\partial R},
    \end{equation}
    which is the metallicity balance of an open box model, where the three terms on the r.h.s.\ are, respectively, metal production by stellar nucleosynthesis, dilution by gas accreted from the CGM, and the advection of metals by the radial flows derived above.\footnote{The outflows implied by our mass loading factor $\eta$ leave at the ambient ISM abundance and therefore do not change $\tilde X_i$ and hence do not appear explicitly in eq.~\eqref{eq:ODEmetallicity}.}

    In eq.~(\ref{eq:ODEmetallicity}), $\tilde{X}_{i,\mathrm{CGM}}$ is the abundance of the gas accreted from the CGM, in the same yield-normalised units. In this model we keep it as a fixed value in radius and time. Since the accreted gas sets a floor to the abundance of the disc, it mainly affects the outer regions, where in-situ enrichment is weak, and hence flattens the profile there \citep[see also][]{pezzulli_angular_2016}. We discuss the implications of assuming different values for $\tilde{X}_{i,\mathrm{CGM}}$ in \autoref{sec:free_params_obs}.
    
     We stress again that our model, and eq. \eqref{eq:ODEmetallicity} in particular, assumes instantaneous recycling. This means that $\tilde X_i$ is only a valid description for elements that are produced in short lived stars shortly after star formation, such as $\alpha$-elements.

    Eq.~\eqref{eq:ODEmetallicity} can be solved by the method of characteristics (see section 4 in \PF). This reduces it to an ordinary differential equation along the curves
      \begin{equation}
      \label{eq:characteristics}
          \frac{\mathrm{d}R_{\mathrm{char}}}{\mathrm{d}t}=u_{R,\mathrm{g}}\!\left(t,R_{\mathrm{char}}\right),
      \end{equation}
      which are set by the radial velocity of the gas.
      We start the characteristics at $t_0$, equidistantly spaced in radius across the whole disc, each with $\tilde X_i(t_0) = 0$, i.e., from a disc of primordial gas, and integrate them numerically with an adaptive LSODA solver (using the scipy implementation \citealt{virtanen_scipy_2020}, based on \citealt{petzold_automatic_1983}). Since the flows are directed inwards at all times, the characteristics move inwards over time. In order to make sure that the disc stays resolved, we insert additional characteristics at the disc edge over time (see \autoref{sec:mu_suppression} for more details).

    Then, defining as in \PF
      \begin{equation}
          \sigma(t)\coloneqq \int_{t_0}^{t}
          \frac{\dot\Sigma_{\mathrm{acc}}(R_{\mathrm{char}},t')}{\Sigma_\mathrm{g}(R_{\mathrm{char}},t')}\,\mathrm{d}t',
      \end{equation}
      the solution along each characteristic is
      \begin{equation}
      \label{eq:final_metallicity}
      \tilde X_i(t)=e^{-\sigma(t)}\int_{t_0}^{t} e^{\sigma(t')}
      \left[
      \frac{\dot\Sigma_\mathrm{SF}(R_{\mathrm{char}},t')}{\Sigma_\mathrm{g} (R_{\mathrm{char}},t')}
      +\tilde{X}_{i,\mathrm{CGM}}\,\frac{\dot\Sigma_{\mathrm{acc}}(R_{\mathrm{char}},t')}{\Sigma_\mathrm{g}(R_{\mathrm{char}},t')}
      \right]\mathrm{d}t'.
      \end{equation}

    \subsection{Free parameters and observables}
    \label{sec:free_params_obs}
      In the setup used in this work, the model has five free parameters: the relative temperature of the CGM $f_\mathrm{temp}$, the final total baryonic mass of the galaxy $M_\mathrm{bar}$, the gas accretion timescale $\tau_\mathrm{acc}$, and the initial and final scalelengths of the  gas disc $R_\mathrm{g, 0}$ and $R_\mathrm{g, f}$, respectively.
      Depending on the application, this list could be extended to include other galactic parameters, such as the virial mass of the galaxy   $M_{200}$, a mass loading factor $\eta$, or the initial metallicity of the CGM $\tilde{X}_\mathrm{CGM}$. Conversely, quantities such as the baryonic mass or the final gas scalelength could instead be fixed to comply with a specific observation of a given galaxy.

      In turn, the model predicts a number of quantities that can be compared with observations, where the exact set can be chosen depending   on the data available for the galaxy one is trying to fit. Some of these are readily understood and can be used directly, such as the stellar and gas mass, the circular speed of the disc $v_\mathrm{D}$ and the circular speed of the accreting material $v_\mathrm{CGM}$. As we fix only the shape of the gas disc over time by imposing it to be exponential (section \ref{sec:discMorph}), one can also use properties of the stellar disc as constraints, such as the scalelength  (which needs to be fitted over an appropriate radial range, as our stellar disc is not a perfect exponential) or the half-mass radius.
      
      An important observable for this work is the metallicity gradient in the ISM, which can be calculated directly, e.g. by fitting an exponential to eq. \eqref{eq:final_metallicity}. Note that here the fitting window should scale with the size of the disc, as a fixed radial range would extend beyond the star-forming  disc at early times. We find a range of $1 - 3 R_\mathrm{g}(t)$ to work well, though this choice might have to be  adjusted for different galaxies or different scale length prescriptions.

      Finally, many GCE models fit not just the gradient but also the absolute metallicity in certain galactic regions, such as the  metallicity around the solar radius in the case of the Milky Way. This is in principle possible here as well, but with two caveats. 
      First,  our formulation assumes instantaneous recycling. While this approximation is adequate for $\alpha$-elements that mainly stem from  short-lived high-mass stars, it will not hold for elements such as iron that are produced with longer delay times after star formation.
      Second, our model uses yield-normalised metallicities, so comparing it to observations, such as observed oxygen abundances,  requires assuming a yield prescription. Specifically, we need an IMF-integrated yield $y_O$, describing how much oxygen is returned to the  ISM per unit mass locked up in stars, such that
      \begin{equation*}
      Z_O = y_O \tilde{X}\quad \mathrm{and} \quad \mathrm{[O/H]} = \log(Z_O) - \log(Z_{O,\odot}),
      \end{equation*}
      where $Z_{O,\odot}$ is the oxygen mass fraction of the Sun.
    \citet{vincenzo_modern_2016} derived these population integrated yields for recent yield grids from \citet{nomoto_nucleosynthesis_2013} and \citet{romano_quantifying_2010}, based on different IMF prescriptions, and found a range of oxygen yields $y_O$ between 0.007 for a \citet{kroupa_distribution_1993} IMF and 0.037 for a \citet{chabrier_galactic_2003} one. This leads to an uncertainty span of $\sim 0.6 \dex$ in the resulting metallicities.
    
    Absolute abundances should therefore be compared with caution. Crucially, though, this uncertainty affects only the metallicity offset and  not the value of the gradient: since $y_O$ enters as a constant multiplicative factor, it shifts $\mathrm{[O/H]}$ by a constant and leaves  the slope unchanged.
    
    The uncertain yields, however, enter our model in a second way, as they also affect on the value of $\tilde{X}_\mathrm{CGM}$. Throughout this work, we fix $\tilde{X}_{i,\mathrm{CGM}} = 0.1$, which for the yield range from  \citet{vincenzo_modern_2016} corresponds to $0.12 - 0.50\,Z_\odot$, where a Salpeter IMF \citep{salpeter_luminosity_1955} gives $\approx 0.3\,Z_\odot$. This value was mainly chosen to be consistent with the lower limit of \citet{miller_constraining_2015}. However, substantially lower abundances, as found by \citet{ponti_abundance_2023}, would require a smaller value. Because the accreted gas sets a floor to the disc abundance, it matters most in the outer disc, where in-situ enrichment is weak: a higher $\tilde{X}_{i,\mathrm{CGM}}$ raises the abundance there and flattens the gradient. Lowering the value significantly would hence steepen the predicted gradient. Nevertheless, we do not fit $\tilde{X}_{i,\mathrm{CGM}}$ in this work, since without an independent constraint on $y_O$ such a fit would largely absorb the uncertainties of the yield and IMF choice instead of actually constraining the metallicity of the CGM. We show the effect of different $\tilde{X}_{i,\mathrm{CGM}}$ in \autoref{fig:xcgm}.

\begin{table}[]
    \centering
    \caption{Model parameters used for the plots throughout this section. This model has a final virial temperature $T_{200} = 8.7\times 10^{5}
\mathrm{K}$.}
    \label{tab:model_values}
    \begin{tabular}{p{4cm}cc}
    \toprule
    Parameter & Symbol & Value \\
    \midrule 
    Total baryonic mass at $t_\mathrm{final}$         & $M_\mathrm{bar}$             & $6\times10^{10} \Msun$ \\
    Gas accretion timescale                           & $\tau_\mathrm{acc}$          & $11 \Gyr$ \\
    Initial gas scalelength at $t_\mathrm{min}$       & $R_\mathrm{g,0}$             & $2.0 \kpc$ \\
    Final gas scalelength                             & $R_\mathrm{g,f}$             & $4.5 \kpc$ \\
    Relative CGM temperature $T_\mathrm{CGM}/T_{200}$ & $f_\mathrm{temp}$            & 0.7 \\
    Mass loading factor                               & $\eta$                       & 0 \\
    \midrule
    Halo virial mass at $t_\mathrm{final}$            & $M_{200}$                    & $1.3\times10^{12} \Msun$ \\
    CGM abundance                                     & $\tilde{X}_\mathrm{CGM}$     & 0.1 \\
    \bottomrule
    \end{tabular}
\end{table}

\begin{figure*}
    \includegraphics[width=0.99\linewidth]{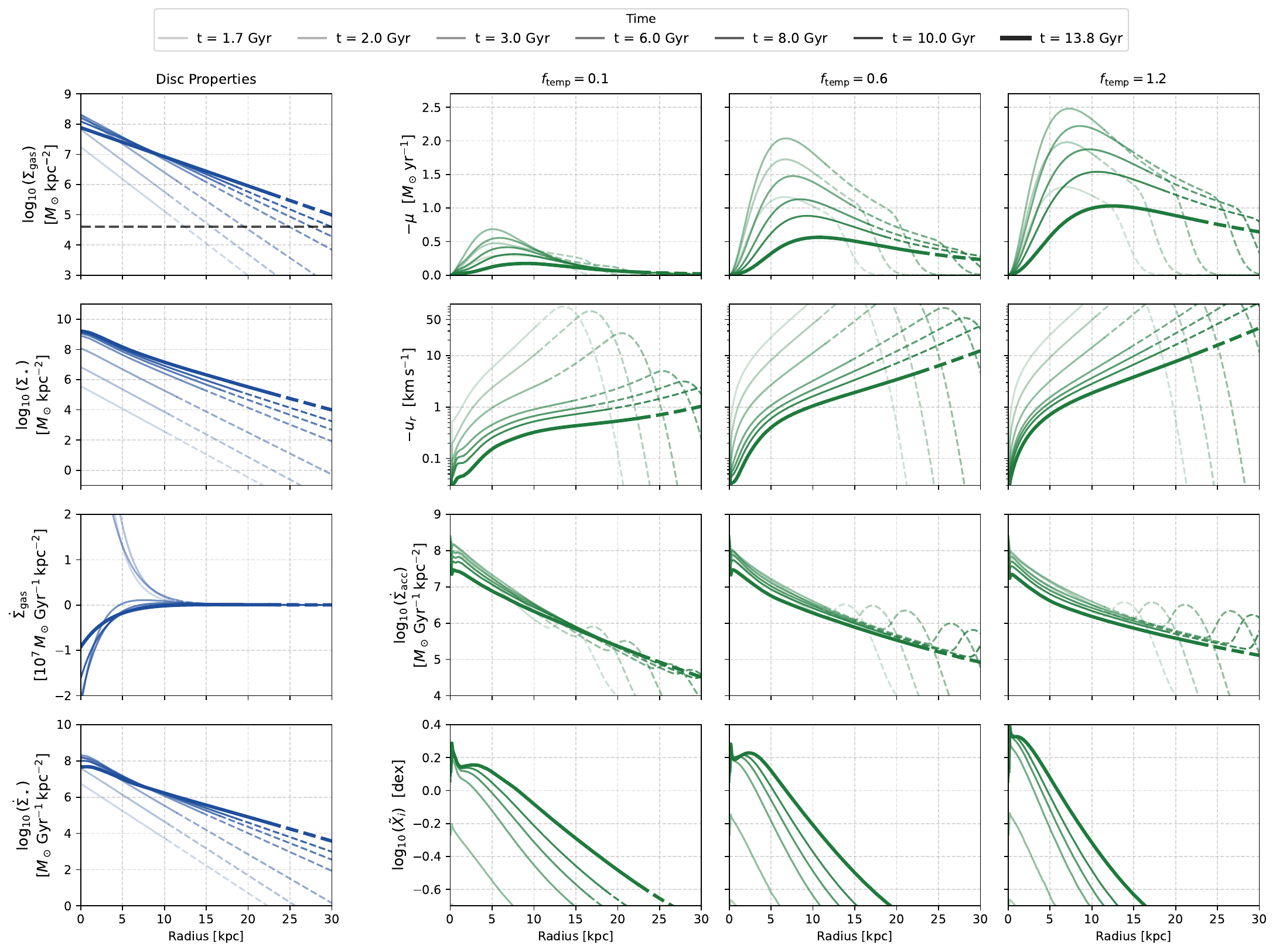}
    \caption{Evolution of three models with different CGM temperatures. The left column shows the evolution of the shape of the gas and stellar discs. These are not affected by the accretion prescription and are, hence, shared between the three models (see text). The values for the underlying model can be found in \autoref{tab:model_values}. 
    In the first two rows on the right, we plot $-\mu$ and $-u_r$ (as defined in eqs.\eqref{eq:mu} and \eqref{eq:ur}, such that positive values correspond to inward transport.
    The line style in all panels turns dashed outside $5R_\mathrm{g}(t)$ and the black dashed horizontal line in panel a) shows the threshold at which $\mu$-regularisation starts (see \autoref{sec:mu_suppression}).}
    \label{fig:full_model}
\end{figure*}

\begin{figure*}
    \centering
    \includegraphics[width =0.8\linewidth]{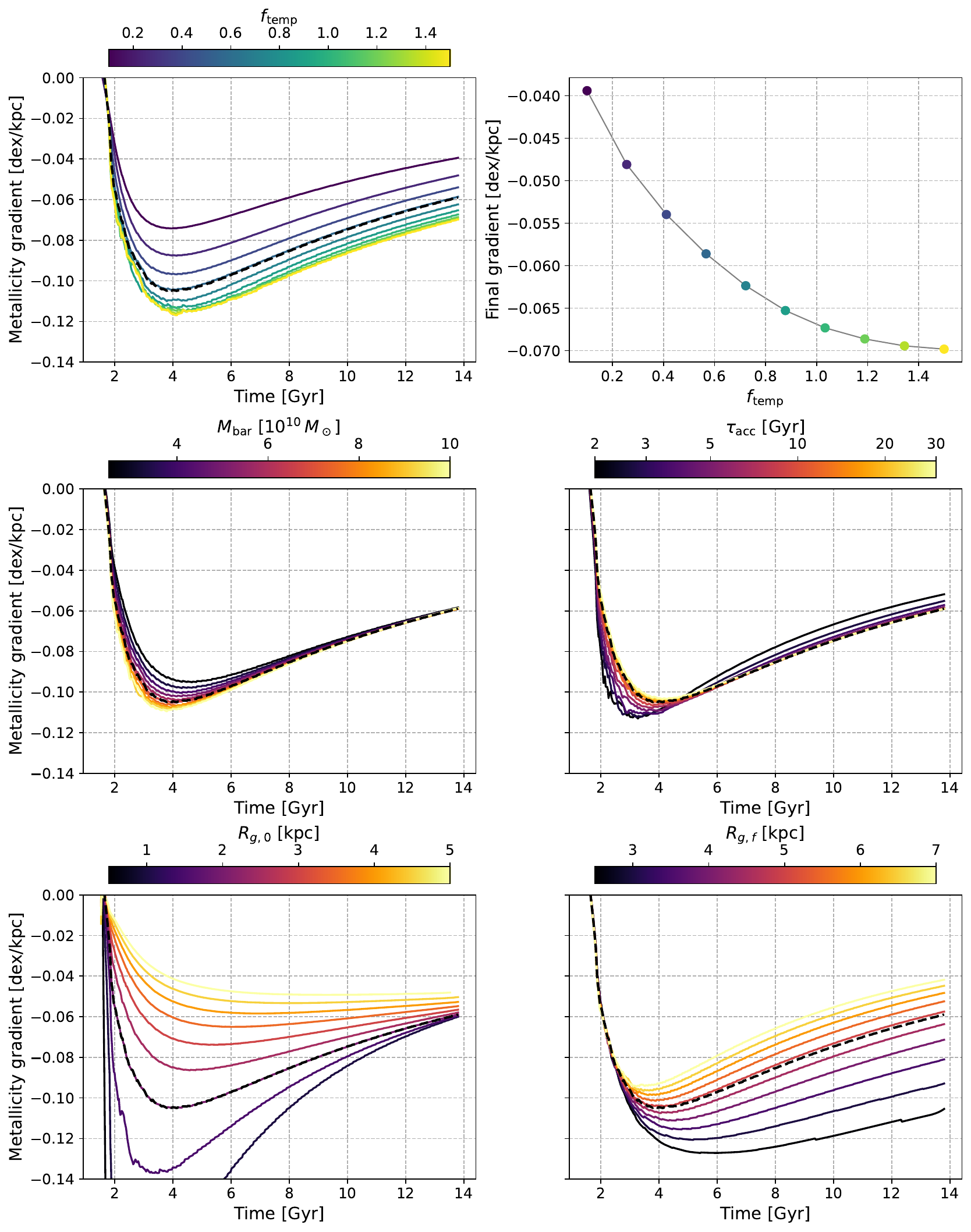}
    \caption{Effect of different values of the free parameters of the model on the metallicity gradient over galactic time. For each panel, we vary one parameter, while keeping the others fixed at the values listed in \autoref{tab:model_values}. In the top right panel, we also show the final gradient for different $f_\mathrm{temp}$ values. In all panels, the dashed black line shows the gradient for the model in \autoref{tab:model_values}.}
    \label{fig:parameter_effects}
\end{figure*}

\subsection{Model behaviour and parameter effects}
Putting all of this together, we can trace the full evolution of a galaxy over its lifetime. As an example, we consider here a set of parameters roughly consistent with a MW-like galaxy,  listed in \autoref{tab:model_values}, and show the evolution of the most relevant physical parameters of the model in  \autoref{fig:full_model}. A complete list of all model inputs, including the numerical grid, is given in \autoref{tab:full_model_params}.

  In the left column of \autoref{fig:full_model}, we plot the evolution of the stellar and gas disc. We can clearly see in the top panel how the gas disc builds up quickly in the beginning and then more slowly at later times, while flattening out as the scalelength grows. In all panels, we switch the linestyle from solid to dashed outside $5R_\mathrm{g}(t)$. %for gas densities below $5.4\times 10^5\Msun\kpc^{-2}$. 
  As our gas disc has no fixed cut-off, this value serves as a rough measure for the outer edge of the disc and generously encompasses the region of the galaxy that has an  effect on our observables at any point in time.
  
  The gas evolution is reflected in the evolution of the stellar disc (in the panel  below), where the inner disc is largely in place within the first $\approx 4.5 \Gyr$ of  galactic evolution, while late-time evolution mainly affects the outer  disc.\footnote{We remind the reader here that all the time values in this work are  cosmic time. As we start the evolution of the galaxy 1.5 Gyr after the Big Bang, the  galaxy is 4.5 Gyr old at $t = 6 \Gyr$.}
  
  The two panels below show the change in gas and stellar surface density. As expected,  the net gas change is only positive in the beginning when the galaxy rapidly builds up;  at later times, star formation consumes the gas faster than accretion replenishes it.
  We also see that $\dot{\Sigma}_\star$ is not a perfect exponential, as a pure  Kennicutt-Schmidt law without inside-out growth (eq. \eqref{eq:KSlaw}, \cite{kennicuttjr._global_1998}) would imply. This is partly due to inside-out growth (contrary to the gas, the stellar disc retains a memory of the shorter scalelength of the gaseous disc at earlier times) and partly to the radial redistribution of stars, as explained in \autoref{sec:stellar_redistribution}: as the disc grows, the rotation curve evolves,  and stars conserving their specific angular momentum $R\,v_\mathrm{rot}$ shift their  guiding radii accordingly. This leads to a slight net movement of stars towards the  centre in the inner $\sim 10 \kpc$ at late times.

  Given that including the effect of a changing potential on the stellar component is a novelty of this work, we also show the radial flows of the stars and the resulting change in stellar surface density in \autoref{fig:stellar_redistribution}. Overall, the flows are low ($\leq 0.5\kms$ after $t=6\Gyr$), but not negligible and lead to a measurable change in stellar surface density.

  In the right three columns of \autoref{fig:full_model}, we then show the evolution of the accretion surface density, radial flows, and metallicity. Each of the columns uses a different  $f_\mathrm{temp}$ to show the effect that a different CGM temperature and, hence, CGM rotation speed has on the galaxy. These three models all share the same evolution of  the stellar and gas disc as plotted on the left: while a higher mismatch in angular momentum affects the ratio between gas that flows in from further out in  the disc and the gas that gets accreted from the CGM at each radius, the change in gas mass at this  radius is solely set by $\dot{\Sigma}_\mathrm{eff}$ (see eq. \eqref{eq:sigma_eff} and  \eqref{eq:sig_eff_vs_sig_acc}), which is independent of the mismatch.
  
  In the top panels on the right side, we show the radial mass flux $\mu$ of the gas, as  defined in eq. \eqref{eq:mu}.  This can be compared directly to Fig. 7 in \PF. We see that for all values of  $f_\mathrm{temp}$, the peak of $\mu$ moves further out as the disc grows, while the  amplitude is strongest early on when the galaxy is growing fastest and decreases steadily thereafter. We also see a strong effect of $f_\mathrm{temp}$: a hotter corona is more pressure supported and rotates more slowly, so the local angular momentum of the accreting gas decreases, and stronger radial flows are needed to conserve overall angular momentum.
  This effect can be seen directly in the panels below, showing the radial velocity of the gas: for all panels, the inflow speed is strongest in the beginning and slower at later times. However, while the model with $f_\mathrm{temp} = 0.1$ barely reaches inflow  velocities of $1\kms$ at the present day, $f_\mathrm{temp} = 1.2$ predicts flows of around $10 \kms$ even today.
  
  In the third row of panels on the right side of \autoref{fig:full_model}, we show the surface density of the accretion, calculated according to eq.~\eqref{eq:ur_and_sigma_acc}. For the model with $f_\mathrm{temp} = 0.1$, where  the circular speed of the accreting gas is close to that of the disc,  $\dot{\Sigma}_\mathrm{acc}$ is close to exponential, with a scalelength that increases  with time as expected for inside-out forming galaxies.  In our model, we fix the total amount of gas accreted onto the galaxy at each time  through eq. \eqref{eq:Macc}. This means that changing $f_\mathrm{temp}$ can change where  in the galaxy gas gets accreted, but not the total amount. Looking at the panels for higher $f_\mathrm{temp}$, we see that while the shape in the centre remains very  similar, it slightly decreases in magnitude, while the accretion in the outskirts increases. This is a natural consequence of the stronger flows driven by a hotter CGM. All models in \autoref{fig:full_model} share the same structural evolution and hence the same effective accretion, eq.~\eqref{eq:sig_eff_vs_sig_acc}, so if the outer regions of the disc get depleted more strongly due to gas flowing  inwards, the disc compensates by accreting more gas from the CGM.
 
  Beyond the outer edge of the disc, we also see some bumps in the gas accretion. These are artefacts of the  suppression of the radial mass flux at very low gas densities (see \autoref{sec:mu_suppression}).

  Finally, in the bottom row, we plot the (yield normalised) metallicity, as calculated in eq.\ \eqref{eq:ODEmetallicity}. As expected, we observe an overall monotonically increasing metallicity with time (again, with the evolution being strongest at early times) and a negative metallicity gradient. Increasing $f_\mathrm{temp}$ makes the gradient steeper, increases the absolute metallicity in the inner few $\kpc$, and depletes the  outskirts. This is again a consequence of the flows: the total amount of metals produced in the three models here is the same; the radial flows just redistribute the metals and  move them towards the centre, building up a strong gradient. In addition, the models with stronger flows accrete less gas in the centre and more in the outskirts, further diluting the outer disc and
  steepening the gradient.
  We also note that even  though the $f_\mathrm{temp} = 0.1$ model only has very moderate radial inflows, it still shows a significant metallicity gradient built up through inside-out formation. 
  In all three models, the profiles also flatten in the inner few $\kpc$ at late times. This plateau reflects a local balance between the metals produced by star formation and the dilution by newly accreted gas, the level of which is set by the ratio of the two rates alone; as that ratio varies little across the inner disc, the central metallicity saturates rather than continuing to rise with the rest of the disc. These panels also show spikes in the very centre of the metallicity profile. These are artefacts stemming from numerically integrating over the $1/R$ singularity in eq.~\eqref{eq:muPDE} and do not affect our results. 

  One of the main applications of our model is to constrain the properties of the accreting gas through the metallicity gradient. For this to be meaningful, we need to know how strongly the gradient responds to the properties of the CGM from where the gas is accreting (encapsulated here in $f_\mathrm{temp}$) as opposed to the other properties of the galaxy. In \autoref{fig:parameter_effects}, we therefore vary each free parameter in turn and follow the evolution of the gradient over cosmic time. Again, this is based on the values in \autoref{tab:model_values}, whose gradient we show as a black dashed line in each panel.

  We can immediately see that all models share the same qualitative behaviour: the gradient steepens rapidly over the first few $\Gyr$, reaches a maximum steepness at $t \sim 2 - 6 \Gyr$ depending on the parameter, and flattens again towards the present day. This flattening out can, at first order, be understood as a direct consequence of the growth of the disc. In Appendix~\ref{app:scalelength}, we show (for the simplified case of a model without flows) that this behaviour is expected on theoretical grounds: the equilibrium gradient depends on radius only through $R/R_\mathrm{g}(t)$, so that the same profile is stretched over a wider disc as it grows. This is consistent with observations, such as from Califa by \citet{sanchez_characteristic_2014}, who find a uniform gradient across their sample when expressing the gradient in terms of effective radii and not absolute size. 

  Starting from the CGM temperature in the top row, a hotter corona produces a steeper gradient at all times due to the stronger radial flows discussed above. In the top right panel, we see that changing $f_\mathrm{temp}$ between 0.1 and 1.5 (i.e., between a nearly corotating corona and a stationary one; see \autoref{fig:cgm_rotation} for a comparison of the CGM rotation curves) gives a difference of about $0.03 \dex\kpc^{-1}$ at the present time, showing that there is predictive power in metallicity gradients for the temperature and thus the kinematics of the CGM. We also see that the gradient responds less towards higher $f_\mathrm{temp}$, where the corona becomes close to stationary. 
  
  Below, in the middle row, we see that both $M_\mathrm{bar}$ and $\tau_\mathrm{acc}$ only have a mild effect on the gradient and only at early times, when the evolution is strongest. This might be slightly surprising, as e.g. \citet{belfiore_sdss_2017} have shown that galaxies with lower stellar mass tend to have flatter gradients than more massive ones. However, for this parameter evaluation, we a) only vary the baryonic mass within one order of magnitude and b) keep the mass of the dark matter halo constant between these models. We leave a systematic study of the model behaviour as a function of halo and stellar mass to future work.

  In the bottom two panels, we see the effect of different scalelength prescriptions, following eq.~\eqref{eq:R_gas}. As expected, the initial scalelength $R_\mathrm{g,0}$ mainly affects the early evolution, where shorter initial scalelengths lead to strong early star formation, while $R_\mathrm{g,f}$ shapes the disc at the end. The effect of $R_\mathrm{g,f}$ is, for the most part, the same one that flattens the gradient with time: a larger final scalelength spreads a similar profile over a larger radius, and therefore gives a shallower gradient in $\dex\kpc^{-1}$ (we refer the reader again to \autoref{app:scalelength} for a larger exploration of this effect).

\begin{table}[h]
\centering
\caption{Prior ranges for the MCMC parameters.}
\label{tab:priors}
\begin{threeparttable}
\begin{tabular}{p{4cm}cc}
\toprule
Parameter & Symbol & Prior range \\
\midrule
Relative CGM temperature $T_\mathrm{CGM}/T_{200}$ & $f_\mathrm{temp}$            & $0.1$--$1.5$ \\
Total baryonic mass at $t_\mathrm{final}$         & $M_\mathrm{bar}$             & $2.5$--$10\times10^{10} \Msun$ \\
Gas accretion timescale\tnote{a}                  & $\tau_\mathrm{acc}$          & $2$--$30 \Gyr$ \\
Initial gas scalelength at $t_\mathrm{min}$       & $R_\mathrm{g,0}$             & $0.5$--$5.0 \kpc$ \\
Final gas scalelength                             & $R_\mathrm{g,f}$             & $2.5$--$7.0 \kpc$ \\
Mass loading factor\tnote{b}                      & $\eta$                       & $0.0$--$3.0$ \\
\bottomrule
\end{tabular}
\begin{tablenotes}
\item[a] Sampled in log-space
\item[b] Only included in some MCMC runs.
\end{tablenotes}
\end{threeparttable}
\end{table}

\begin{table*}
\centering
\caption{Observational constraints used in the MCMC to fit the Milky Way.}
\label{tab:constraints}
\begin{threeparttable}
\begin{tabular}{lp{5.5cm}cccl}
\toprule
Observable & Description & Value & Uncertainty & Units & Source\\
\midrule
$M_{\star,\rm final}$            & Present-day stellar mass              & $5.4\times10^{10}$  & $0.57\times10^{10}$ & M$_\odot$      & \multirow{4}{*}{\cite{mcmillan_mass_2017}} \\
$M_{\rm gas,\rm final}$          & Present-day gas mass                  & $1.2\times10^{10}$ & $0.3\times10^{10}$  & M$_\odot$      &                                              \\
$R_{\star \rm,final}$         & Present-day stellar thin disc scalelength         & $2.53$              & $0.14$              & kpc            &                                              \\
\midrule
$\rm [O/H]_{\odot}$              & oxygen abundance at the solar annulus           & $0.05$               & $0.4$ \tnote{a}     & dex            & \multirow{2}{*}{\cite{mendez-delgado_gradients_2022}} \\
$\nabla \mathrm{[O/H]}_{\rm final}$           & Present-day oxygen gradient      & $-0.059$            & $0.012$             & dex kpc$^{-1}$ &                                              \\
\bottomrule
\end{tabular}
\begin{tablenotes}

  \item[a] The error here is not a measurement error but comes from the uncertainty in yields that we need to assume to convert from our dimensionless metallicity $\tilde{X}_i$ to a physical [O/H] scale (see \autoref{sec:free_params_obs}).
\end{tablenotes}
\end{threeparttable}
\end{table*}

\section{Fitting procedure}
\label{sec:fitting}
In this section, we describe our setup to use our model in a Markov Chain Monte Carlo (MCMC) analysis in order to fit the parameters we left free in \autoref{sec:model}. We then apply it to the Milky Way in section \ref{sec:results}. Here, we deliberately use summary statistics as constraints, all of which can be observed at a single redshift, so that the same analysis can be done for other nearby galaxies and potentially also high-$z$ galaxies, without needing to be able to e.g., resolve individual stars or measure their kinematics. 

In general, the setup can easily be adjusted to the type of data available: the metallicity gradient could be fit to individual \ion{H}{II}-region metallicities rather than to a single fitted slope, and where the evolution of e.g. the disc size is known, that can be included as well. For this work, we keep to the minimal set presented above and discuss the results from these.

\subsection{Setup}
We are using the \texttt{emcee}\footnote{\url{https://emcee.readthedocs.io/en/stable/}} python package for our MCMC fitting \citep{foreman-mackey_emcee_2013}. This package implements the affine-invariant ensemble sampler for an MCMC algorithm described in \citet{goodman_ensemble_2010}. We simultaneously fit galaxy evolution parameters and the parameters of the CGM driving the accreting gas. 

In our fiducial model, we fit five free parameters (see \autoref{sec:free_params_obs}): the relative CGM temperature $f_\mathrm{temp}  = \frac{T_\mathrm{CGM}}{T_\mathrm{200}}$ of eq.~\eqref{eq:f_temp}, the total baryonic mass $M_\mathrm{bar}$ and the accretion timescale $\tau_\mathrm{acc}$, which together set the mass accretion history of eq.~\eqref{eq:Macc}, and the initial and final gas scalelengths, respectively, $R_\mathrm{g,0}$ and $R_\mathrm{g,f}$ of eq.~\eqref{eq:R_gas}, which determine the speed of the inside-out growth. Later, we additionally fit the mass loading factor $\eta$ of eq.~\eqref{eq:Mgas} (discussed in \autoref{sec:discussion}).

The fitting parameters and the priors we chose for the Milky Way are listed in \autoref{tab:priors}. The latter are flat within the given range, except for $\tau_\mathrm{acc}$, which enters the mass history in eq.~\eqref{eq:Macc} as $\propto \frac{1}{\tau_\mathrm{acc}} e^{-t/\tau_\mathrm{acc}}$. Hence, a uniform prior would bias our fit towards large values, where the model is least sensitive to it. Therefore, we sample it with a log-uniform prior.

 We use a classical likelihood, assuming independent observables and Gaussian errors. The total log-likelihood is then $\ln\mathcal{L}(\theta) = \sum_k \ln\mathcal{L}_k(\theta)$, where $\theta$ is the vector of free parameters listed in \autoref{tab:priors} and the index $k$ runs over the constraints of \autoref{tab:constraints}. The posterior we sample is, up to a normalisation constant,
\begin{equation}
  \ln p(\theta\,|\,\{d_k\}) = \ln p(\theta) + \ln\mathcal{L}(\theta),
\end{equation}
where $p(\theta)$ is the prior described above.

In principle, not every point in the prior volume has to yield a usable model. We assign $\ln p = -\infty$ (i.e., reject the parameter set) when the integration of the characteristic lines of eq.\eqref{eq:ODEmetallicity} does not complete, so no metallicity distribution can be calculated.
Especially for extreme parameter space regions with very fast radial flows, the calculation of the characteristic lines becomes stiff and can take very long to complete (if at all). Hence, we also include a 30 second timeout for this. As a normal characteristics integration takes about 0.1 seconds, this only happens when the solver has collapsed to arbitrarily small timesteps. In the MCMCs runs for this work, this never happened, but we include it as a safeguard to ensure the entire chain does not stop.
Finally, we reject the parameter set if any other predicted observable is not finite; though that does not happen anywhere in the parameter space discussed here. 

For the MCMC runs, we use 50 walkers evaluated for 2000 iterations each, with a mean acceptance fraction of $0.45$, consistent with efficient sampling for an ensemble sampler \citep[][]{foreman-mackey_emcee_2013}. The individual chains sample consistently from the posterior distribution after $\sim$80 iterations. We conservatively discard 250 steps as burn-in, which, given a maximum autocorrelation time across parameters of $\sim$ 75 steps, leaves us with $\sim$1150 independent samples per parameter.

\subsection{Data}
\label{sec:data}

\begin{figure}
    \centering
    \includegraphics[width=1\linewidth]{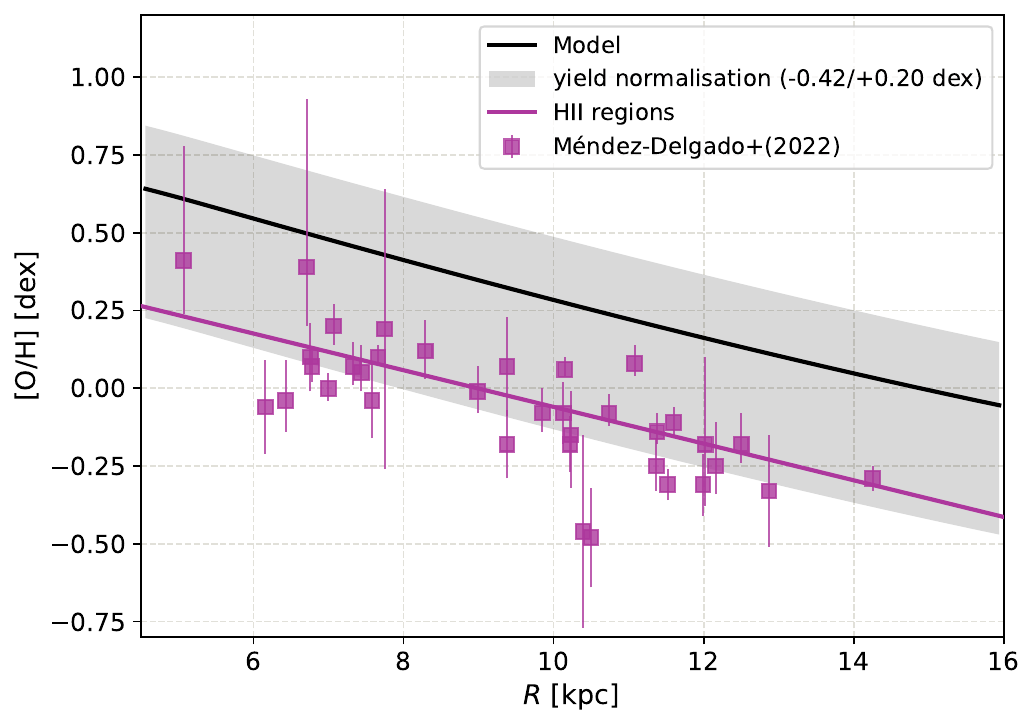}
    \caption{Comparison of the metallicity gradients from one of our models (parameters in \autoref{tab:model_values}) and the best fit oxygen gradient from \cite{mendez-delgado_gradients_2022}. As our model internally uses yield normalised metallicities, for a data comparison a yield scale has to be assumed (see \autoref{sec:free_params_obs}). The possible offset in [O/H] caused by a different yield/IMF choice can be seen in the gray band.}
    \label{fig:hii_region_comparison}
\end{figure}

The data we use in our likelihood function are shown in \autoref{tab:constraints}. We constrain our model using a number of kinematic and morphological data points, as well as the final metallicity gradient. In order to test our method and remove possible contamination from measurement systematics, we chose to use a consistent source for stellar and gas mass, as well as a measure for the extent of the stellar disc today. We adopt the Milky Way model from \citet{mcmillan_mass_2017} as a reliable base for this. It includes a best-fit stellar mass that we can use directly, and both an \ion{H}{I} and an H$_2$ gas disc. The gas discs in \citet{mcmillan_mass_2017} are inputs to the mass model and are not fitted, so they do not come with an error. We use their total cold gas mass of $1.2\times10^{10}M_\odot$ and assume an error of $0.3\times10^{10}M_\odot$. Note that \citet{mcmillan_mass_2017} states explicitly that this value is the total gas mass, including helium and heavier elements, so no additional correction needs to be applied.  As a final constraint, we use their best-fit thin-disc stellar scale length, which we compare to the slope of an exponential fit to our final $\Sigma_\star$ in the same window we use to measure the metallicity gradient, $1-3 R_\mathrm{g}$.\footnote{Since two of these constraints are fitted parameters of the same model, they are not independent, as our likelihood assumes. The posteriors of the parameters they constrain, mainly $M_\mathrm{bar}$ and $R_\mathrm{g,f}$, are therefore likely slightly too narrow.} We chose to compare to the thin, star-forming disc as it most closely resembles the characteristics of the stellar disc in our model, whereas the thick disc is mainly made up of very old, metal-poor stars. While there is no clear consensus about the thick disc origin, the mechanisms that have been proposed, such as radial migration \citep[e.g.][]{schonrich_chemical_2009, loebman_genesis_2011}, inside-out formation \citep[e.g.][]{minchev_formation_2015}, or heating through minor mergers \citep[e.g.][]{quinn_heating_1993, villalobos_simulations_2008}, all trace effects that our model does not include.
\citet{mcmillan_mass_2017} also quotes a best-fit $M_{200} = 1.3 \pm 0.3 \times 10^{12} \Msun$\footnote{Note that while \citet{mcmillan_mass_2017} refers to this value as "virial mass" or $M_\mathrm{v}$ throughout the paper, it is specified in the text that it refers to $M_{200}$}, which we adopt in our model.

For the ISM metallicity gradient at the present time, we chose recent measurements of the 42 MW \ion{H}{ii} regions from \citet{mendez-delgado_gradients_2022}, who re-analysed measurements from \citet{mendez-delgado_helium_2020, arellano-cordova_galactic_2020, arellano-cordova_radial_2021} with Gaia EDR3 distances \citep{gaiacollaboration_Gaia_2021} and accounted for the impact of temperature fluctuations within an \ion{H}{ii} region on the oxygen abundance \citep{peimbert_temperature_1967}. The resulting final oxygen gradient is $-0.059 \pm 0.012\dex\kpc^{-1}$.

We note that, regarding the abundance gradient, there are, however, several other tracers and determination methods. 
Another option to measure the ISM $\alpha$-element gradient is to use young open clusters, as done, e.g., by \citet{magrini_gaia_2023}, who obtain an $\alpha$-gradient for the inner 11.2 kpc between $-0.051 \pm 0.009\dex\kpc^{-1}$ for $\nabla\mathrm{[O\,I/H]}$ and $-0.096 \pm 0.008\dex\kpc^{-1}$ for $\nabla\mathrm{[Mg\, I/H]}$. They attribute the significant difference between oxygen and magnesium to the different stages of nucleosynthesis in which they are produced in heavy stars; see their section 3.3 and \citet{mcwilliam_evolution_2008}.

Classical Cepheids (CCs) are another a well-established tracer of this gradient. \citet{luck_distribution_2011} arrived at values of $\nabla\mathrm{[O/H]} = -0.056 \pm 0.003$ and $\nabla\mathrm{[Mg/H]} = -0.048 \pm 0.004\dex\kpc^{-1}$ from samples of 313 and 303 stars, respectively, out to a Galactocentric distance of $\approx 17$~kpc. \citet{trentin_cepheid_2023} used a somewhat larger sample, supplemented with CCs at greater distances and lower metallicities, and recovered a shallower oxygen gradient of $\nabla\mathrm{[O/H]} = -0.046 \pm 0.002$, but a steeper magnesium gradient of $\nabla\mathrm{[Mg/H]} = -0.056 \pm 0.002\dex \kpc^{-1}$. \citet{dasilva_Oxygen_2023} did a similar analysis and recovered a quite shallow oxygen gradient of $-0.029 \pm 0.006\dex\kpc^{-1}$, but noted that their sample is overall better fit using logarithmic instead of linear regression, and stressed that their line list differs from that used by \citet{trentin_cepheid_2023} and \citet{luck_distribution_2011}. Finally, \citet{nunnari_classical_2026} performed a non-local thermodynamical equilibrium spectroscopic analysis of 401 CCs, where 379 of them used the same spectra as \citet{dasilva_Oxygen_2023}. They found that especially the oxygen triplet around $7770$\r{A} is affected by these corrections, substantially flattening the oxygen gradient (measured in  CCs out to $\approx 14$~kpc) to $-0.018 \pm 0.015\dex \kpc^{-1}$, while keeping the Mg gradient at $-0.061 \pm 0.005\dex\kpc^{-1}$ (with CCs out to $\approx 30\kpc$). They also note that a logarithmic gradient or at least a broken powerlaw fit (with a break in the outer disc between roughly 8 and 15~kpc), improves the fit.

With all these uncertainties, we decided for this work to adopt a single tracer that can also be applied directly to nearby galaxies: the \ion{H}{ii}-region fit from \citet{mendez-delgado_gradients_2022}. 
We show these data in \autoref{fig:hii_region_comparison}, where we also plot, for comparison, the model discussed in the last section with parameters from \autoref{tab:model_values}. Here, we rescaled the values from \citet{mendez-delgado_gradients_2022} to solar scale, assuming a solar $12+\log(\mathrm{O/H})$ value of $8.69$ \citep{asplund_chemical_2021}.

Finally, we also fit the average oxygen abundance around the solar annulus. Using the best fit gradient and oxygen abundance in the Galactic centre from \citet{mendez-delgado_gradients_2022}, this value is [O/H] = 0.046 dex.
However, as shown in \autoref{sec:free_params_obs}, comparing this quantity to our model depends strongly on the choice of yields and the IMF. For this fit, we chose an oxygen yield of $y_O$ of 0.018, which \citet{vincenzo_modern_2016} derived for \citet{romano_quantifying_2010} yields and a \citet{salpeter_luminosity_1955} IMF, and is consistent with our return factor of $\mathcal{R} =0.3$.
As stated before, the uncertainties are quite large. In \autoref{fig:hii_region_comparison}, we show them as a band around the model gradient, where the low end of the band is set by \citet{nomoto_nucleosynthesis_2013} yields with a \citet{kroupa_distribution_1993} IMF and the high end by \citet{romano_quantifying_2010} yields and a \citet{chabrier_galactic_2003} IMF, both at a metallicity of $Z = 5.0 \times 10^{-2}$.\footnote{Note that a different choice of IMF also changes the return fraction $\mathcal{R}$. While \citet{vincenzo_modern_2016} derived oxygen yields per stellar mass locked up in low mass stars (see their eq. (2)), for a perfect comparison, we would need to rerun our model with a different return fraction. For now, to calculate the uncertainty in figure~\ref{fig:hii_region_comparison}, we just include a factor $\frac{1-\mathcal{R}_\mathrm{IMF}}{1-\mathcal{R}_\mathrm{model}}$, where $\mathcal{R}_\mathrm{IMF}$ is the return fraction implicated by a different IMF choice and $\mathcal{R}_\mathrm{model}=0.3$ is the return fraction adopted in this work.}

This uncertainty means that the absolute value of [O/H]$_\odot$ can only be used as a weak constraint. For the MCMC, we adopt an error on the constrain on the larger side at $0.4\dex$.

\subsection{Recovery tests}
\begin{figure*}
    \centering
    \includegraphics[width=0.49\textwidth]{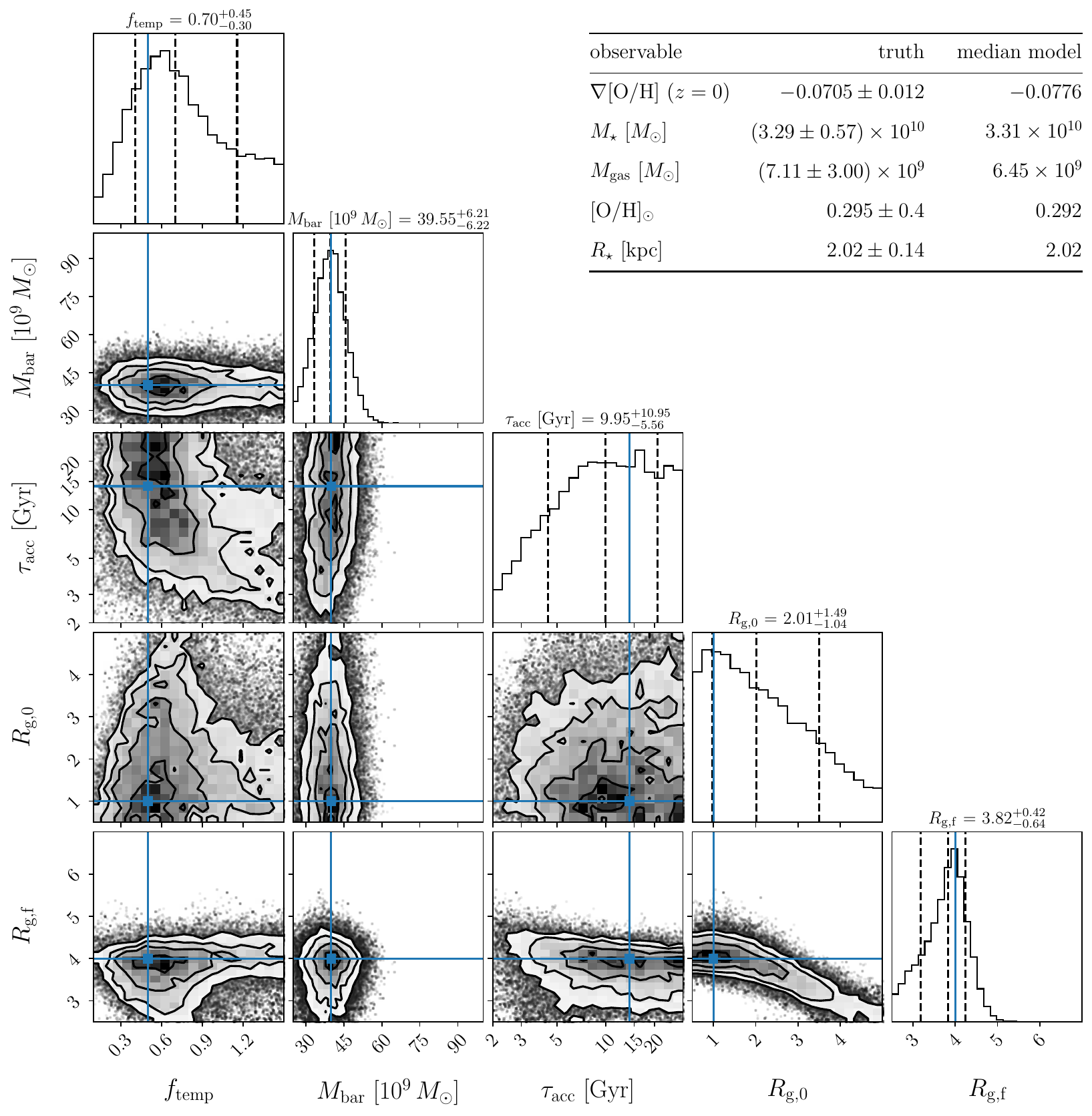}
    \includegraphics[width=0.49\textwidth]{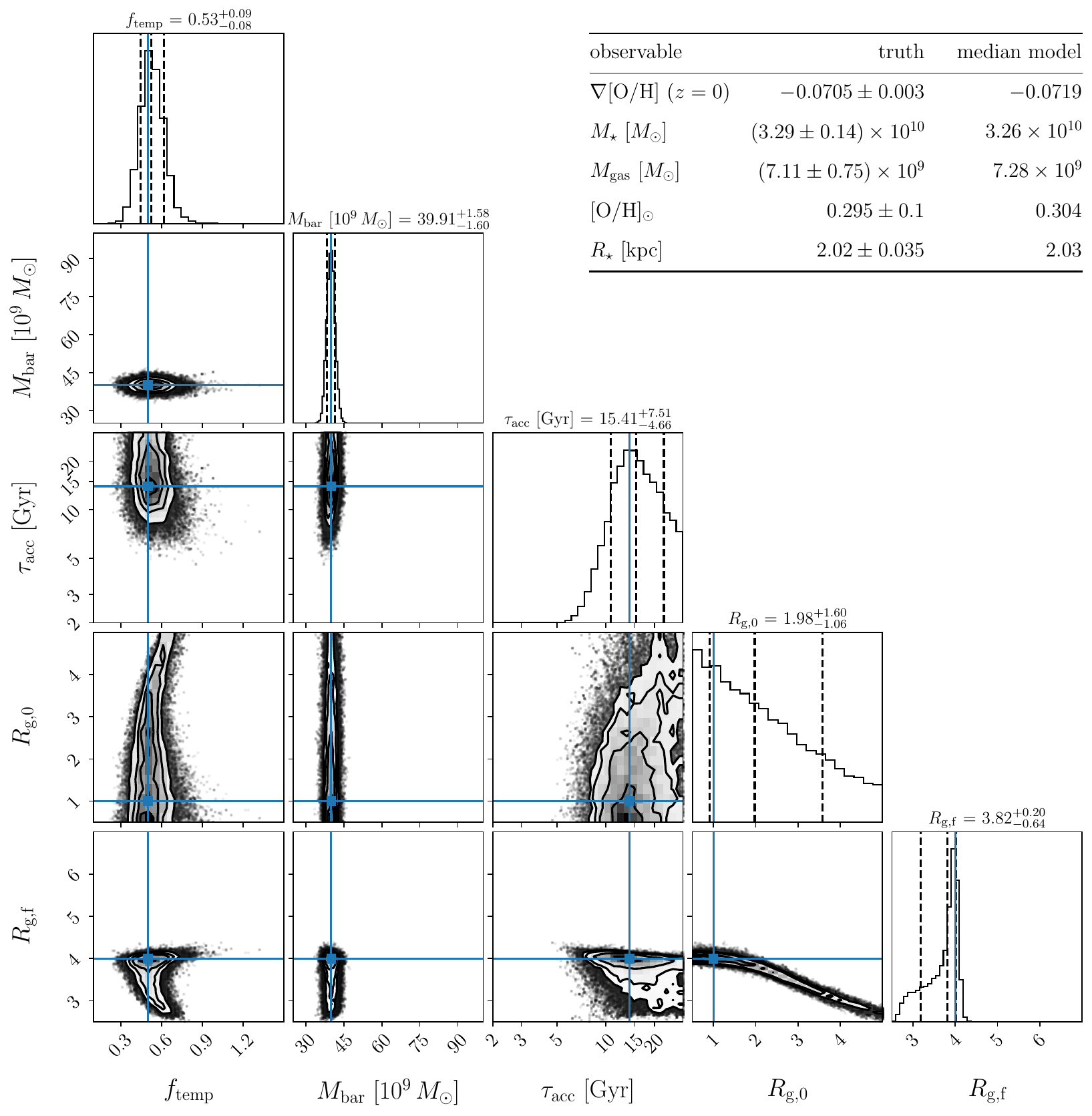}
    \caption{Recovery test for a model with the same errors as we use for data and an MCMC run with 1/4th of the fiducial errors. The dashed vertical lines in the plots on the diagonal show the median and 1$\sigma$ intervals each, the solid blue lines show the positions of the parameters used to run the test model.
    The input values are $f_\mathrm{temp} = 0.5$, $M_\mathrm{bar} = 4\times 10^{10} \Msun$, $\tau_\mathrm{acc}= 14 \Gyr$, $R_{\mathrm{g,0}} = 1\kpc$ and $R_\mathrm{g,f} = 4 \kpc$.}
    \label{fig:recovery_test}
\end{figure*}
Before applying the fit to real data, we test whether the fitting procedure is actually sensitive to the input data. For this, we chose a set of true parameters $\theta_\mathrm{true}$ that are deliberately a bit removed from the expected MW parameter values to run the forward model and treat its output as a synthetic data set, assigning the same uncertainties as the corresponding real measurements. The left panel of \autoref{fig:recovery_test} shows the results of these fits using the absolute errors from \autoref{tab:constraints}.
  
We can see that at the uncertainty level of our data, we recover the input of $M_\mathrm{bar}$ and $R_\mathrm{g,f}$ well and obtain a decent constraint on $f_\mathrm{temp}$, where the true value lies slighly below the mode of the distribution. We also see that both $\tau_\mathrm{acc}$ and $R_\mathrm{g,0}$ are not well constrained and only provide a lower and upper limit, respectively. Our true values, however, lie soundly in the allowed region, close to the peak of the posterior.
 
To estimate the change of the posteriors depending on the data uncertainties, we ran the same recovery test again, but reducing all errors by a factor of 4. Indeed, all the constraints become significantly tighter, and the posterior on $\tau_\mathrm{acc}$ peaks around the input value. However, the $1\sigma$ interval remains quite wide. Finally, the initial scalelength remains largely unconstrained, only providing an upper limit of about $3.5\kpc$.

We conclude that with this set of observables measured at redshift zero, $R_\mathrm{g,0}$ has too small an effect on the present-day disc to be constrained. This is consistent with what we saw in \autoref{fig:parameter_effects}: the initial scalelength affects the evolution of the gradient mainly at early times. Constraining $R_\mathrm{g,0}$ therefore requires observables that carry information about the early disc. In subsection~\ref{sec:gradient_evolution} we show that including the observed evolution of the metallicity gradient over the past $6\Gyr$ does indeed sharpen the posterior on $R_\mathrm{g,0}$.

We note that the model overall favours substantial inside-out formation. However, a degeneracy between $R_{\mathrm{g,f}}$ and $R_{\mathrm{g,0}}$ persists even with artificially lowered uncertainties, allowing, albeit at lower probability, for little or no scalelength evolution over cosmic time.

\section{Results}
\label{sec:results}

\subsection{Posteriors}
\begin{figure}
    \centering
    \includegraphics[width=1\linewidth]{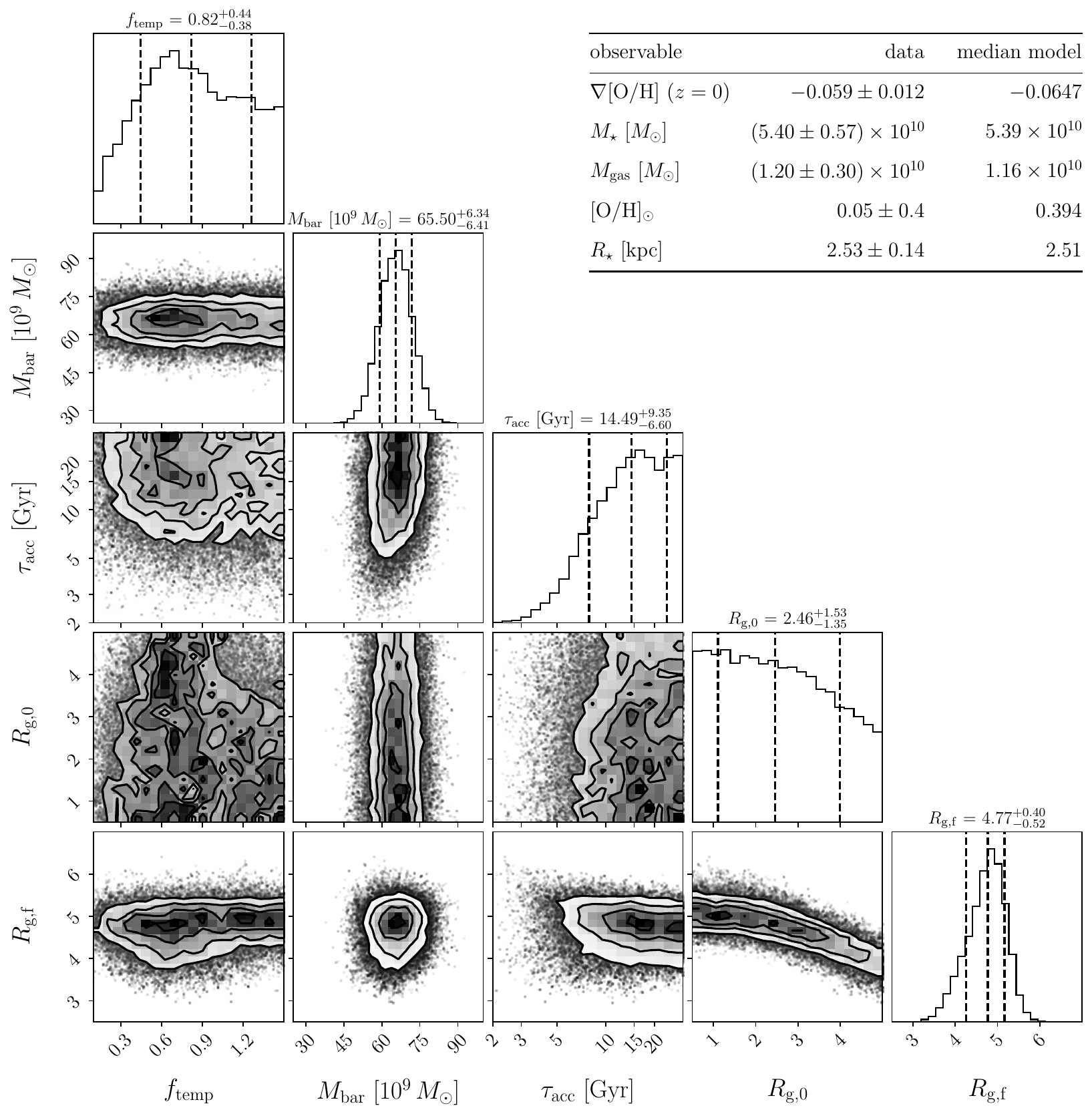}
    \caption{Marginalised posteriors from our fiducial MCMC run on the Milky-Way constraints from \autoref{tab:constraints}.}
    \label{fig:corner_plot_fiducial}
\end{figure}

\begin{table}[]
    \centering
   \begin{tabular}{p{4cm}cc}
    \toprule
    Parameter & Symbol & Value \\
     \midrule
    Total baryonic mass at $t_\mathrm{final}$         & $M_\mathrm{bar}$             & $65.5 \times 10^9 \Msun$\\
    Gas accretion timescale                           & $\tau_\mathrm{acc}$          & $14.49 \Gyr$ \\
    Initial gas scalelength at $t_\mathrm{min}$       & $R_\mathrm{g,0}$             & $2.46\kpc$\\
    Final gas scalelength                             & $R_\mathrm{g,f}$             & $4.77\kpc$\\
    Relative CGM temperature $T_\mathrm{CGM}/T_{200}$ & $f_\mathrm{temp}$            & 0.68 \\
    Mass loading factor                               & $\eta$                       &0    \\
    \bottomrule
    \end{tabular}
    \caption{Fiducial model parameters, based on the fit in \autoref{fig:corner_plot_fiducial}.}
    \label{tab:fiducial values}
\end{table}

\begin{figure}
    \centering
    \includegraphics[width=1\linewidth]{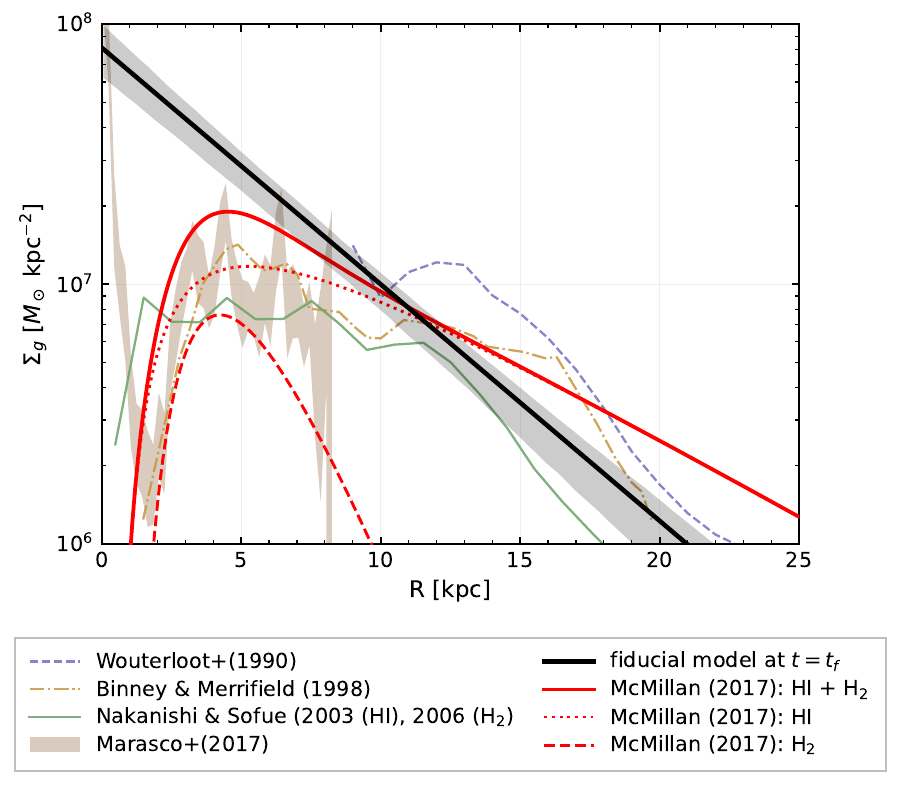}
    \caption{Comparison of the final surface density profile of our fiducial model compared to \citet{mcmillan_mass_2017} and literature data for the total gas surface density from \citet{wouterloot_iras_1990, binney_galactic_1998, nakanishi_threedimensional_2003,nakanishi_threedimensional_2006} and \citet{marasco_distribution_2017}. The shaded band around the fiducial model shows, the 16th--84th percentile of the gas surface density across models drawn from the posterior}
    \label{fig:sigma_gas}
\end{figure}

\begin{figure*}
    \centering
    \includegraphics[width=0.9\linewidth]{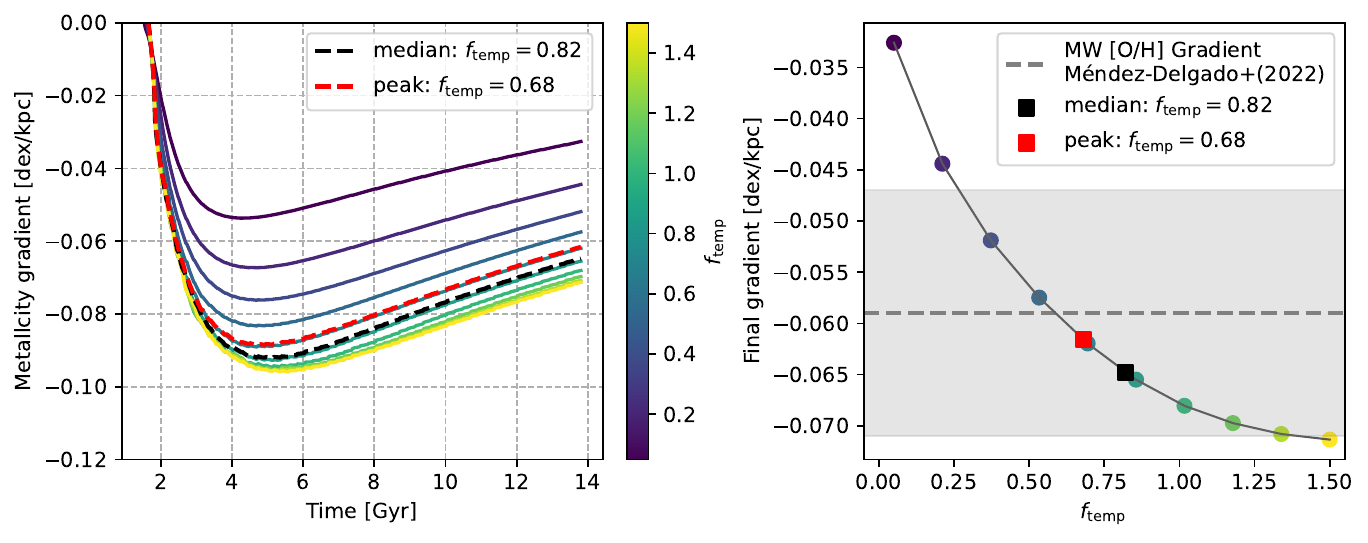}
    \caption{ Comparison of the evolution of the metallicity gradient (left panel) and final gradient (right panel) for our fiducial model when varying $f_\mathrm{temp}$. The red line (and square respectively) belong to the peak of the posterior in \autoref{fig:corner_plot_fiducial}, the black one to the median. On the right panel, the horizontal dashed line shows the \ion{H}{II}-region metallicity gradient from \citet{mendez-delgado_gradients_2022} and the shade region the quoted error.}
    \label{fig:evolution_vs_t_frac}
\end{figure*}

We now fit the five free parameters of our model reported in \autoref{tab:priors} to the Milky-Way observational constraints of \autoref{tab:constraints}. The resulting marginalised posteriors are shown in \autoref{fig:corner_plot_fiducial}.

Both the total baryonic mass and the final gas scalelength are well constrained. $M_\mathrm{bar}$ is recovered as $65.50^{+6.34}_{-6.41} \times10^9 \Msun$, in close agreement with \citet{mcmillan_mass_2017}, whose best fitting model gives a mass of $66.2\pm 5.7 \times10^9 \Msun $, where the error comes solely from the stellar component. However, this is hardly surprising: we constrain both final stellar and gas mass with the \citet{mcmillan_mass_2017} model. Nevertheless, it confirms that our model can reproduce these constraints simultaneously, without any single observable pulling the posterior away from the expected range.

Secondly, the final gas scalelength is also well constrained to $4.77^{+0.40}_{-0.52}\kpc$, with a slight tail towards lower scalelengths. Comparing this value to the expectations for the Milky Way is a bit more involved. For a single scalelength, often $3.75 \kpc$ from \citet{kalberla_global_2008} is cited. However, this fit was only done for surface densities at radii $R\gtrsim 12\kpc$ as the Galactic \ion{H}{I} component is observed to be relatively flat in the inner $\sim 10\kpc$ (as confirmed in more recent works; e.g. \cite{marasco_distribution_2017}). Additionally, the two main components of gas in the Milky Way, molecular $\mathrm{H_2}$ and atomic \ion{H}{I} gas, are distributed differently. The molecular gas is more centrally concentrated, declining outside the very centre of the Milky Way with a scalelength of $\sim 2\kpc$ \citep[e.g.][measuring the density of molecular clouds]{miville-deschenes_physical_2017}. Often, models giving exponential surface densities set up two gas discs with different scalelengths \citep[e.g.][giving $R_{H_2} = 3\kpc$ and $R_{HI} = 10\kpc$]{sysoliatina_fully_2022}.
In \autoref{fig:sigma_gas}, we show a comparison of the final surface density of our fiducial model with the description of \citet{mcmillan_mass_2017} (who uses $R_{H_2} = 1.5\kpc$ and $R_{HI} = 7\kpc$ with an inner dip) as well as some observational data. Due to the assumption of an exponential gas disc made in \autoref{sec:model}, we cannot reproduce the flat part of the gas distribution observed in the central MW. However, our model diverges most from the observations in the central $\sim 4-5\kpc$, the region dominated by the Galactic bar. Bars are well known to shock the gas along their leading edges and funnel it towards the centre \citep[e.g.][]{athanassoula_existence_1992}, evacuating the region they sweep through.  As our model contains no bar, we cannot make claims on this region, restricting our analysis instead to the region outside $\sim 5\kpc$. As a test, we also reran the MCMC, fitting only the gas mass outside $R = 5\kpc$ to check that our fit is not dominated by our surface density peak in the centre. We found that this did not significantly change our results.

\autoref{fig:sigma_gas} also shows that our outer disc scalelength is shorter than that of \citet{mcmillan_mass_2017} (our $R_\mathrm{g,f}= 4.77^{+0.40}_{-0.52}\kpc$ against their $R_\mathrm{g} = 7\kpc$). However, we remind the reader that we do not fit our model to the observed gas scalelength (which is not well constrained by data), but instead to that of the stars $R_\star$. As the stellar and gas discs in our model are connected by a Kennicutt-Schmidt law (eq.~\eqref{eq:KSlaw}), a non-inside-out forming disc would always have a scalelength ratio of $R_\mathrm{gas}/R_\star = n$, where $n = 1.4$ for our model. Inside-out growth, where a large fraction of stars forms early on concentrated in the centre, can increase this ratio (the fiducial model has a ratio of 1.9), but not enough to reconcile the final scalelength of $R_\star=2.53\pm 0.14 \kpc$ from \citet{mcmillan_mass_2017} with a final gas scalelength of $R_\mathrm{g} = 7\kpc$ of \citet{mcmillan_mass_2017} (which would require a ratio of 2.8). Indeed, our model can only reproduce $R_\mathrm{g} = 7\kpc$ when assuming impossibly fast accretion timescales of $\lesssim 2\Gyr$, which also leads to a significantly too low gas fraction at late times. 
Considering that we do not fit our model on the observed gas distribution, the comparison of our model with observations, outside the bar region, is encouraging. Our surface density profile lies between the combined profile of Nakanishi \& Sofue (\citealt{nakanishi_threedimensional_2003} for the \ion{H}{I}-component and \citealt{nakanishi_threedimensional_2006} for H$_2$), and the measurements of \citet{wouterloot_iras_1990} and \citet{binney_galactic_1998}.

Returning to \autoref{fig:corner_plot_fiducial}, we can see that the initial scalelength $R_\mathrm{g,0}$ is less well constrained, with the posterior only giving it a preference towards $R_\mathrm{g,0}\lesssim 3\kpc$. As mentioned, $R_\mathrm{g,0}$ mainly affects the early evolution of the metallicity gradient and loses a lot of its influence at late times (as shown in \autoref{fig:parameter_effects}). Or model favours inside out growth with a median growth of about  $0.2 \kpc \Gyr^{-1}$. We do, however, find a tail of the $R_\mathrm{g,f}$ distribution consistent with a non-growing disc.

We also only get weak constraints on the accretion timescale $\tau_\mathrm{acc}$. The main observable constraining $\tau_\mathrm{acc}$ is the final gas mass, as it traces how quickly gas needs to be turned into stars over the galaxy lifetime. However, $M_\mathrm{gas}$ also has the largest relative error of our observables, giving only a weak constraint. We can, however, note that our model strongly prefers a long accretion timescale with $8 \Gyr \lesssim \tau_\mathrm{acc} \lesssim 24 \Gyr$, while excluding timescales of less than $5\Gyr$.

Finally, we can look at $f_\mathrm{temp}$, the parameter setting the temperature of the CGM and effectively the kinematics of the accreting gas. Here, the constraints are also not extremely tight, but there is a clear peak of the posterior around 0.68. The overall shape is, however, not Gaussian, so the median and the peak do not coincide, with the median lying at a higher value of  $f_\mathrm{temp} 0.82 ^{+0.44}_{-0.38}$.
The reason for this becomes clear when looking at \autoref{fig:evolution_vs_t_frac}. In the right panel, we vary $f_\mathrm{temp}$ while keeping all other parameters fixed at their fiducial values, and plot the resulting final metallicity gradient.  As a higher $f_\mathrm{temp}$ means a larger angular momentum mismatch between the CGM and the disc, and hence stronger radial flows, these models also have a steeper final gradient. As the CGM gets closer to the halo virial temperature, the gradient also becomes less sensitive to the shift. 
In the same panel, we also we plot the value for the MW gradient and its error, taken from  \citet{mendez-delgado_gradients_2022}. While the peak of the posterior with $f_\mathrm{temp}=0.68$ matches the gradient better, the error is so large that it encompasses a substantial region of our parameter space. Hence, the median model, which gives a final gradient $0.003\dex\kpc^{-1}$ lower, is still in agreement. This also means that as gradient measurements improve, we will also be able to improve our constraints on $f_\mathrm{temp}$ and the CGM kinematics.

From this fit to the Milky-Way data, we now also decide on some fiducial model parameters. The values can be found in \autoref{tab:fiducial values}. We chose the values of the median of the posterior everywhere, except for $f_\mathrm{temp}$, where, for the reasons discussed above, we chose the peak value of 0.68 instead. We note that this is mainly a matter of convenience in order to be able to plot a model in our figures and that the real results are the posteriors we recover in \autoref{fig:corner_plot_fiducial}.

\subsection{Radial flows and the CGM versus other determinations}

\begin{figure}
    \centering
        \includegraphics[width=1\linewidth]{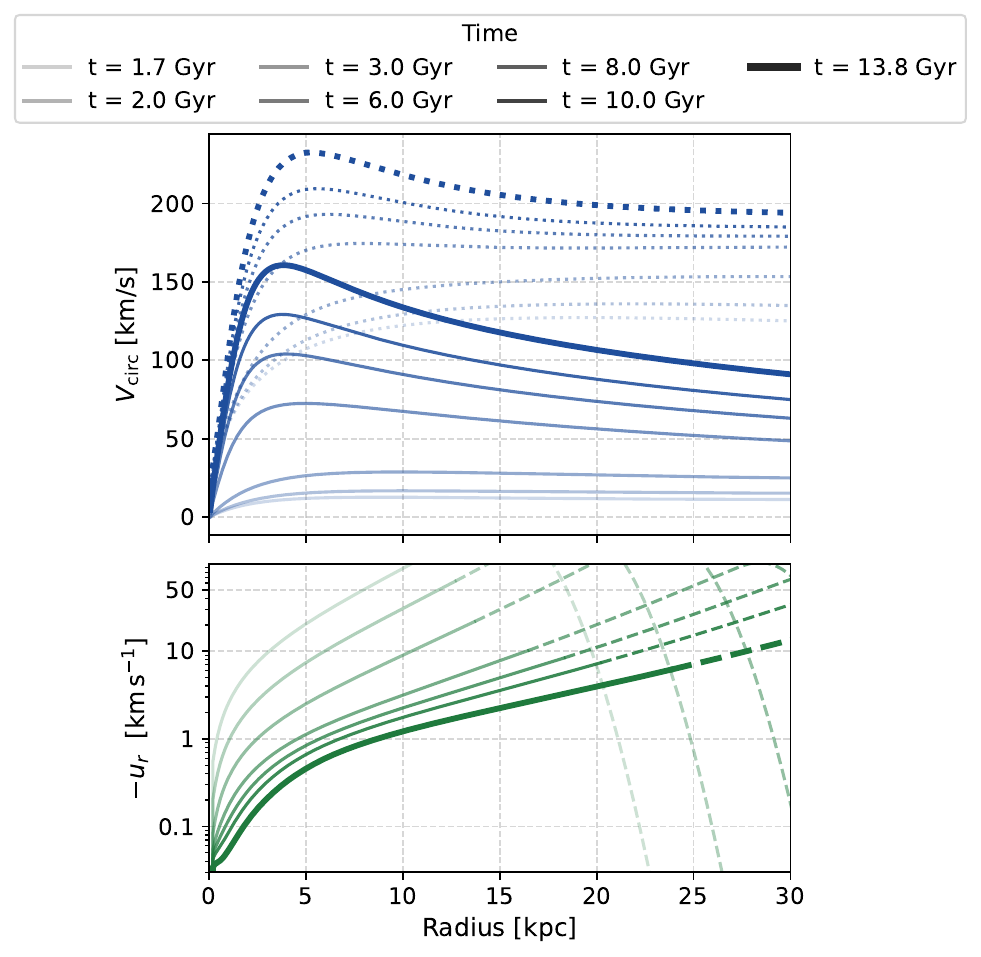}
    \caption{Top panel: Rotation curve for both CGM (solid) and disc (dotted) over time for the fiducial model as defined in \autoref{tab:fiducial values}.
    Bottom panel: The radial inflow velocity of the gas. As in \autoref{fig:full_model}, the linestyle turns dashed outside $R= 5\,R_\mathrm{g}(t)$.}
    \label{fig:vrot}
\end{figure}

\begin{figure}
    \centering
    \includegraphics[width=1\linewidth]{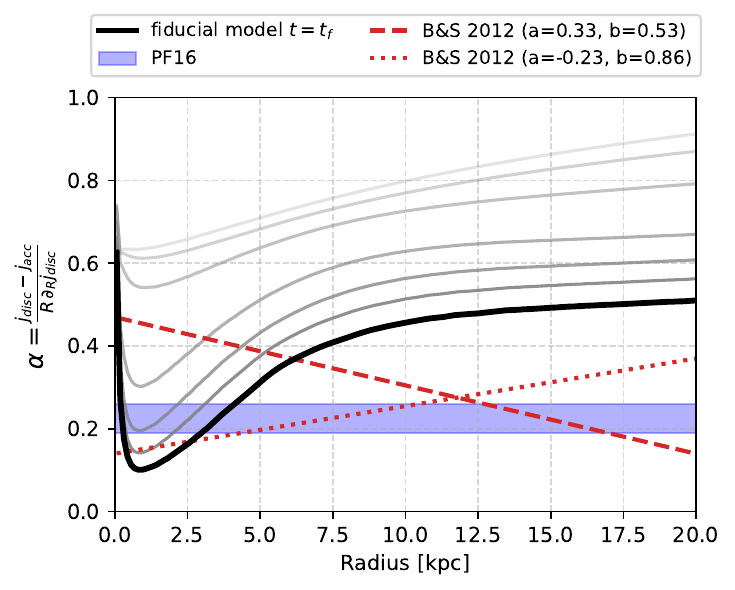}
    \caption{Comparison of the mismatch between CGM and disc rotation curve. The deficit in angular momentum is herby defined as in eq.~\eqref{eq:alpha} and reduces to $\alpha = 1- v_\mathrm{acc}/v_D$ in case of a flat a rotation curves.  We show our fiducial model as well as the best fit parameters from \citet{bilitewski_radial_2012} (who quote two best fit models depending on whether they only fit the [Fe/H] gradient or also the absolute iron abundance) and PF16. The light opaque lines show the mismatch at previous times, at the same intervals as \autoref{fig:vrot}}
    \label{fig:compare_alpha}
\end{figure}

The angular momentum mismatch between the disc and the gas accreting from the CGM drives the radial flows in our model. We can hence check whether both the flows it produces and the CGM properties are comparable to what we expect from the literature.

\subsubsection{Radial flows of the gas}

In our model, the flows are directed inwards throughout the disc; they are always fastest at the outer edge of the gas disc, and they weaken steadily over time as the accretion rate declines (see Figure \ref{fig:vrot}). While the flows at early stages reach $\sim 21 \kms$ (at $t=3 \Gyr$ and $R=5\,R_\mathrm{g}(t)$), at the present day they have dropped to $\approx 6 \kms$ there, with the disc interior experiencing flows of $2.0 \kms$ at $14 \kpc$ and less than $1 \kms$ inside the solar annulus. 

%This general behaviour is qualitatively consistent with the inside-out models by \PF, although \PF found lower velocities (less than 1 km s$^{-1}$ throughout the disc), possibly due to their different prescription for inside-out growth. 

\citet{bilitewski_radial_2012} find a similar qualitative behaviour: flows that increase outward with overall decreasing flow magnitudes over time and their best-fit model has, at present, comparable radial velocities to ours. Notably, they adopted a similar inside-out growth scheme as in this work (time-increasing exponential gas scale-length). %As expected the overall flow magnitudes depend directly on the overall angular momentum mismatch. 

%We can also compare our values to other chemical evolution models built on similar premises. 
\PF's model also shows qualitatively similar behaviour, with inward velocities of the order of $1 \kms$ peaking at $\lesssim 1.4 \kms$ for their inside-out formation model. Compared to them, our flows are a bit on the large side. This could be due in part to the larger values of $\alpha$ implied by our fiducial model, compared to \PF,  at large radii (Figure \ref{fig:compare_alpha}), and in part also to the different scheme for inside-out growth (radially increasing effective accretion time-scale) adopted by \PF.  

\citet{johnson_constraints_2025} also invoke angular momentum mismatch to drive radial inflows (their 'angular momentum dilution' scenario), with a constant ratio of accreted to disc circular velocity of $\beta_{\phi,\mathrm{acc}} = 0.7$, i.e. a mismatch of $0.3$ in our convention. Their flows likewise slow down as the accretion rate drops and reach $\approx 0.2-1 \kms$ at the present day, in line with what we recover. Note, though, that they state they cannot fully recover the observed properties of the Milky Way (especially its metallicity evolution) from radial flows (constant or induced by angular momentum mismatch) alone. They do, however, keep the mismatch constant over time, while in our model it is considerably higher at early times and decreases thereafter (as we show below), so the results might not be directly comparable.

\citet{wang_origin_2022} built a model of coplanar inflow from the outer edges of the galaxy discs. 
Their model is significantly different from ours as they do not impose angular momentum conservation but invoke magnetic viscosity to transport angular momentum within the disc towards the outer edge, beyond which it must be transferred to some other unspecified medium.
They also predicted maximum radial velocity at the edge of the disc, but their radial flows are significantly larger (20-40 km/s at the outer edge) compared to ours or to those in \PF; this is because, in the absence of direct accretion onto the disc, all the effective accretion  must be supplied through radial flows (equation \ref{eq:sig_eff_vs_sig_acc}).

The results of \citet{wang_origin_2022} are presented as analogous to those from cosmological simulations such as those of \citet[][]{trapp_gas_2022}, who claim to have gas accretion in the outer disc and important transport effects.
Once looked at in more detail, however, \citet[][]{trapp_gas_2022}'s simulations show strong radial flows only before the accreting gas joins the disc, i.e., when it is still in the CGM.
In-disc motions are on the order of a few km/s, closer to our predictions than to \citet{wang_origin_2022}.

Finally, we can look at radial motions in observations.
As measuring radial inflows at high redshift remains difficult, we restrict ourselves to the present day. \citet{diteodoro_radial_2021} measured radial flows in nearby spiral galaxies and found the presence of both in- and out-flows of up to $20\kms$, but most of their flows are only on the order of a few $\kms$. They concluded that there is no strong evidence for substantial radial inflow to feed the inner star formation in nearby galaxies. This result was challenged by \citet{wang_similar_2023}, who pointed out that radial motions are degenerate with a change in the position angle (warp) and could be missed if this degeneracy is not properly accounted for. Such a degeneracy was recently investigated by He et al. (2026, submitted), who showed that fitting kinematic data simultaneously for PA and radial flows returns general agreement with \citet{diteodoro_radial_2021}.

\subsubsection{CGM kinematics}

Turning to the CGM itself, our fiducial model has $T_\mathrm{CGM} = 0.68\, T_{200}$, which, at the present time, gives $T_{200} = 8.7 \times 10^5$\,K and hence $T_\mathrm{CGM} = 5.9 \times 10^5$\,K. The resulting present-day CGM circular velocity falls from $\approx 160 \kms$ at $4 \kpc$ to $\approx 110\kms$ at $20 \kpc$ so that it lags behind the disc by $\approx 70-95 \kms$ (see \autoref{fig:vrot}). The posterior on $f_\mathrm{temp}$ is, however, broad: its median is $0.82^{+0.44}_{-0.38}$, corresponding to $T_\mathrm{CGM} = 7.1^{+3.8}_{-3.3}\times 10^5$\,K, and $35\%$ of the posterior is supervirial. Even at the upper end of this range, our corona remains cooler than the $\sim 2 \times 10^6$\,K inferred from X-ray observations \citep[e.g.][]{henley_xmmnewton_2013}, which would require $f_\mathrm{temp} \approx 2.3$. Part of this discrepancy may stem from our assumption of an isothermal corona. If the CGM instead exhibits a temperature gradient, as proposed in \citet{miller_constraining_2015}, this could potentially reconcile our models with the observations. We return to this problem in \autoref{sec:varying_the_cgm}.

The CGM kinematics is currently poorly constrained, mostly due to the insufficient spectral resolution of X-ray data.
\citet{hodges-kluck_rotation_2016} measured the Doppler shifts of \ion{O}{VII} absorption lines towards background AGNs and found a best-fitting rotational velocity of $183 \pm 41 \kms$ for an extended halo, disfavouring a static corona and, less significantly, a fully corotating one. Our fiducial model fits this estimate in the inner region, dropping below their $1\sigma$ range outside of about $12 \kpc$.

\citet{marinacci_galactic_2011}, using hydrodynamical simulations, considered a \emph{galactic fountain} scenario, in which cold gas clouds, ejected by the disc as a consequence of supernova feedback, interact with the corona, transferring to the latter part of their own angular momentum, up to a saturation or equilibrium point where the angular momentum transfer stops being efficient and the relative velocity between the two phases is of $\approx 75\kms$.
Following this argument, they estimate a rotation velocity of the corona of about $100-150\kms$, similar to what we find in our fiducial model. 
We emphasise, however, that great caution is required in this comparison. This is because our model constrains the rotation of the CGM \emph{prior} to any interaction with the fountain (see the appendix of \PF), while the estimate of \citet{marinacci_galactic_2011} applies to the corona \emph{after} the angular momentum has been transferred. With this caveat in mind, the numerical agreement between their results and ours may be interpreted as an indication that the corona is, coincidentally, in dynamical equilibrium with the fountain gas already before the interaction, so that no angular momentum exchanges is effectively needed between the two phases.

Finally, we can compare the angular momentum mismatch we recover to the estimates of other GCE models. These usually assume a flat rotation curve, for  which the mismatch is simply given by the ratio of the circular speed of the accreting gas to that of the disc. However, for a realistic rotation curve as used in our model, the quantity that sets the radial flows is instead $\alpha$ (see eq.~\eqref{eq:alpha}). A direct comparison between our model, \citet{bilitewski_radial_2012} and \PF can be found in \autoref{fig:compare_alpha}.

The blue band given for \PF corresponds to the mismatch they derive when comparing their inside-out formation model (see their section 5) to \citet{luck_distribution_2011} Cepheid data, where the lower (upper) bound is set by comparing to the oxygen (magnesium) gradient.
In general, our fiducial model tends to prefer a larger mismatch for $R>4 \kpc$.

The two lines for \citet{bilitewski_radial_2012} come from their two best-fit models: they parametrised the angular momentum mismatch by two parameters such that $\frac{v_\mathrm{CGM}}{v_\mathrm{disc}} = b + a \frac{R}{20 \kpc}$, and then fitted their model to reproduce the iron gradient from \citet{luck_distribution_2011} (note that their model does not assume instantaneous recycling and hence includes a proper treatment of the delayed iron enrichment from SNIa). 
The two models in the plot differ in whether they fit both gradient and absolute abundance (dot-dashed) or just the gradient (dashed) of the CCs, as they note the high uncertainties on the assumed yields and absolute abundance measurements (similar to the point we made in \autoref{sec:free_params_obs}). Interestingly, their less-favoured model (a=-0.25, b=0.86, when fitting also the absolute Cepheid [Fe/H]-abundances shown as a dotted line in \autoref{fig:compare_alpha}) fits our final mismatch profile better than their preferred model. 
They further find that inside-out formation alone does not remove the need for radial flows: for a scalelength growing linearly from $2.8$ to $3.7 \kpc$, the change in average mismatch needed to reproduce the observed gradient is only $\Delta \bar{\alpha}\approx 0.1$, about $20 \kms$. Their gas disc is, however, considerably smaller than ours. Adopting their scalelength prescriptions in our model shows a similar but even smaller difference between the constant and the growing disc, corresponding to a change in the mismatch of $\Delta \bar{\alpha} \approx 0.02$ (however, direct comparisons are difficult due to the different shapes of $\alpha(R)$ between the models).

Both \citet{bilitewski_radial_2012} and \PF assume a mismatch that is constant over time, while in our model it decreases substantially over time and is lowest in the present day. As can be seen from the faint grey lines, the evolution slows down at later times. Our mismatch also has a minimum within the inner $3 \kpc$ and increases monotonically outwards from there. While the radial and temporal shapes hence differ, the magnitude of our recovered mismatch, $\approx 0.1-0.5$ across the disc, is comparable to the values found by \PF and \citet{bilitewski_radial_2012}.
This is likely at least in parts due to the different gradient measurements these authors use: both \PF and \citet{bilitewski_radial_2012} fit their models on Cepheid data from \citet{luck_distribution_2011}, which give an $\alpha$-element gradient between $-0.056 \dex \kpc^{-1}$ (for oxygen) and $-0.048 \dex \kpc^{-1}$ (for magnesium), shallower than the value of $-0.059 \pm 0.012  \dex \kpc^{-1}$ from \citet{mendez-delgado_gradients_2022} we adopt here (also note that the fiducial model slightly overshoots this value and recovers a gradient of $-0.062  \dex \kpc^{-1}$). Additionally, \citet{bilitewski_radial_2012} use a model without the instantaneous recycling approximation and fit on the [Fe/H] gradient instead, which is given at $-0.062 \pm 0.002  \dex \kpc^{-1}$. Since both models are calibrated on the same Cepheid sample, which gives a shallower $\alpha$-element gradient than the one we adopt, it is unsurprising that they recover a somewhat smaller angular momentum mismatch than we find. 

\section{Discussion} 
\label{sec:discussion}

\subsection{Metallicity gradient evolution}
\label{sec:gradient_evolution}
\begin{figure*}
    \centering
    \includegraphics[width=0.9\linewidth]{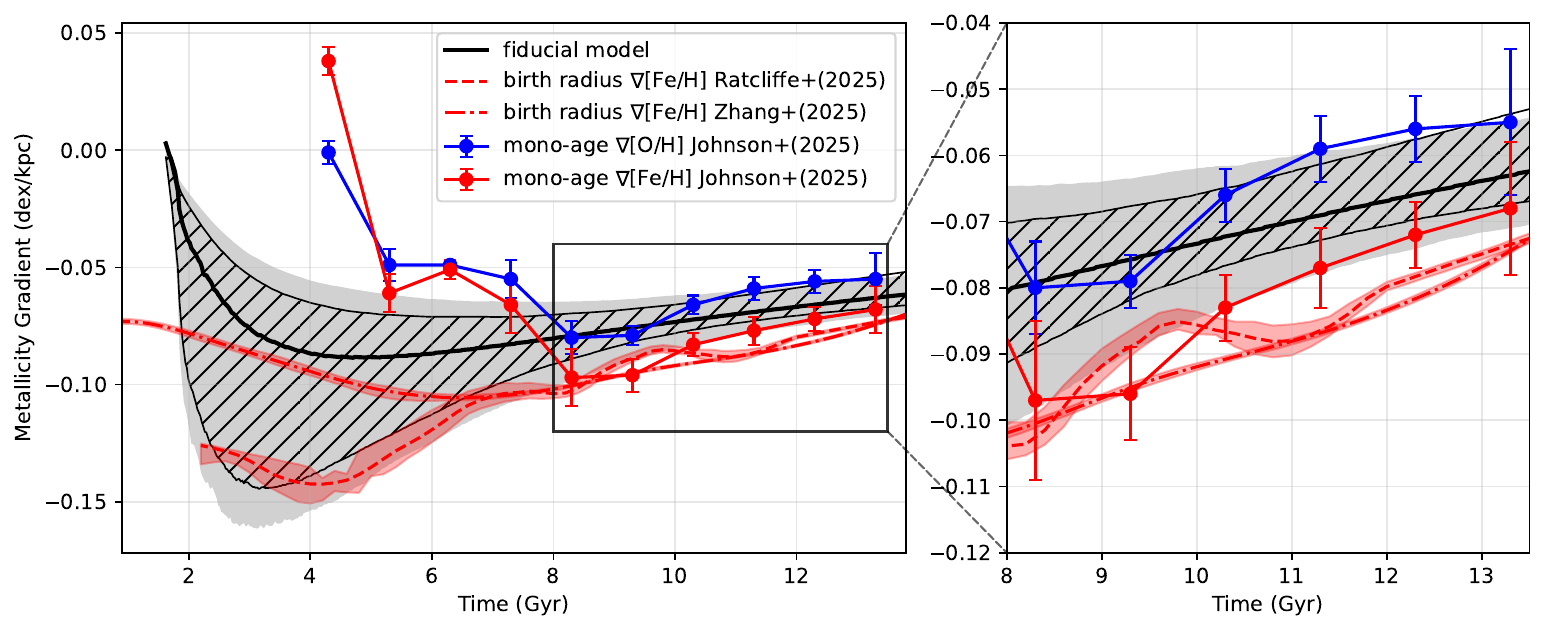}
    \caption{Comparison of the gradient evolution of our fiducial MW model with the gradients of \citet{ratcliffe_evolution_2025} and \citet{zhang_orbital_2025}. We also include the observed gradient evolution from \citet{johnson_milky_2025}. The shaded region here shows the $16-84\%$ quartile from the fiducial fit in \autoref{fig:corner_plot_fiducial} and the hashed region from the fit in \autoref{fig:corner_plot_past_metgradient}, where we are additionally fitting to the mono-age oxygen gradient from \citet{johnson_milky_2025} within the boxed region.   Note that all but the [O/H]-gradient from \citet{johnson_milky_2025}  are iron gradients and hence not directly comparable to our work.}
    \label{fig:metgradient_comparison}
\end{figure*}
\begin{figure}
    \centering
    \includegraphics[width=1\linewidth]{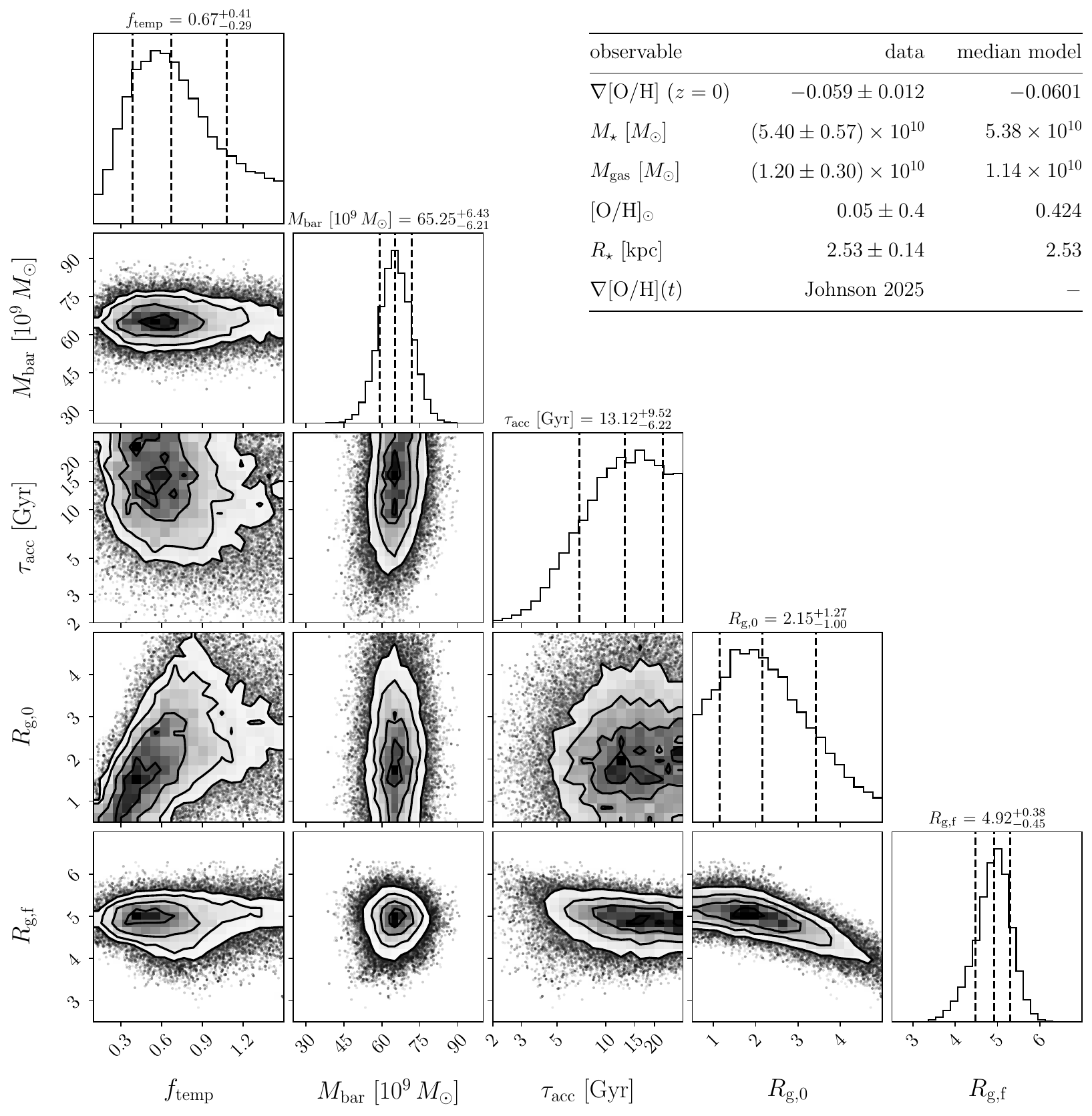}
    \caption{Corner plot with fit to the same data as \autoref{tab:constraints}, but with an additional constraint on the evolution over the past 6 Gyr fit to the [O/H]-gradient observed by \citet{johnson_milky_2025}}
    \label{fig:corner_plot_past_metgradient}
\end{figure}
In the previous section we fitted our model to a set of present-day observables of the Milky Way, i.e. at a single point in time. As shown in  \autoref{fig:parameter_effects}, our free parameters shape the metallicity gradient throughout the evolution of the disc, with $R_\mathrm{g,0}$ affecting mainly the early times. Fitting to the evolution of the gradient, rather than to its final value alone, could hence constrain our parameters better.  As nearly all measurements of the gradient evolution in the Milky Way use [Fe/H], we discuss these here as well, even though they cannot be compared to our model directly.

Measuring this evolution has become possible in recent years mainly through better stellar ages, with the advent of APOGEE \citep{majewski_apache_2017} and of asteroseismic age measurements, though ages remain hard to determine precisely. \citet{anders_red_2017} measured the mono-age gradients using asteroseismic and spectroscopic observations of 418 red-giant stars and found a metallicity gradient that flattens out with stellar age. Present-day mono-age gradients are, however, affected by radial migration, which moves stars away from their birth radius, and they  claim that once this is taken into account, their measurement is consistent with a non-evolving ISM gradient.

Building on these improved age measurements, some studies have recently also tried to undo the effect of radial migration by estimating the birth radii of individual stars, producing birth-time gradients that we can compare to our models directly. \citet{ratcliffe_unveiling_2023} estimated the birth radius of stars from their ages and their iron abundance and found a substantially steeper [Fe/H] gradient with a maximum of $\approx -0.15 \dex\kpc^{-1}$ about 9 Gyr ago (see their Fig.~5). They also derived a magnesium gradient which follows the same evolution but is on average $\approx-0.01$ to $-0.03\dex$ flatter. They re-examined this in \citet{ratcliffe_evolution_2025} using ages from \citet{anders_spectroscopic_2023}, which left the peak of the gradient roughly intact but led to an earlier flattening (their Fig.~5). They attributed the discrepancy between the measured gradient and their inferred gradient fully to radial migration and heating. 
In \autoref{fig:metgradient_comparison}, we plot the evolution of our fiducial MW model (solid black line) compared to this gradient, with the shaded band showing the $16-84\%$ range of the gradient evolution across our posterior. We also plot the iron gradient from \citet{zhang_orbital_2025}, who tried to reverse radial migration by explicitly fitting the heating and migration history of the Galaxy to LAMOST sub-giants in a non-parametric way. Their gradient agrees well with that of \citet{ratcliffe_evolution_2025} over the last 8 Gyr, but flattens out earlier at higher ages.

There are also studies doubting the paradigm that radial migration affects present-day gradients so strongly. Studying open clusters, \citet{carbajo-hijarrubia_occaso_2024} found an evolution of the iron gradient with age directly in the observations, without correcting for migration, reaching $\approx -0.12 \dex\kpc^{-1}$ for clusters older than 3 Gyr (see their Fig.~17).
Additionally, \citet{johnson_milky_2025}, using APOGEE stars with age measurements from two different machine learning algorithms, conclude that measuring the metallicity gradient based on the mode of the metallicity distribution function at different radii is much less affected by radial migration than when using the mean or median. They explain this by a quick, early evolution of the overall metallicity at different radii, up to an equilibrium between mass loaded outflows and star formation. They state that the evolution of the gradient is relatively flat over the last 8-10 Gyr (see their Fig.~4 and Tab.~1).
Looking at the cutout in \autoref{fig:metgradient_comparison}, the [Fe/H] gradient over the last 8 Gyr is remarkably similar between the one measured using the MDF-mode of mono-age groups by \citet{johnson_milky_2025} and that from inferred birth-gradients by \citet{ratcliffe_evolution_2025} and \citet{zhang_orbital_2025}. We also show the fiducial model (in black), where the shaded region gives the $16-84\%$ range of the gradient at each timestep across $1200$ samples drawn from the posterior in \autoref{fig:corner_plot_fiducial}.

Given that our model cannot be directly compared to [Fe/H]-gradients, and that the inferred birth gradients depend both on the adopted age scale and on the migration model used, we can only draw tentative conclusions. First, our models naturally recover the shape of the evolution, with an initial steepening and a flattening out over the last $\approx 8\Gyr$, even though the models do not quite agree on when the turn-over happens. Second, the [Fe/H] gradient measured by \citet{johnson_milky_2025} traces the inferred iron birth gradient well, i.e. their MDF-mode measurement appears to be largely unaffected by radial migration. If the same holds for oxygen, their [O/H] gradient can be compared to our model directly. In fact, our fiducial model, which was fit only to present-day observables, matches the [O/H] gradient of \citet{johnson_milky_2025} well over the last $\sim 8\Gyr$ (see \autoref{fig:metgradient_comparison}).

As a test, we also ran an MCMC fitting the gradient evolution in the last 6 Gyr of our model to the gradient observed in \citet{johnson_milky_2025}, where we weigh the individual data points in the probability function by their total number (6) in order not to overestimate their impact. The results can be found in \autoref{fig:corner_plot_past_metgradient}. The posteriors do not move very much, though the distributions of both $R_\mathrm{g,0}$ and $f_\mathrm{temp}$ become more peaked, effectively excluding a non inside-out forming disc and also disfavouring a supervirial corona. This tightening also shows in \autoref{fig:metgradient_comparison}, where the hatched $16-84\%$ region of this fit is narrower than the shaded one throughout. We also can finally see the expected degeneracy between $R_\mathrm{g,0}$ and $f_\mathrm{temp}$: lower initial scalelengths and hence stronger initial inside-out formation require weaker radial flows
and vice versa to build the same gradient.

\subsection{Mass loading factor}
\label{sec:mass_loading}
\begin{figure}
    \centering
        \includegraphics[width=1\linewidth]{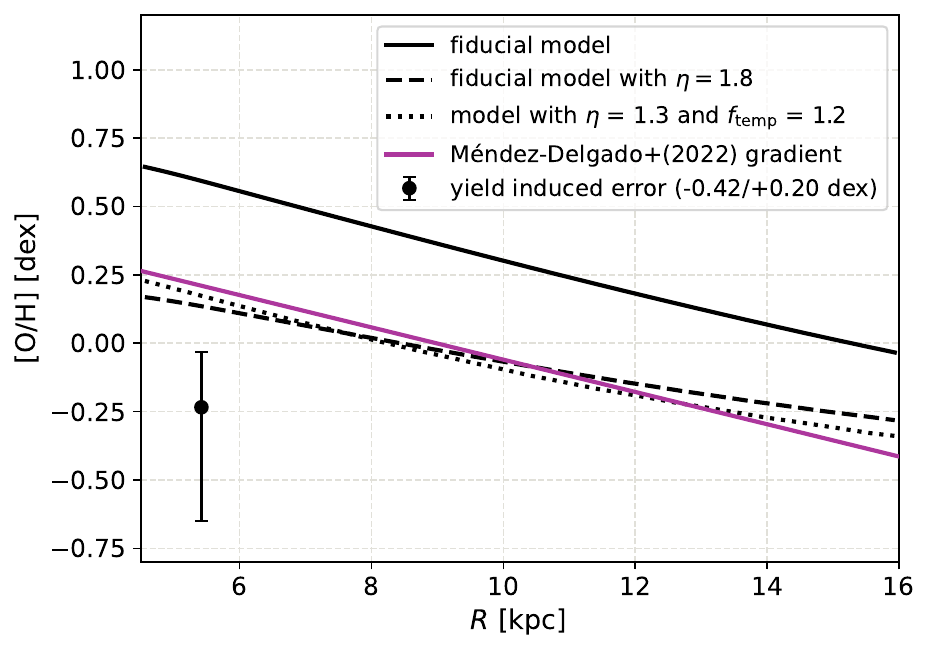}
        \begin{tabular}{p{3cm}cc}
        \toprule
        Model & $\nabla$[O/H] [dex\,kpc$^{-1}$] & [O/H]$_\odot$ \\
        \midrule
        fiducial & $-0.062$ & $0.415$ \\
        $\eta = 1.8$ & $-0.042$ & $0.011$ \\
        $\eta = 1.3$, $f_\mathrm{temp} = 1.2$ & $-0.053$ & $0.002$ \\
        \midrule
        \cite{mendez-delgado_gradients_2022} & $-0.059$ & $0.046$ \\
        \bottomrule
        \end{tabular}
    \caption{Comparison of the metallicity gradients from our fiducial model and two models with a non-zero massloading factor with \ion{H}{II}-regions in the Milky Way and the derived best fit oxygen gradient from \cite{mendez-delgado_gradients_2022}. As in \autoref{fig:hii_region_comparison}, we show the possible offset in [O/H] caused by a different yield and IMF choice in the bottom left corner.
    Below the plot we also show the final gradients of the models and the oxygen abundance at $R=8.2\kpc$ compared to  \cite{mendez-delgado_gradients_2022}.}
    \label{fig:massloading_OH_comparison}
\end{figure}

\begin{figure}
    \centering
    \includegraphics[width=1\linewidth]{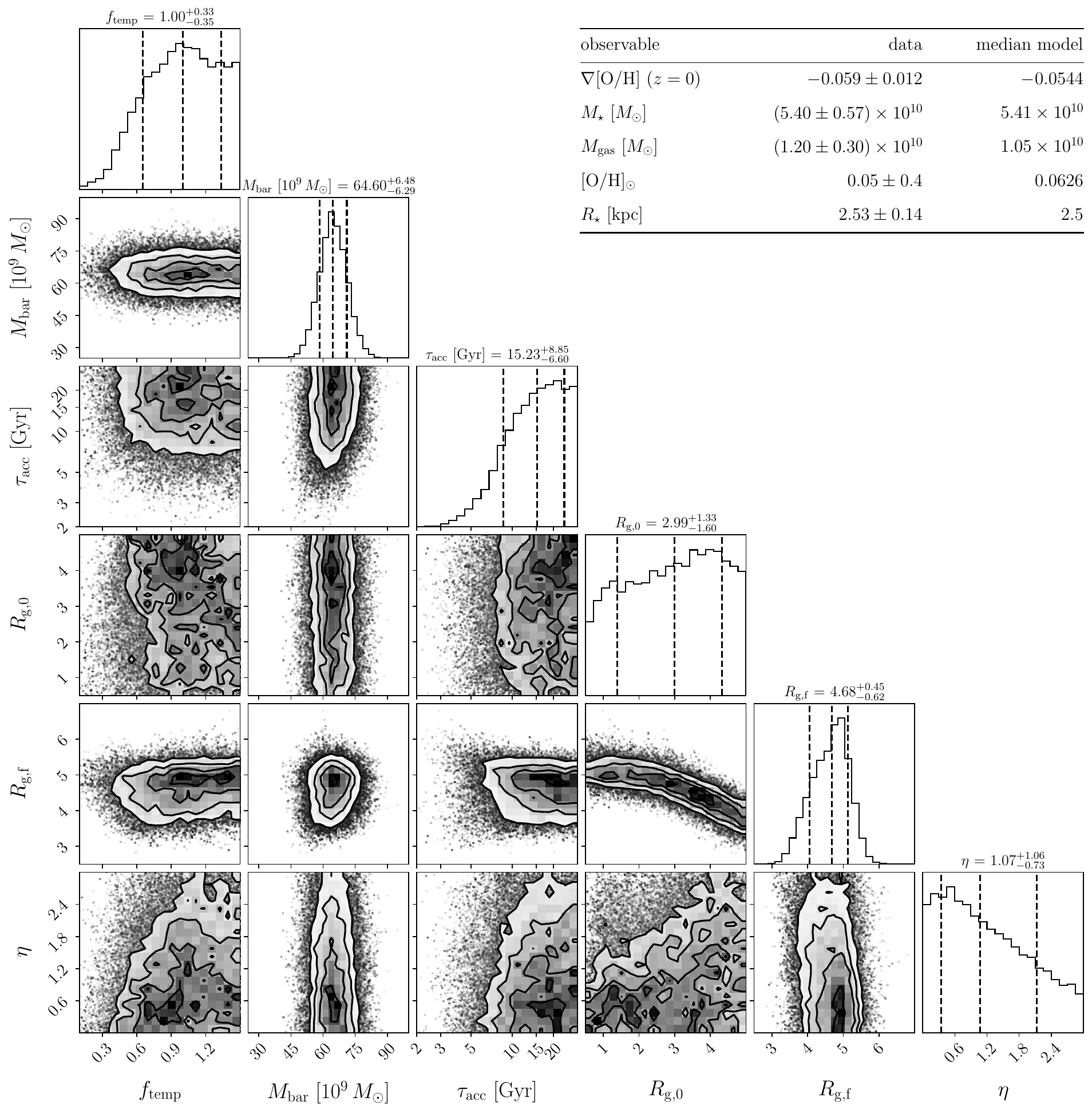}
    \caption{MCMC fit to the same data as in \autoref{tab:constraints}, but adding the mass loading factor $\eta$. }
    \label{fig:corner_plot_eta}
\end{figure}
In \autoref{fig:massloading_OH_comparison}, we show the final metallicity profile of the fiducial model (solid black line), compared to the gradient measured by \citet{mendez-delgado_gradients_2022} in purple. Comparing these two, it looks like the main shortcoming of our model is an overprediction of the [O/H]-abundance. We recover the gradient, but retain too much of the yields. 

A non-zero mass loading factor could rectify this, as it naturally removes gas from the disc. Removing gas does not, by itself, lower the metallicity, if, as in our model, the outflow has the ambient ISM metallicity. However, as our models prescribe the structural evolution of the disc, any mass lost through feedback has to be compensated for by additional accretion, which dilutes the disc gas and lowers the overall metallicity. This can be seen in the dashed line in \autoref{fig:massloading_OH_comparison}, where we plot the fiducial model but with a mass-loading factor $\eta = 1.8$. As expected, this brings the metallicity offset downwards into better agreement with the overall oxygen abundance at the solar annulus. However, it also significantly flattens the gradient (by $0.02\dex\kpc^{-1}$, as seen in the table in \autoref{fig:massloading_OH_comparison}). This is a direct consequence of our feedback prescription, where the outflow is proportional to the star formation rate at an annulus. We can counteract this by strongly increasing $f_\mathrm{temp}$ to super-virial temperatures and effectively bringing the CGM close to non-rotating (see the dotted line in \autoref{fig:massloading_OH_comparison}). However, even then, the resulting gradient is on the flatter side. 

We follow this up in \autoref{fig:corner_plot_eta} where we perform a whole MCMC fit including $\eta$. The model has a preference for a low but non-zero mass loading factor, though with only weak constraints. On the other hand, the posterior on $f_\mathrm{temp}$ moves significantly towards higher values and hence stronger flow; i.e., stronger flows are necessary to build up the observed metallicity gradient for a non-zero mass loading factor. This can also be seen in the degeneracy between high $f_\mathrm{temp}$ and high $\eta$ \autoref{fig:corner_plot_eta}.

Two caveats limit what we can conclude here. The first is that the main effect currently driving the need for feedback is the very high [O/H] around $8.2\kpc$. However, this is also exactly the most uncertain quantity; as can be seen in the bottom left corner of \autoref{fig:massloading_OH_comparison} (and was explained in \autoref{sec:free_params_obs}), the effect of choosing different yields or IMF can drastically change the [O/H]-offset and bring it into agreement with the data. The second is our feedback prescription itself: as the outflows are proportional to the star formation rate, they remove gas preferentially from the inner disc, and hence any feedback always flattens the gradient.

Hence, at this point, we can draw no strong conclusions on feedback; it is not directly needed to reproduce the observables, though it would lessen the tension with the high overall metallicity. In the case of significant outflows, we likely need a different feedback prescription that does not remove as much metal enriched gas from the centre. An option to explore in future work could be a mass loading factor increasing in radius as, e.g., proposed by \citet{johnson_milky_2025}.

\subsection{Extensions of the CGM model}
\label{sec:varying_the_cgm}

In this work, following \citet{pezzulli_angular_2017}, we linked the rotation velocity of the accreting gas to the physical properties (thermal state and angular momentum distribution, or AMD) of the hot CGM (corona), assumed to be in rotating equilibrium in the gravitational potential of the galaxy. In particular, in this first exploration, we made the assumptions that the CGM is isothermal and has an exponential AMD. Both of these assumptions are likely simplifications that are worth investigating.

First, it is often assumed that the AMD follows exactly that of the dark matter, usually described by an exponential \citep{bullock_universal_2001}. However, this does not  take into account alterations due to feedback. 
It has been shown that outflows from supernova feedback or active galactic nuclei can lead to the expulsion of low angular momentum gas from the central regions of discs \citep[e.g.][]{governato_bulgeless_2010}. This gas can then mix with CGM gas with a higher angular momentum \citep[e.g.][]{brook_hierarchical_2011}, leading to a change in the overall shape of the AMD. In extreme cases, low angular momentum gas could be expelled into the IGM, leading to a further change in the shape of the AMD. 
\citet{pezzulli_angular_2017} considered the effect of a feedback-induced alteration of the AMD while addressing a similar problem to what we have encountered in this work: a super-virial corona (as suggested by observations for the Milky Way) has most of its angular momentum stored far away from the galactic disc, requiring large radial flows to bring it where it is needed to sustain star formation. They found that an altered AMD (due to either outflows or mixing) could partially compensate for the high-temperature effect, bringing more high-angular momentum gas towards smaller radii. Including such a modified AMD in our chemical evolution model would therefore be a natural extension of this work and will be implemented in a future study. We emphasise, however, that even with an exponential AMD, we have already found an encouraging result in the required direction: an increase in the outflow loading factor brings our fit to favour higher gas temperatures, in closer agreement with X-ray observations (Section \ref{sec:mass_loading}).

Second, the assumption of an isothermal corona is likely too simplistic as well. \citet{fukugita_massive_2006} argue that the corona has to be hotter than the virial temperature of the dark matter in order to reproduce the observed central densities, and \citet{miller_constraining_2015} found some indications of a slight temperature gradient within the Milky Way corona. In their section 7, \citet{pezzulli_angular_2017} tested this by replacing the isothermal equation of state with a polytropic one, leading to a corona with a supervirial centre and a temperature that decreases with radius. They concluded that a small deviation from the isothermal assumption improves the ability of a supervirial corona to supply the disc with high angular momentum gas, but barely enough to sustain inside-out growth. 
A declining temperature profile could also help in our models: the corona could be supervirial in the centre and hence fulfil the observational constraints, while the gas accreting onto the disc further out is cooler, rotates faster, and hence induces weaker radial flows.  However, it is unclear whether this effect would be strong enough, within the relevant region of the disc, to actually change our results: the temperature would have to drop below $T_{200}$ already within the first few kiloparsecs in order to drive the flows necessary to build up the observed metallicity gradient.

Finally, all the adjustments discussed so far still assume a barotropic halo, i.e., a halo where the pressure and density are stratified along the same surfaces, which, as  described in \autoref{sec:radial_flows}, forces it to rotate cylindrically. The more general solution would be to instead consider a baroclinic equilibrium. In baroclinic solutions, the rotation velocity is allowed to change with the height above the disc, as shown, for instance, by \cite{barnabe_hydrostatic_2006} in the context of the modelling of the extra-planar gas of disc galaxies. A proper treatment of baroclinic equilibrium would let us probe the CGM specifically at the height at which the accretion occurs, as the AMD close to the disc might be different from that further above the plane. In this context, \citet{sormani_models_2018} presented a family of baroclinic solutions with the pressure stratified on elliptical surfaces within a spherically symmetric gravitational potential.

The latter assumption is justified when investigating the overall shape of the corona, but it severely breaks down in the region close to the disc, where the accretion occurs and is therefore crucial for the study of accretion-induced radial flows and chemical evolution. A proper treatment of accretion flows from the condensation of a baroclinic corona therefore requires a dedicated study and we leave it for future work.

\section{Conclusions}
\label{sec:conclusions}
The metallicity gradient of a galaxy carries information about the history of the gas that flows into it and within it. Metals are produced locally, wherever stars form, but they are diluted by the gas accreting onto the disc and advected by the flows within it, so the shape of the gradient reflects both where the accreting gas lands and how it moves afterwards. In this work, we have taken a first step towards disentangling the different mechanisms at play in order to address the long-standing question: what are the properties of the gas that accretes onto a galactic disc? Answering this question opens the possibility of understanding the gas accretion mechanism(s).

For this, we have built and presented a semi-analytic chemical evolution model with a careful treatment of the radial flows that drive the metallicity gradient. We account self-consistently for two sources of such flows: the deepening of the potential well over cosmic time, and the accretion of gas from the corona with a lower local specific angular momentum than the disc.

The first effect was first pointed out by \citet{pitts_chemical_1989} and recently included in a chemical evolution model by \citet{johnson_constraints_2025}, although with a simplified treatment. The second effect has been studied more extensively (e.g.\ \citealt{pezzulli_accretion_2016} and references therein), but always with simple parametric forms of the rotation curve of the accreting gas and no dependence on time.
Here, we build on the treatment presented in \citet{pezzulli_accretion_2016} and add critical improvements. 

First, for the first time, we self-consistently take into account the detailed evolution and spatial distribution of dark matter, stars, and the interstellar medium. 

Second, we build a chemical evolution model where the kinematics of the accreting gas is directly derived from a physical model of the circumgalactic medium. In this  work, we model the CGM as an isothermal corona in rotating equilibrium in the combined, time-evolving, gravitational potential of stars, gas and dark matter. The angular momentum mismatch then follows from the physical state of the corona, most notably its temperature and angular momentum distribution (AMD). Contrary to previous work, we find it to be a strong function of both time and galactocentric distance. As in previous work, we allow the disc to grow inside-out, which we model here as a time-increasing exponential scale-length of the gaseous disc. 

In our fiducial model, the evolving gravitational potential drives slow but non-negligible stellar flows of about $0.2 \kms$ at late times.
In agreement with previous work, our models predict radial gas inflows that increase towards the edge of the disc and weaken over time. Secondly, as expected but never self-consistently shown before, we find that a cooler corona rotates faster, drives weaker radial flows, and hence produces a flatter metallicity gradient.

We set up a Bayesian inference framework and applied it to the Milky Way. We fit the model's free parameters to the observed stellar and gas mass, the extent of the stellar disc and the oxygen gradient of the Galaxy measured from \ion{H}{II} regions. The gradient mainly constrains the temperature of the corona, and hence the kinematics of the accreting gas, while the masses and the disc extent constrain the structural evolution of the disc.
Our models favour a disc growing inside-out at $\sim 0.2 \kpc \Gyr^{-1}$, and have a mild preference for a subvirial corona, although statistically consistent with a supervirial one, with  $T_\mathrm{CGM}/T_{200} = 0.82^{+0.44}_{-0.38}$ (and a formal peak at 0.68) and hence an inferred CGM temperature of $T_\mathrm{CGM} = 7.1^{+3.8}_{-3.3}\times 10^5$\,K.

Our fiducial model does not include outflows and correctly reproduces the observed metallicity gradient and its evolution over the past $\approx 8 \Gyr$, as well as the observed absolute metallicity of the Milky Way, although the latter requires integrated stellar yields on the low side of the theoretically expected range. We also explored a modified version of our model including outflows, which lower the absolute metallicity scale and flatten the abundance gradient. Our median model with outflows reproduces the data with higher stellar yields, a higher temperature for the corona ($T_\mathrm{CGM} = 8.6 ^{+3.0}_{-2.9}\times10^5$ K at the present day), and a moderate mass loading factor ($\eta = 1.01^{+1.04}_{-0.69}$).

The best-fit CGM temperatures we recover are in tension with X-ray measurements \citep[e.g.][]{henley_xmmnewton_2013}, only partially attenuated by the introduction of outflows. This may in part reflect the simplifying assumptions we made about the corona, namely an exponential AMD and an isothermal temperature profile.
Feedback from star formation and an active galactic nucleus could promote both angular momentum mixing (altering the shape of the AMD) and a temperature gradient within the corona, both of which should help bringing our results in better agreement with observations. Including these effects is a natural extension of our model and will be explored in future work.

Our models also do to not reproduce exactly the current observed surface density profile of the gaseous disc of the Milky Way, predicting too much (little) gas at small (large) galactocentric distances. In addition to large systematic observational uncertainties, unavoidably associated to our view from inside of our own Galaxy, these discrepancies could be, in part, due to the fact that we do not model additional gas flows due to the presence of the bar \citep[see e.g.][]{athanassoula_existence_1992} and in part to the fact that we neglect a decline in star formation efficiency in the outer regions as a consequence of the disc flaring \citep{bacchini_volumetric_2019a,bacchini_volumetric_2019}. Both these limitations could be overcome in future work by exploring different prescriptions for inside-out growth and star formation, as well as applying our formalism to external galaxies.

In general, the application of our framework to external galaxies will allow us to profit from better constrained galactic structure, make more general statements about how accretion takes place in galaxy discs and test whether and how it varies with galaxy mass and size.

\begin{acknowledgements}
We would like to thank the FLOWS group in Groningen for useful discussions and Cecilia Bacchini for providing us with the \ion{H}{I}-data shown.
This work has received funding from the European Research Council (ERC) under the Horizon Europe research and innovation programme (Acronym: FLOWS, Grant number: 101096087).
\end{acknowledgements}

\section*{Data availability}
  All data in this work is published and available in the cited references. The code used to produce the models presented here is available from the corresponding author upon reasonable request.

\bibliographystyle{aa}
\bibliography{references}

\appendix
\onecolumn

\section{Additional Information}
\begin{figure}[h]
    \centering
    \includegraphics[width=1\linewidth]{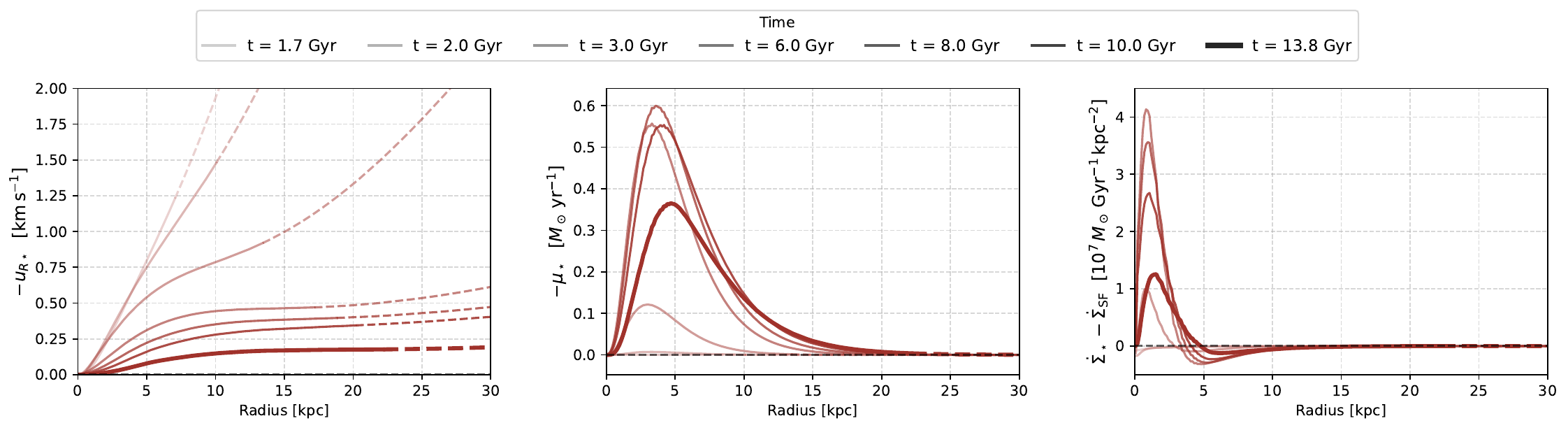}
    \caption{Effect of the redistribution of stars in the centre due to the building up galactic potential (see \autoref{sec:stellar_redistribution}).
    We show from left to right: the inflow velocity of the stars $-u_\mathrm{R,\star}$, the stellar radial mass flux $\mu_\star$ and the change of stellar surface denisty due to this redistribution.\newline
    The model parameters are the same as in \autoref{fig:full_model} and taken from \autoref{tab:model_values}. Again, we switch the linestyle to dashed at $R= 5 R_\mathrm{g}$ to indicate the edge of the disc.}
    \label{fig:stellar_redistribution}
\end{figure}

\begin{figure}[h]
    \centering
    \includegraphics[width=0.8\linewidth]{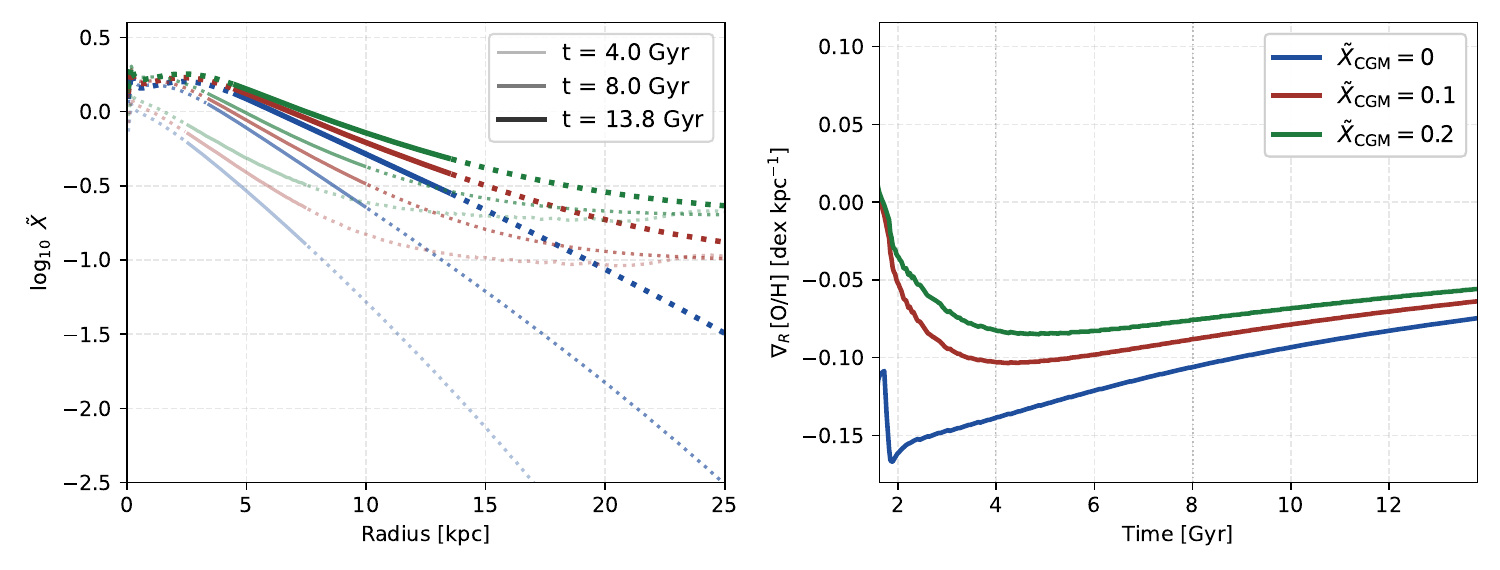}
    \caption{Effect of changing the metallicity of the accreting gas $\tilde{X}_\mathrm{CGM}$. The left plot shows the metallicity vs radius for three times and 3 different values of $\tilde{X}_\mathrm{CGM}$. Here, the lines are dashed outside the window in which we fit the gradient, which is $1-3 R_\mathrm{g}$. On the right we show the resulting effect on the measured gradient over time.}
    \label{fig:xcgm}
\end{figure}

  \begin{table*}[ht]
  \centering
  \caption{Parameters of each model component as we use them for our MCMC fitting. }
  \label{tab:full_model_params}
  \begin{tabular}{lllp{7cm}}
  \toprule
  \textbf{Component} & \textbf{Parameter} & \textbf{Value} & \textbf{Description} \\
  \midrule
  \multirow{6}{*}{Evaluation Grid}
    & $t_{f}$        & $13.8\,\mathrm{Gyr}$   & Final time\\
    & $t_{0}$        & $1.5\,\mathrm{Gyr}$    & Model start time \\
    & $N_t$          & $500$                  & Number of timesteps \\
    & $R_{\max}$     & $50.0\,\mathrm{kpc}$   & Outer radial boundary \\
    & $N_R$          & $400$                  & Number of radial rings \\
    & --             & linear                 & Radial grid spacing \\
  \midrule
  \multirow{5}{*}{Dark matter halo}
    & $M_{200}(t_f)$ & $1.3\times10^{12}\,M_\odot$ & Present-day virial mass \\
    & $\beta$        & $0.1$                  & MAH power-law index, eq.~\eqref{eq:McBrideMass} \citep{mcbride_mass_2009} \\
    & $\gamma$       & $0.7$                  & MAH exponential decay, eq.~\eqref{eq:McBrideMass} \citep{mcbride_mass_2009} \\
    & $\lambda_\mathrm{spin}$ & $0.035$       & Halo spin parameter, eq.~\eqref{eq:j_CGM} \\
    & cosmology      & Planck18               & \citet{planckcollaboration_Planck_2020} \\
  \midrule
  \multirow{3}{*}{Accretion history}
    & $M_\mathrm{g,0}$ & $10^{6}\,M_\odot$    & Initial gas mass, eq.~\eqref{eq:Mgas} \\
    & $M_\mathrm{bar}(t_f)$ & MCMC parameter  & Final baryonic mass, eq.~\eqref{eq:Macc} \\
    & $\tau_\mathrm{acc}$ & MCMC parameter    & Accretion timescale, eq.~\eqref{eq:Macc} \\
  \midrule
  \multirow{4}{*}{Star formation}
    & $n$            & $1.4$                  & Kennicutt--Schmidt index, eq.~\eqref{eq:KSlaw} \citep{kennicuttjr._global_1998} \\
    & $A_\mathrm{KS}$ & $0.1625$ & KS normalisation in $M_\odot$, pc$^2$, Gyr; including Helium correction, eq.~\eqref{eq:KSlaw} \\
    & $\eta$         & fixed to 0 or MCMC parameter   & Mass-loading factor, eq.~\eqref{eq:Mgas} \\
    & $\mathcal{R}$  & $0.3$                  & Instantaneous return fraction, eq.~\eqref{eq:KSlaw} \\
  \midrule
  \multirow{2}{*}{Inside-out growth}
    & $R_{g,0}$      & MCMC parameter                   & Gas scale length at $t_0$, eq.~\eqref{eq:R_gas} \\
    & $R_{g,f}$      & MCMC parameter                   & Gas scale length at $t_f$, eq.~\eqref{eq:R_gas} \\
  \midrule
  \multirow{2}{*}{Galactic potential}
    & $h_{z,\mathrm{gas}}$ & $0.3\,\mathrm{kpc}$ & Scale height of the gas disc, eq.~\eqref{eq:vertical_density} \\
    & $h_{z,\star}$  & $0.3\,\mathrm{kpc}$       & Scale height of the stellar disc, eq.~\eqref{eq:vertical_density} \\
  \midrule
  \multirow{9}{*}{Isothermal CGM}
    & $f_\mathrm{temp}$ & MCMC parameter                & CGM temperature, $T_\mathrm{CGM}/T_{200}$, eq.~\eqref{eq:f_temp} \\
    & $f_\mathrm{j}$ & 1.0        & relative specific CGM angular momentum  $f_\mathrm{j} = \bar{j}_\mathrm{CGM}/\bar{j}_\mathrm{DM}$,
  eq.~\eqref{eq:j_CGM} \\
  & $\bar{j}_\mathrm{max}/\bar{j}_\mathrm{DM}$          & $3.9$              & Relative maximal angular momentum of the CGM\\
      & $R_c$          & $0.1\,\mathrm{kpc}$    & Inner core radius of the CGM solution \\
    & $\mu_\mathrm{mol}$          & $0.59$                 & Mean molecular weight, eq.~\eqref{eq:f_temp} \\
    & $\tilde X_\mathrm{CGM}$ & $0.1$         & Metallicity of the accreted gas (yield-normalised), eq.~\eqref{eq:ODEmetallicity} \\
  \bottomrule 

  \end{tabular}
  \end{table*}

In this appendix, we show some extra material that might be interesting to the reader but is not directly relevant to the conclusions we draw. 

First, in \autoref{fig:stellar_redistribution}, we show the effect of building up the potential and the resulting radial flows on the stellar disc. We see in the left panel that while the flows are small, they are not negligible. Especially at early times, they can reach speeds of up to $1.5\kms$ at the outer edge of the disc and still drive a flow of about $0.2\kms$ across the disc at late times. Especially in the inner $5\kpc$ of the disc, this leads to a significant redistribution of the stars and a peak mass flux of $\sim 0.35\Msun\yr^{-1}$ in the centre even at late times (middle panel). In the right panel, we also show the difference between the total change in stellar surface density and that caused by star formation. This measure also shows the locations in the disc that experience a net increase in stellar surface density ($\sim 0-4\kpc$) and those experiencing a net depletion ($\sim 4-9\kpc$) due to this effect.

Then, figure \autoref{fig:xcgm} plots the final metallicity (left panel) and the evolution of the metallicity gradient over time (right panel) for three different values of $\tilde{X}_\mathrm{CGM}$, the yield-normalised abundance of the CGM (see section \autoref{sec:free_params_obs}). Here, the blue lines, showing $\tilde{X}_\mathrm{CGM}=0$, indicate primordial accreting gas (which leads to a nearly perfect exponential profile, as seen in the left panel), whereas the lines indicating abundances of $\tilde{X}_\mathrm{CGM} = 0.1$ (which is the value adopted in this work) and $\tilde{X}_\mathrm{CGM}=0.2$ have metallicity profiles that flatten towards larger radii.  The evolution of the gradient for the three models happens largely in parallel, especially at late times, when they all lie within $\sim 0.02\dex\kpc^{-1}$ of one another. At early times, the model assuming primordial accreting gas builds up the quickest and also reaches the steepest maximal gradient before flattening out.

Lastly, \autoref{tab:full_model_params} contains a summary of all model parameters that were used to build the models used in this work. Free parameters of the MCMC fit are marked as such in the ``Value'' column. The first section ``Evaluation Grid'' describes the grid on which we perform the majority of our calculations. We use an even and rectangular grid, spaced linearly in time and radius, that extends outwards to $50\kpc$. For the calculation of the full galactic potential (as explained in \autoref{sec:potential}), we impose a boundary condition at $R_{200}$. We hence need to pad the grid until $R_{200}$ for this part of the calculation and trim it back down to $50\kpc$ after the calculation of $j_\mathrm{CGM}$.

\section{Low mass flux suppression}
\label{sec:mu_suppression}

\begin{figure*}[h]
    \centering
    \includegraphics[width=1\textwidth]{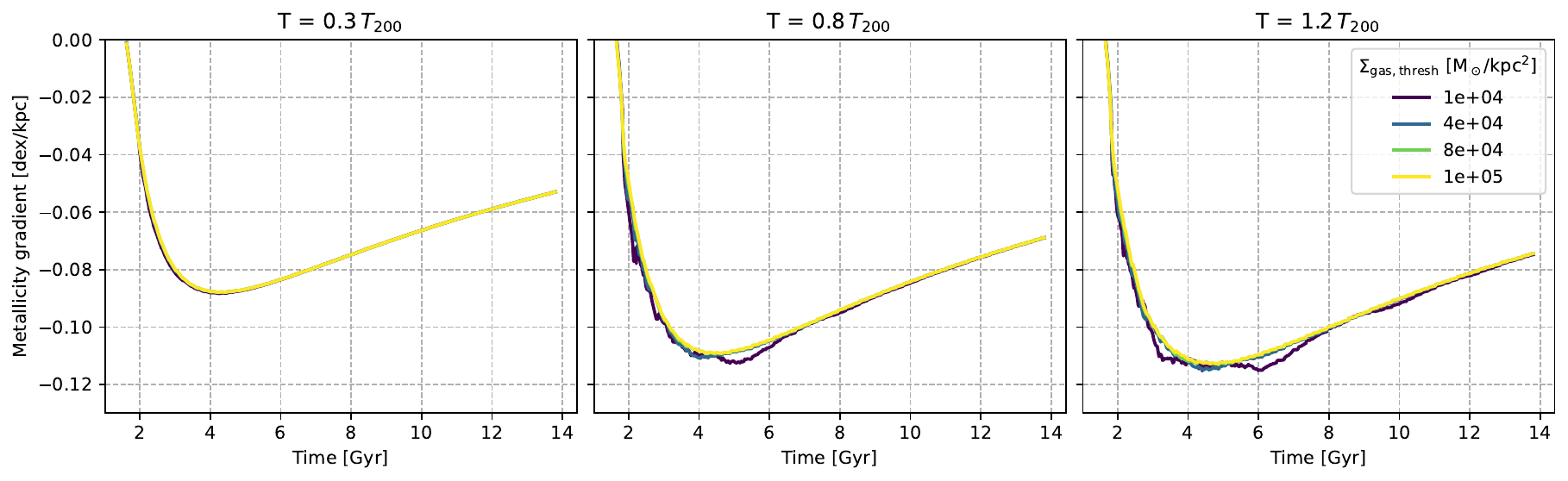}
    \caption{Effect of the low density suppression of $\mu$ on the metallicity gradient over time. The three panels show the effect on different relative CGM temperatures.}
    \label{fig:sig_threshold}
\end{figure*}
\begin{figure}[h]
    \centering
    \includegraphics[width=0.8\linewidth]{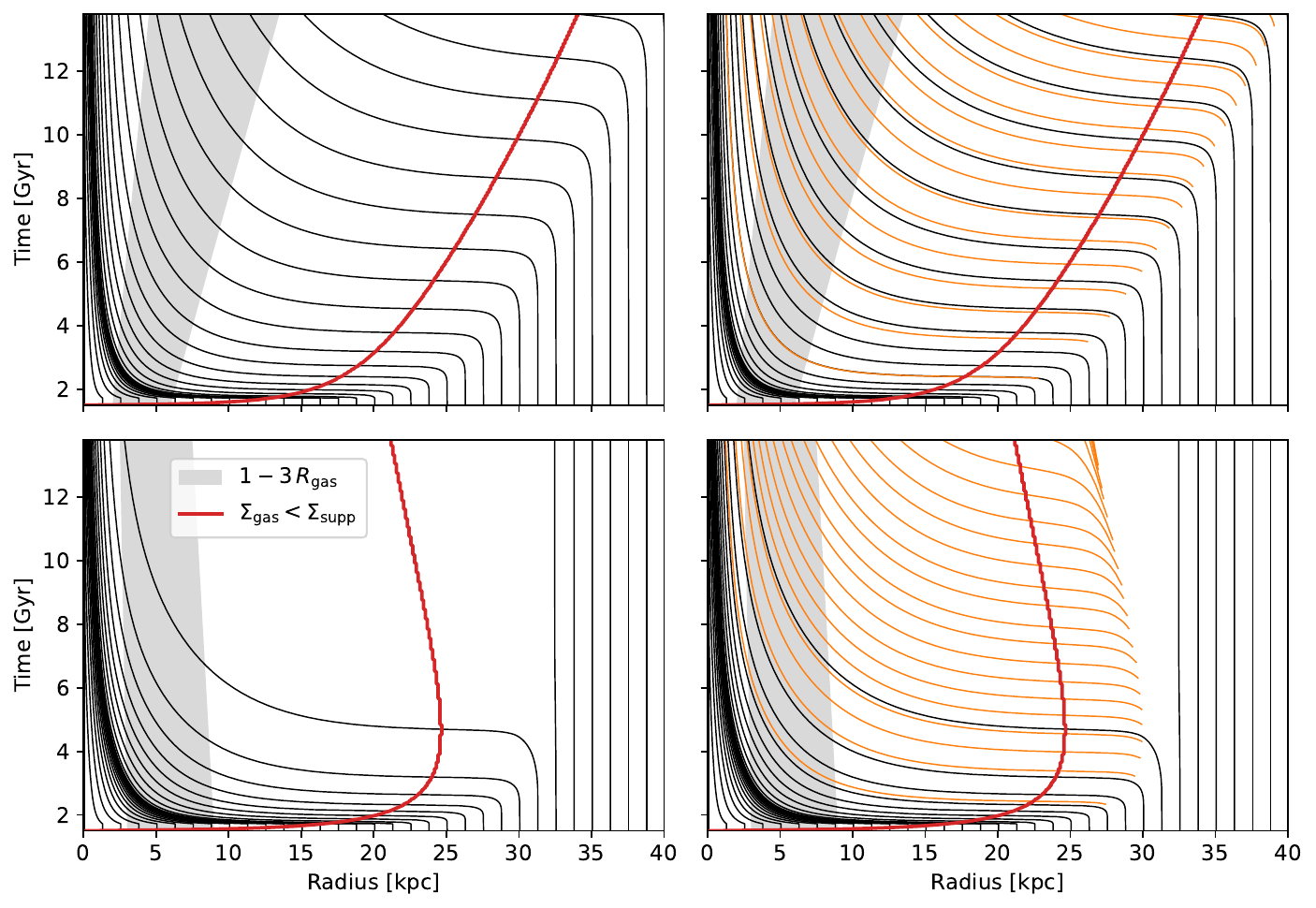}
    \caption{Characteristic lines $R_\mathrm{char}(t)$ (see \autoref{sec:metallicity}) for a growing (top) and shrinking (bottom) disc, without (left) and with (right) the insertion of additional characteristics (in orange). The red line marks $R_\mathrm{start}(t)$, the grey band the region between one and three gas scale lengths within which we measure the metallicity gradient.}
    \label{fig:characteristics_injection}
\end{figure}

    We added a suppression of $\mu_g$ for gas surface density below a certain threshold. This is necessary, as otherwise even the small radial flux in the outskirts, together with a low gas surface density, leads to increasingly large      radial flows. For the models discussed here, this leads to a severe violation of the Courant condition \citep{courant_Uber_1928}: the implied Courant number on our time grid reaches values of order $10^{12}$ in the outer disc. This means that gas there has a radial velocity causing it to cross the entire disc within a fraction of a timestep. The characteristic lines needed in order to calculate the metallicity evolution (see eq.~\eqref{eq:characteristics}) are then swept inwards
      essentially instantaneously.

      We therefore suppress the mass flux once the gas surface density drops below a threshold value of $\Sigma_\mathrm{supp} = 4\times 10^{4}\,\Msun\kpc^{-2}$.
      The suppression starts at
      \begin{equation}
          R_{\mathrm{start}}(t)=\min\left\{R \;:\; \Sigma_{\mathrm{gas}}(t,R')<\Sigma_{\mathrm{supp}}
          \;\;\forall\, R'\geq R\right\},
      \end{equation}
      the innermost radius beyond which the gas surface density stays below the threshold at all larger radii, so that an isolated dip in the gas profile does not trigger the suppression. For the suppression form, we take
      \begin{align*}
          s(t,R)&=\frac{1}{2}\left[1+\tanh\!\left(\frac{R-R_{\mathrm{start}}(t)}{w}\right)\right]\\
          \Delta(t,R)&=w\,\ln\!\left[1+\exp\!\left(\frac{R-R_{\mathrm{start}}(t)}{w}\right)\right],
      \end{align*}
      such that
      \begin{equation*}
        \mu_\mathrm{g,supp}(t,R)=\mu_\mathrm{g}(t,R)\left\{\left(1-s(t,R)\right)+s(t,R)\exp\!\left[-\left(\frac{\Delta(t,R)}{l}\right)^2\right]\right\}.
      \end{equation*}
      Here, $s(t,R)$ blends between the unsuppressed flux inside $R_{\mathrm{start}}$ and the suppressed flux outside, and $\Delta$ is a smoothed version of $\max\left(R-R_{\mathrm{start}},0\right)$. This ensures that the suppression acts only outside $R_{\mathrm{start}}$ while keeping $\mu_\mathrm{g,supp}$
      differentiable everywhere. We use a width of the suppression turn-on of $w = 1.0 \kpc$ (to avoid kinks) and a scale length of the actual suppression $l = 2.5\kpc$.
    
    This regularisation introduces an accretion peak around the threshold radius, effectively making up for the lack of accretion further out. This is more pronounced for models with larger $f_temp$ which drive stronger radial flows. 
    However, this gas comes from regions of barely any star formation (and hence is effectively primordial gas). Additionally, $\dot{\Sigma}_\mathrm{acc}$ is still clearly dominated by central disc accretion (see e.g. \autoref{fig:full_model}).
    Hence, there is only a negligible effect on the final metallicity gradient, as can be seen in \autoref{fig:sig_threshold}, while ensuring that the gradient calculation stays numerically stable. 
    For the fiducial models and MCMC runs, we use a threshold of $4\times 10^4 M_\odot \mathrm{kpc}^{-2}$ which corresponds to a column density of approximately $0.4\times 10^{19} \mathrm{cm}^{-2}$. With that, the threshold reaches never within $5\,R_\mathrm{g}(t)$.
    
    A side effect of this suppression is that reducing the flux at the edge of the disc can split the family of characteristics used to calculate the metallicities into two, where one side gets advected to the centre and the other stays stationary. For models forming inside-out this is not problematic, as the growing disc, and hence growing $R_\mathrm{start}$, causes previously stationary characteristics to flow in and close the gap. However, for non-growing or even shrinking discs, especially in the case of strong radial flows, this can lead to a gap that keeps widening and eventually empties out the whole disc region, causing the metallicity calculation to fail (see the left two panels of \autoref{fig:characteristics_injection}).
    
    To avoid this, we insert additional characteristics whenever $R_\mathrm{start}$ moves less than the width of a radial grid cell in a timestep, placing them at the radius where the suppression factor has dropped to $10^{-2}$ ($5.36\,\kpc$ beyond $R_\mathrm{start}$ for the parameters above). These characteristics carry a starting abundance of $\tilde{X}_\mathrm{CGM}$, matching the ambient gas metallicity. This procedure is effectively the "boundary extension" method discussed in section 4.4 of \PF, where we inject the new  characteristics not at the grid edge but at the radius where the mass flux is vanishing. We show this in the right panels of \autoref{fig:characteristics_injection}: for the growing disc on top, the extra characteristics simply fill in between the existing ones and leave the final gradient unchanged, whereas for the shrinking disc below, they keep the outer disc populated, allowing us to calculate the metallicity profile over the whole disc.

\section{Effects of the growth of the scalelength on the metallicity gradient}
\label{app:scalelength}
\begin{figure}[h]
    \centering
    \includegraphics[width=0.9\linewidth]{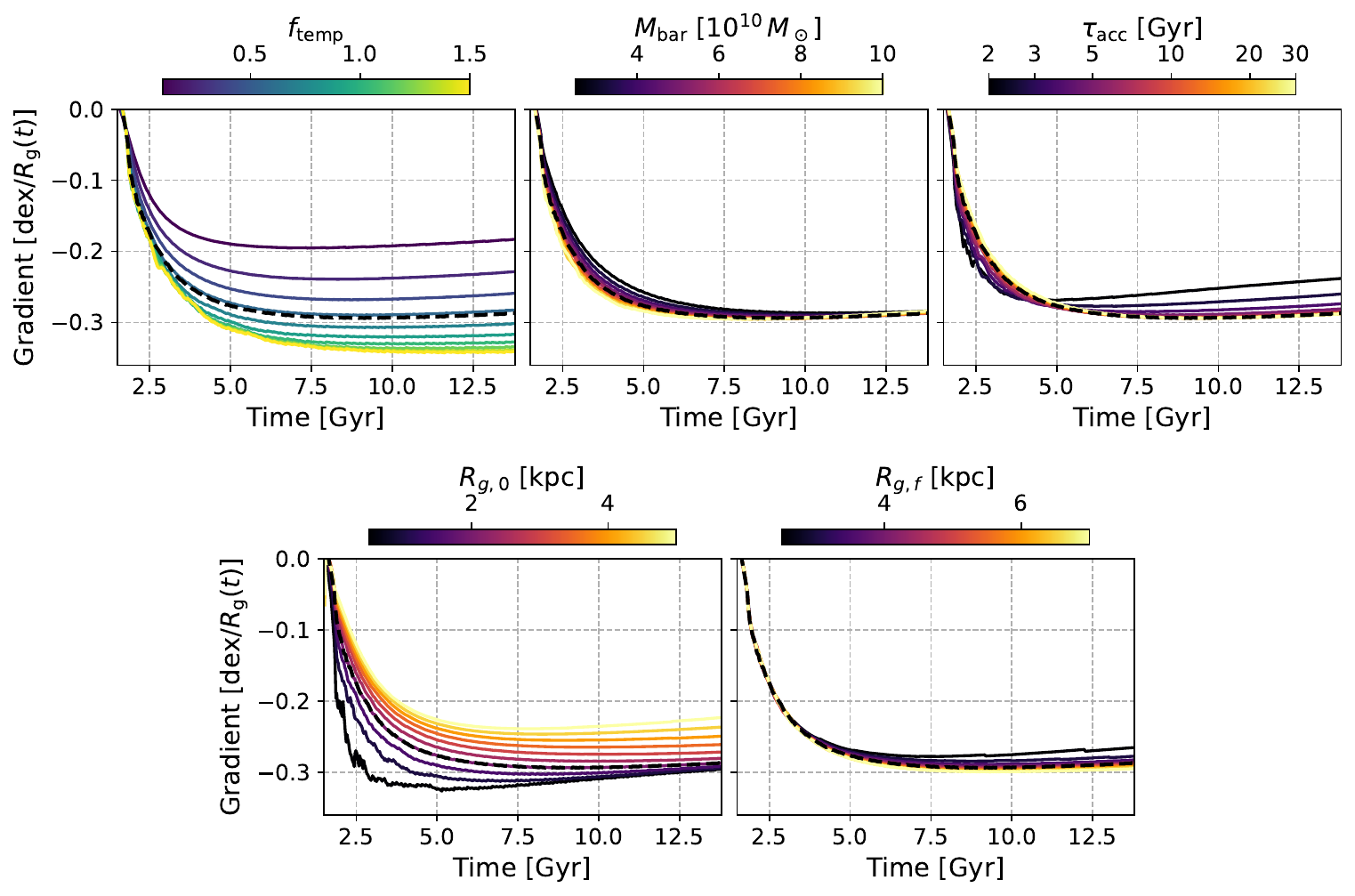}
    \caption{Evolution of the metallicity gradient over time when varying the model free parameters. The plotted models are exactly the same as in \autoref{fig:parameter_effects} with parameters from \autoref{tab:model_values}, except that the gradients are rescaled by the scalelength at that time.}
    \label{fig:gradient_per_scalelength}
\end{figure}

In this section, we provide a brief explanation of why the late-time flattening of the metallicity gradient depends on the growth of the disc. The analytical argument below holds for a disc without radial flows whose metallicity has reached equilibrium. 
While our models satisfy neither condition exactly (except in the very centre of the disc), we can see in \autoref{fig:gradient_per_scalelength} that also the evolution of the gradient in our models is mainly set by the changing scalelength. The one parameter that still changes this evolution at all times is $f_\mathrm{temp}$, in the top left panel of the plot.

In our model, the only radial transport of gas is by the radial flows, which move the characteristics themselves, and there is no exchange of gas between neighbouring characteristics. 
Hence, along each characteristics, the chemical evolution is that of an independent open box model \citep[e.g.][]{tinsley_chemical_1978}, which at each timestep has an equilibrium solution at
\begin{equation}
    \tilde{X}_\mathrm{eq}(t) = \frac{\dot{\Sigma}_\mathrm{SF}}{\dot{\Sigma}_\mathrm{acc}} + \tilde{X}_\mathrm{CGM}
    \label{eq:app_Xeq}
\end{equation}
(derived by setting $\dot{\tilde{X}} = 0$ in eq.~\eqref{eq:ODEmetallicity}). This is also the equilibrium metallicity of the `bathtub' model as introduced by \citet{finlator_origin_2008}.

Let us now define the dimensionless radius
\begin{equation}
    x \coloneqq \frac{R}{R_\mathrm{g}(t)} .
\end{equation}
With our definition of the gas disc in eq.~\eqref{eq:sigma_gas},
\begin{equation}
    \Sigma_\mathrm{g} = \Sigma_0(t)\, e^{-x},
    \qquad \mathrm{with} \quad\Sigma_0(t) \coloneqq \frac{M_\mathrm{g}(t)}{2\pi R_\mathrm{g}^2(t)},
    \label{eq:app_sigma_g}
\end{equation}
so that the star formation rate becomes
\begin{equation}
    \dot{\Sigma}_\mathrm{SF} = A_\mathrm{KS}\,\Sigma_\mathrm{g}^{\,n} = \frac{\Sigma_0(t)}{\tau_\mathrm{dep}(t)}\, e^{-nx}, 
    \qquad \mathrm{with} \quad \tau_\mathrm{dep}(t) \coloneqq \frac{1}{A_\mathrm{KS}\,\Sigma_0^{\,n-1}(t)},
    \label{eq:app_sf}
\end{equation}
where $\tau_\mathrm{dep}$ is the gas depletion time at the centre of the disc.

In the case without radial flows, all gas necessary to sustain the evolution of the gas disc is accreted directly; hence, the effective accretion is exactly equal to the ``actual'' accretion, i.e., eq.~\eqref{eq:sigma_eff} becomes
\begin{equation}
    \dot\Sigma_\mathrm{acc} = \dot\Sigma_\mathrm{eff} = \ell\, \dot\Sigma_\mathrm{SF} + \dot\Sigma_\mathrm{g}, \qquad \mathrm{where} \quad  \ell \coloneqq 1 + \frac{\eta}{1-\mathcal{R}} .
\end{equation}
Taking the time derivative of eq.~\eqref{eq:app_sigma_g} at fixed $R$ gives
\begin{equation}
    \dot{\Sigma}_\mathrm{g} = \left[\frac{1}{\tau_M} + \frac{x-2}{\tau_R}\right] \Sigma_\mathrm{g},
    \qquad \mathrm{with} \quad \tau_M \coloneqq \frac{M_\mathrm{g}}{\dot{M_\mathrm{g}}}, 
    \quad \mathrm{and} \quad
    \quad \tau_R \coloneqq \frac{R_\mathrm{g}}{\dot{R_\mathrm{g}}},
    \label{eq:app_dsigma}
\end{equation}
with $\tau_M$ and $\tau_R$ the growth times of the gas mass and of the scale length, and hence
\begin{equation}
    \dot\Sigma_\mathrm{acc} = \ell\, \frac{\Sigma_0}{\tau_\mathrm{dep}}\, e^{-nx} + \Sigma_0 \left[\frac{1}{\tau_M} + \frac{x-2}{\tau_R}\right] e^{-x}.
    \label{eq:app_acc}
\end{equation}
For our model, the quantities $\Sigma_0$, $\tau_\mathrm{dep}$, $\tau_M$ and $\tau_R$ are functions of time alone and the entire radial dependence of eqs.~\eqref{eq:app_sf} and \eqref{eq:app_acc} is in $x$.

Inserting these into eq.~\eqref{eq:app_Xeq}, both $\Sigma_0/\tau_\mathrm{dep}$ and the factor $e^{-nx}$ cancel, and the equilibrium metallicity can be expressed as
\begin{equation}
    \tilde{X}_\mathrm{eq} = \tilde{X}_\mathrm{CGM} + \left\{ \ell + \left[\frac{\tau_\mathrm{dep}}{\tau_M} + \frac{\tau_\mathrm{dep}}{\tau_R}\left(x - 2\right)\right] e^{(n-1)x} \right\}^{-1},
    \label{eq:app_selfsimilar}
\end{equation}
 which holds no dependence on radius, except in terms of $x$.
The shape of the profile is set by the Kennicutt-Schmidt index $n$, the loss factor $\ell$, and the two ratios of timescales $\tau_\mathrm{dep}/\tau_M$ and $\tau_\mathrm{dep}/\tau_R$; the radius enters only through $x$. The metallicity gradient, which for a model in equilibrium is just the radial derivative of eq.~\eqref{eq:app_selfsimilar}, is therefore
\begin{equation}
    \frac{\partial \tilde{X}_\mathrm{eq}}{\partial R} = \frac{1}{R_\mathrm{g}(t)}\,\frac{\partial \tilde{X}_\mathrm{eq}}{\partial x} .
    \label{eq:app_gradient}
\end{equation}
As the disc grows, $R_\mathrm{g}$ increases and the gradient measured in $\dex\,\kpc^{-1}$ flattens, even though the profile in units of $x$ is unchanged. Note that this is a direct consequence of the self-similarity of the gas disc and holds for any scalelength prescription.

\end{document}